\documentclass[12pt]{article}
\pdfoutput=1

\usepackage{cite}
\usepackage{epsfig}
\usepackage{amsmath}
\usepackage{amssymb}
\usepackage{cancel}
\usepackage{pstricks}
\usepackage{afterpage}
\usepackage{color}

\def\mathswitch#1{\relax\ifmmode#1\else$#1$\fi}
\newcommand{\msbar}{\mathswitch{\overline{\text{MS}}}\ }
\newcommand{\tev}{\,\, \mathrm{TeV}}
\newcommand{\gev}{\,\, \mathrm{GeV}}
\newcommand\g{\gamma}

\newcommand\p{\partial}
\newcommand\inte{\frac{1}{2}\int_{-\pi R}^{\pi R} d x^5\,}
\newcommand\summ{\sum_{n=1}^{\infty}}
\newcommand\summt{\sum_{m,n,l=1}^{\infty}}
\newcommand\summf{\sum_{m,n,l,k=1}^{\infty}}

\newcommand{\mycaption}[1]{\caption{\sl #1}}

\makeatletter
\def\section{\@startsection {section}{1}{\z@}{+3.0ex plus +1ex minus
  +.2ex}{2.3ex plus .2ex}{\large\bf\boldmath}}
\def\subsection{\@startsection{subsection}{2}{\z@}{+2.5ex plus +1ex
minus +.2ex}{1.5ex plus .2ex}{\normalsize\bf\boldmath}}
\def\subsubsection{\@startsection{subsubsection}{3}{\z@}{+3.25ex plus
 +1ex minus +.2ex}{1.5ex plus .2ex}{\normalsize\it}}

\graphicspath{{.}{plots/}}

\begin{document}
\thispagestyle{empty}

\def\thefootnote{\fnsymbol{footnote}}

\begin{flushright}
\end{flushright}

\vspace{1cm}

\begin{center}

{\Large {\bf Radiative corrections to masses and couplings in Universal Extra
Dimensions}}
\\[3.5em]
{\large
Ayres~Freitas$^1$, Kyoungchul Kong$^2$ and Daniel Wiegand$^1$
}

\end{center}

\vspace*{1cm}

{\sl\noindent
$^1$ Pittsburgh Particle-physics Astro-physics \& Cosmology Center
(PITT-PACC),\\ \phantom{$^1$} Department of Physics \& Astronomy, University of Pittsburgh,
Pittsburgh, PA 15260, USA\\[1em]
$^2$ Department of Physics and Astronomy, University of Kansas, Lawrence, KS 66045, USA
}

\vspace*{2.5cm}

\begin{abstract}

Models with an orbifolded universal extra dimension receive important
loop-induced corrections to the masses and couplings of Kaluza-Klein (KK)
particles. The dominant contributions stem from so-called boundary terms which
violate KK number. Previously, only the parts of these boundary terms
proportional to $\ln(\Lambda R)$ have been computed, where $R$ is the radius of
the extra dimension and $\Lambda$ is cut-off scale. However, for typical values 
of $\Lambda R \sim 10 \cdots 50$, the logarithms are not particularly large and
non-logarithmic contributions may be numerically important. In this paper, these
remaining finite terms are computed and their phenomenological impact is
discussed. It is shown that the finite terms have a significant impact on the KK
mass spectrum. Furthermore, one finds new KK-number violating interactions that
do not depend on $\ln(\Lambda R)$ but nevertheless are non-zero. These lead to
new production and decay channels for level-2 KK particles at colliders.

\end{abstract}

\setcounter{page}{0}
\setcounter{footnote}{0}

\newpage


\section{Introduction}

\noindent Universal extra dimensions is an attractive concept for physics
beyond the Standard Model (SM), which introduces one or several space-like extra
dimensions that are constrained to a compact volume with periodic boundary
conditions. All fields of the theory can propagate in the extra dimension(s),
and upon compactification they can be decomposed into a tower of Kaluza-Klein
(KK) excitations with increasing mass. In this article, we focus on the minimal
universal extra dimension model (MUED), which introduces one extra dimension
compactified on a circle with radius $R$ but assumes that there are 
no additional operators generated by the UV completion.

An important feature of universal extra dimensions is the
existence of KK parity, which is a remnant of the extra-dimensional Lorentz
symmetry. It helps to satisfy constraints from electroweak precision data and
other low-energy measurements \cite{ued} and leads to the existence of a dark
matter candidate, the lightest level-1 KK particle \cite{ueddm,lev2dma,lev2dm}.

Similarly, KK parity forbids the single production of level-1 KK excitation at
colliders and their decay into SM particles. Instead, they can be produced in
pairs and lead to characteristic missing-momentum signatures from stable dark
matter particles escaping the detector \cite{cms2,uedpheno,Cembranos:2006gt,kkgluon}.
These signatures are reminiscent of supersymmetry (SUSY), so that SUSY-like
analysis techniques  can be used to search for MUED at the LHC. From existing
LHC data, a limit of  $R^{-1} \gtrsim 1.4\tev$ has been derived \cite{uedlhc}.
On the other hand, level-2 KK
excitations can be singly produced through loop-induced vertices
\cite{cms} and decay back into pairs of SM particles. These vertices violate KK
number, but conserve KK parity. As a result, level-2 particles can be
searched for as narrow resonances at colliders \cite{uedlevel2}.

At tree-level, the compactification of the extra dimension(s) generates
almost identical masses for all particles of the same KK level, up to relatively small effects from electroweak symmetry breaking (EWSB). However, loop effects
lead to mass corrections \cite{uedmass} that produce a mildly hierarchical
particle spectrum at each KK level \cite{cms}. Thus these corrections are very
important for understanding the collider and dark matter phenomenology of MUED.
However, in existing calculations \cite{uedmass,cms}, only the corrections
proportional to $\ln(\Lambda/m_n)$ have been computed, where $m_n$ is the
tree-level mass of KK level $n$ and $\Lambda$ is a cut-off scale.
The appearance of $\Lambda$ follows from the fact that the mass corrections are
divergent and must be renormalized. To avoid the regime where the model becomes
strongly interacting, it is usually assumed that $\Lambda < 50\,R^{-1}$
\cite{uedcutoff,ued}.
Therefore, one obtains the bound $\ln(\Lambda/m_n) < 4$,
so that the logarithms are not parametrically large and non-logarithmic
contributions may be numerically important.

The goal of this article is to compute these finite, non-logarithmic terms and study
their phenomenological impact. Even relatively small corrections could have an
impact on the KK mass hierarchy and open or close the phase space for certain
decay channels. There is some level of ambiguity in the definition of the
non-logarithmic part due to the need for renormalizing the divergences in the
mass corrections. In this work, the \msbar scheme is chosen as a well-defined
prescription for this purpose. It corresponds to the assumption that no
Lorentz-symmetry breaking mass terms are generated by the unspecified UV
dynamics at the scale $\mu=\Lambda$, 
{\it i.e.,} there are no boundary localized terms at $\mu=\Lambda$.

In the same vein, for the loop-induced couplings between level-2 KK excitations
and SM (level-0) particles, only terms proportional to $\ln(\Lambda R)$ have
been known so far \cite{cms,lev2dm}. By also including the non-logarithmic
contributions, the effective strength of these couplings can get modified
significantly. More importantly, one can identify new loop-induced couplings
that are independent of $\ln(\Lambda R)$ but non-zero. These lead to new
production and decay channels for various level-2 KK particles, as will be
discussed below.

The paper is organized as follows: After a brief review of MUED and the required
notation in section~\ref{sc:ued}, the calculation of the KK mass corrections and
the KK-number violating vertices is discussed in sections~\ref{sc:mass} and
\ref{sc:int}, respectively. In section~\ref{sc:pheno}, numerical results for the
new corrections are shown and their phenomenological impact for KK particle
production and decay is discussed. Our conclusions are presented in
section~\ref{sc:concl}. The appendix contains detailed formulae for the
tree-level interactions of the KK particles, which are used as an input to our
calculations, and an explicit list of the KK-number violating interactions for
the different fields in MUED.


\section{Brief Review of MUED}
\label{sc:ued}

\noindent
The Universal Extra Dimension scenario \cite{ued} postulates an extension of the
SM, where all fields are permitted to propagate in some number of compact flat
space-like extra dimensions. For a review, see $e.\,g.$ Ref.~\cite{uedreview}.
The minimal model has one extra dimension, which is compactified on a circle
$S^1$ with radius $R$. For energies not too far above $R^{-1}$, this model can
be described by a four-dimensional (4D) theory where each field has a Kaluza-Klein
tower with masses $m_n = n/R$. Here $n$ is called Kaluza-Klein (KK) number and
is a conserved quantum number. The zero modes ($n=0$) are identified with the SM
particles.

However, to accommodate chiral fermion zero modes, an additional breaking of the
5D Lorentz invariance is necessary. The minimal choice for this
purpose is the introduction of orbifold boundary conditions. Specifically, the
Lagrangian of the theory is required to be invariant under the $\mathbb{Z}_2$
transformation $x^5 \to -x^5$, where the index 5 denotes the extra dimension.
Left- and right-handed fermion components can be even or odd under this
transformation, but only $\mathbb{Z}_2$-even fields have a zero mode. The field
content of the 5D SM extension is summarized in Tab.~\ref{tab:mued}.

\begin{table}[t]
\renewcommand{\arraystretch}{1.2}
\centering
\begin{tabular}{|c|ccc|c|}
\hline
Field & SU(3)$_{\rm C}$ & SU(2)$_{\rm L}$ & U(1)$_{\rm Y}$ & $\mathbb{Z}_2$ \\
\hline
$G^M \equiv (G^\mu,G^5)$ & adj. & -- & -- & $(+,-)$ \\
$W^M \equiv (W^\mu,W^5)$ & -- & adj. & -- & $(+,-)$ \\
$B^M \equiv (B^\mu,B^5)$ & -- & -- & adj. & $(+,-)$ \\
\hline
$(Q_{\rm L},Q_{\rm R})$ & {\bf 3} & {\bf 2} & $-1/6$ & $(+,-)$ \\
$(u_{\rm L},u_{\rm R})$ & {\bf 3} & -- & $+2/3$ & $(-,+)$ \\
$(d_{\rm L},d_{\rm R})$ & {\bf 3} & -- & $-1/3$ & $(-,+)$ \\
$(L_{\rm L},L_{\rm R})$ & -- & {\bf 2} & $-1/2$ & $(+,-)$ \\
$(e_{\rm L},e_{\rm R})$ & -- & -- & $-1$ & $(-,+)$ \\
\hline
$H$ & -- & {\bf 2} & $+1/2$ & $+$ \\
\hline
\end{tabular}
\mycaption{Field content of the minimal universal extra dimension (MUED) model and their gauge and
$\mathbb{Z}_2$ quantum numbers. The Latin index $M$ indicates a 5D Lorentz
index, $M=0,1,2,3;5$.
\label{tab:mued}}
\end{table}

The orbifolding leads to a breaking of KK number through loop-induced boundary
terms at the fixed points $x^5 = 0,\, \pi R$ \cite{uedmass}. Nevertheless, a
subgroup called KK parity, which corresponds to even/odd KK numbers, is
still conserved.

Since extra dimension theories are non-renormalizable, there can be additional
operators generated at a cut-off scale $\Lambda$. These are typically small,
since the cut-off scale is estimated to be $\Lambda \sim \text{few} \times 10
\,R^{-1}$ \cite{uedcutoff,ued}. However, in general, the list of UV-induced operators could
also include boundary-localized KK-number violating interactions, and since they
compete with loop-induced boundary terms, they can be phenomenologically
relevant \cite{nmued}. In the MUED scenario it is assumed that KK number is approximately
conserved by the UV theory, so that the UV boundary terms are negligible.

From the usual SM interactions, one can construct the MUED Lagrangian and
Feynman rules, see $e.\,g.$ Refs,~\cite{uedreview,uedcomphep,Belyaev:2012ai}.
For practical calculations, one also needs to introduce gauge-fixing and ghost
terms for the 5D gauge fields $V^M$ ($V=G,W,B$). In this work, a covariant
gauge fixing is employed, which has the form
\begin{align}
{\cal L}_{\rm gf} &= \frac{1}{2} \int_{-\pi R}^{\pi R} dx^5 \,
 \biggl [ -\frac{1}{2\xi} \bigl (\partial^\mu V^a_\mu - \xi\, \partial_5 V^a_5
 \bigr )^2 \biggr
 ]\,, \\
{\cal L}_{\rm ghost} &= \frac{1}{2} \int_{-\pi R}^{\pi R} dx^5 \,
 \Bigl [ \bar{c}^a \bigl (-\partial^\mu \partial_\mu + \xi\, \partial_5^2 
 \bigr ) c^a +
 g^{(5)} f^{abc} \bigl (-\partial^\mu \bar{c}^a G^c_\mu c^b + 
 \xi \, \partial_5 \bar{c}^a G^c_5 c^b \bigr ) \Bigr ] \,.
\end{align}
where $c^a$ and $\bar{c}^a$ are 5D ghost and anti-ghost fields, respectively, 
$a,b,c$ are adjoint gauge indices, $f^{abc}$ is the non-Abelian structure constant,
and $g^{(5)}$ denotes the 5D gauge coupling. For simplicity, the choice
$\xi\equiv 1$ for the gauge parameter is used throughout this paper.

A summary of the relevant 4D Lagrangian terms for the work presented here is
given in Appendix~\ref{sc:feynr}.


\section{Mass corrections}
\label{sc:mass}

The KK modes of fields propagating in a compactified extra dimension receive
loop-induced corrections to the basic geometric mass relation $m_n = n/R$. These
corrections stem from contributions where the loop propagators wrap around the
extra dimension. They can be separated into two categories: bulk and boundary
mass corrections \cite{uedmass,cms}.

While the bulk corrections are present in extra-dimensional models with and without
orbifolding, and they lead to mass terms that are independent of $x^5$ and
conserve KK number, the boundary terms are a consequence of the orbifolding condition, and they lead
to mass terms that are localized in the extra dimension. For example, for a
scalar field, the boundary mass correction has the form
\begin{align}
{\cal L} \supset \frac{\pi R}{2} \bigl [ \delta(x^5) + \delta(x^5+\pi R) \bigr ]
 \Phi^\dagger \partial_5^2 \Phi\,.
\end{align}
These boundary terms break KK number.

Both the bulk and boundary corrections are induced first at one-loop order.
Whereas the bulk corrections are UV finite, the boundary contributions are UV
divergent and must be renormalized. We employ \msbar renormalization for this
purpose, with the \msbar scale set to the cut-off scale $\Lambda$.

\subsection{Approach}

\begin{figure}[t]
\centering
\includegraphics[width=5.5in]{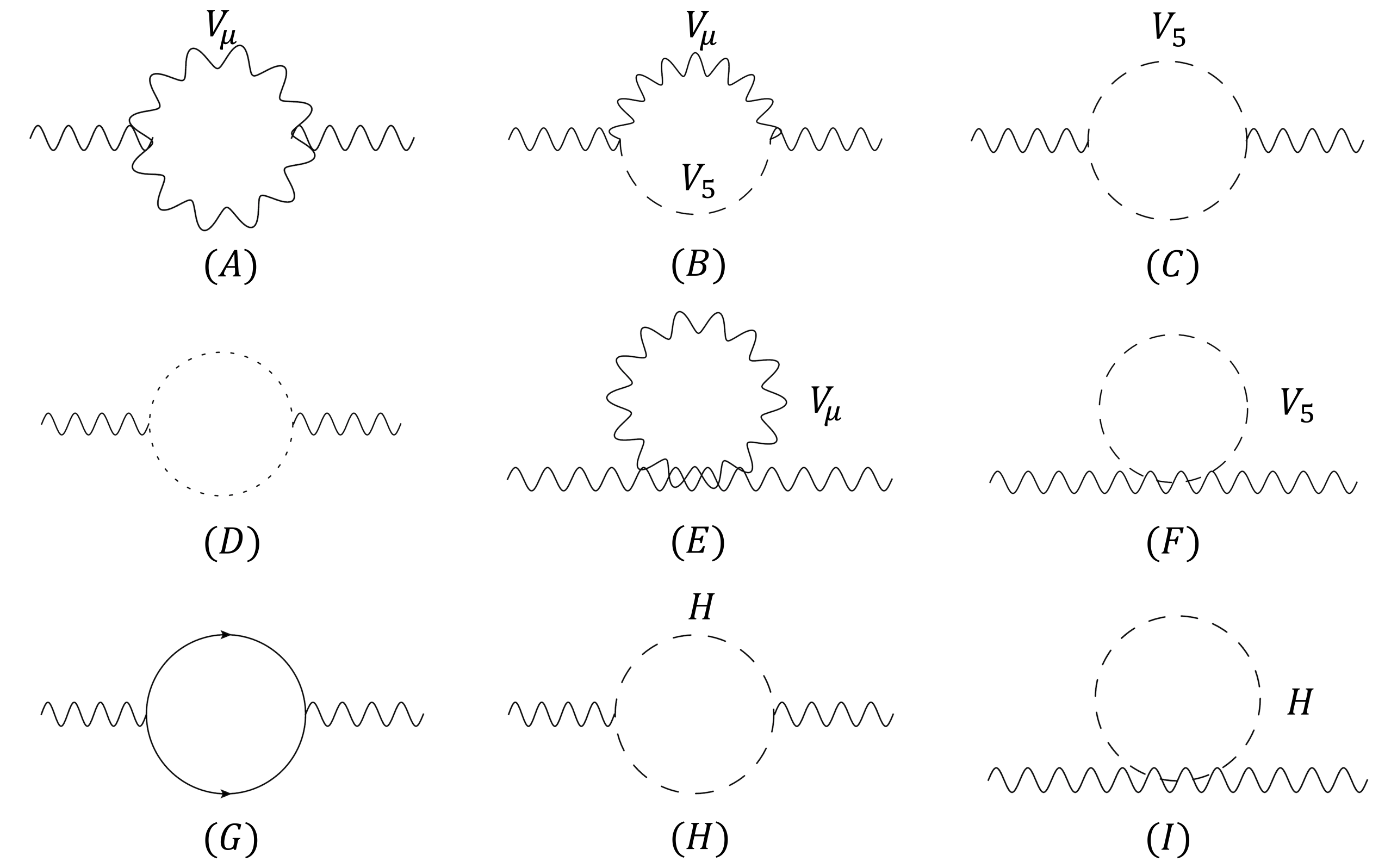}
\vspace{-1ex}
\mycaption{Generic Feynman diagrams for the one-loop contributions to the
self-energies of KK vector bosons. Here wavy, dashed, dotted and solid lines
indicate the KK modes of vector bosons, scalars, ghosts and fermions,
respectively. $V_5$ denotes the scalar degree of
freedom from the fifth component of a 5D gauge field, whereas $H$ stands for
the contribution from a genuine 5D scalar field. 
\label{fig:vv}}
\end{figure}

\noindent
To compute the mass
corrections we choose to work in the effective 4D theory. Every self-energy
diagram contributing to the corrections of a given mode contains the infinite
tower of increasingly heavy KK-modes running in the loop and needs to be treated
in a manner similar to the one described in Ref.~\cite{cms}. We will begin by
describing the general procedure employed on the example of diagram $(A)$ in
Fig.~\ref{fig:vv}.

For a vector boson the self-energy can be decomposed into covariants
according to
\begin{equation}
\Pi^{ab}_{\mu\nu}(p^2)=-\left[g_{\mu\nu}\left(p^2\Pi_{\rm T}^{(1)}{(p^2)}
 +M_n^2\Pi_{\rm T}^{(2)}{(p^2)}\right)-p_\mu p_\nu \Pi_2(p^2)\right]\delta^{ab}\,,
\end{equation}
where additionally to the usual transverse part $\Pi_{\rm T}^{(1)}$ a second
contribution  $\Pi_{\rm T}^{(2)}$ emerges as a proportionality constant to the fifth
momentum component.
The mass correction is then defined as 
\begin{equation}
\delta M^2_{V_n}=-M_n^2\;\Re\text{e}\left\{\Pi_{\rm T}^{(1)}{(M_n^2)}
 + \Pi_{\rm T}^{(2)}{(M_n^2)}\right\}. \label{eq:dmv}
\end{equation}
Both the self-energy and the resulting mass correction are dependent on an
infinite sum over all heavy modes running in the loop, which in turn is
divergent. To find a sensible regularization scheme we first make use of the
Poisson summation identity
\begin{equation}
\sum_{n=-\infty}^{\infty}f(n)
 = \sum_{k=-\infty}^{\infty}\mathcal{F}\left\{f\right\}(k)
\end{equation}
where the Fourier transform $\mathcal{F}$ is defined as
\begin{equation}
\mathcal{F}\left\{f\right\}(k) 
 = \int_{-\infty}^{\infty}{dx \; e^{-2\pi ikx}f(x)}.
\end{equation}
By applying the identity to the divergent mass correction, the sum over
KK-numbers is transformed into a sum of winding numbers in position space about
the fifth dimension. The most straightforward way to define a physical observable
is to subtract the (formally infinite) contribution of the zero winding number
from the sum, since it is equivalent to the diagram in the 5D uncompactified
theory.

For our example we restrict ourselves to the mass correction of the first vector
mode; all higher modes can be found by rescaling the mode number. The example
diagram then amounts to a mass correction described by
\begin{align}
\delta M^2_{V_1} =\, &\frac{g^2}{32\pi^2}C_A\frac{34-41d+11d^2}{(d-3)(d-1)}\,A_0(M^2)\nonumber\\
&+\frac{g^2}{32\pi^2(d-1)}\,C_A\sum_{n=1}^{\infty}\Big\{\left[n(3-2d)+d-1\right]\,A_0(n^2M^2)\nonumber\\
&-\left[(n+1)(3-2d)-d+1\right]\,A_0\left((n+1)^2M^2\right)\nonumber\\
&+M^2(d-1)(5+2n+2n^2)\,B_0\left(M^2,n^2M^2,(n+1)^2M^2\right)\Big\}\,,
\end{align}
where $d=4-2\epsilon$ is the number of space-time dimensions in dimensional regularization,
and the term outside the sum stems from the diagram with a zero mode (massless
SM vector boson) in the loop.

The explicit form of the $A_0$ and $B_0$ functions appearing in the equation are
well-known and can be written as
\begin{align}
A_0(M^2) &= M^2 \left( \frac{1}{\epsilon}+1-\ln\frac{M^2}{\mu^2} \right) , \\
B_0\left(M^2,\left(n+1\right)^2M^2,n^2M^2\right)
 &= \frac{1}{\epsilon}+2-\ln\frac{M^2}{\mu^2}+2\left[n\log\left({n}\right)-(n+1)\log\left({n+1}\right)\right].
\end{align}
Splitting up the first term in order to extend the sum to include $n=0$ and
taking the the limit $d\to 4$ yields
\begin{equation}
\delta M^2_{V_1} = \frac{23 g^2}{96\pi^2}C_A M^2\left(\frac{112}{69}
 +\ln\frac{\Lambda^2}{M^2}\right)+\frac{g^2}{8\pi^2}C_A\sum_{n=0}^{\infty}{n^2\ln{n}}
\end{equation}
where any polynomial terms under the sum have been dropped since their Fourier
transform only amounts to derivatives of delta functions. Note that we
assumed the existence of a \msbar UV-counterterm at the cut-off scale $\mu =
\Lambda$ to cancel the divergence in the remainder.

The remaining sum is now treated as outlined above, starting from the Fourier transform
\begin{equation}
\int_{-\infty}^{\infty}dx\; |x|^2 \ln|x| \, e^{2\pi i kx}
 = \frac{1}{4\pi^2|k|^3}+\frac{\gamma_{\rm E}}{4\pi^2}\delta^{(2)}(k)\,,
\end{equation}
where $\delta^{(n)}$ denotes the n-th distribution derivative of the Dirac
$\delta$-function.

Dropping the zero winding number ($k{=}0$) piece, we can identify the finite rest in terms
of the Riemann $\zeta$-function
\begin{equation}
\sum_{n=0}^{\infty}|n|^2\ln|n|
 =\frac{1}{8\pi^2}\sum_{k=-\infty}^{\infty}\left(\frac{1}{|k|^3}
 +\gamma_E\,\delta^{(2)}(k)\right) \sim \frac{\zeta\left(3\right)}{4\pi^2}
 \,.
\end{equation}
The finite contribution to the mass correction stemming from diagram $(A)$ then is given by
\begin{equation}
\delta M^2_{V_n}=\frac{g^2 M_n^2}{32\pi^2}C_A\left(\frac{23}{3}L_n+\frac{112}{9}+\frac{\zeta\left(3\right)}{\pi^2}\right),
\end{equation}
where $L_n=\ln(\Lambda^2/m_n^2)$. The last term in parenthesis, proportional to
$\zeta(3)$, can be identified as a contribution to the bulk mass correction
\cite{cms},
so that the remaining two terms belong to the boundary mass correction. In a
similar fashion, the contribution from other diagrams in Fig.~\ref{fig:vv} 
to the boundary corrections can be singled out.

\medskip
Analogously we decompose the fermion self-energies displayed in 
Fig.~\ref{fig:ff} according to
\begin{equation}
\Sigma^{ij}_f(p^2)=\left[\cancel{p}P_+\Sigma_{\rm R}{(p^2)}
 +\cancel{p}P_-{(p^2)}\Sigma_{\rm L}(p^2) 
 +M_n\Sigma_{\rm S}{(p^2)}\right]\delta^{ij},
\end{equation}
with fundamental SU($N$) indices $i,j$, and $P_\pm \equiv (1\pm\gamma_5)/2$.
The fermion mass correction is then given by
\begin{equation}
\delta M_{f_n}=\frac{M_n}{2}\;\Re\text{e}\left\{\Sigma_{\rm R}{(M_n^2)}
 +\Sigma_{\rm L}{(M_n^2)}+2\Sigma_{\rm S}{(M_n^2)}\right\},
\end{equation}
and similarly for the scalars in Fig.~\ref{fig:ss}. In all cases, the relevant
diagrams have been generated with the help of {\sc FeynArts 3} \cite{feynarts}.

\medskip

It is interesting to note that the boundary mass corrections can also be
obtained by computing KK-number violating self-energy corrections. In this case,
the bulk contribution is absent, and only one KK level contributes in the loop.

For instance, for the case of a vector boson, the KK-number violating
$V_n$--$V_{n'}$ self-energy (with $n'=n\pm 2$) can be written as \cite{cms}
\begin{align}
\Pi_{\mu\nu}^{ab} = -\Bigl [
 g_{\mu\nu} p^2 \Pi_{\rm T}^{(1)}(p^2) + g_{\mu\nu} \tfrac{1}{2}(n^2+{n'}^2)
  \Pi_{\rm T}^{(2)}(p^2)  - p_\mu p_\nu \Pi_2(p^2) \Bigr ]\,,
\end{align}
as a consequence of the 5D Lorentz symmetry. Thus from this self-energy one can
extract $\Pi_{\rm T}^{(1)}$ and $\Pi_{\rm T}^{(2)}$ and then compute the
KK-number conserving mass correction from \eqref{eq:dmv}.

We have explicitly checked that both approaches lead to the same results for the
boundary mass corrections.

\subsection{Results}

\begin{figure}[t]
\centering
\includegraphics[width=5.5in]{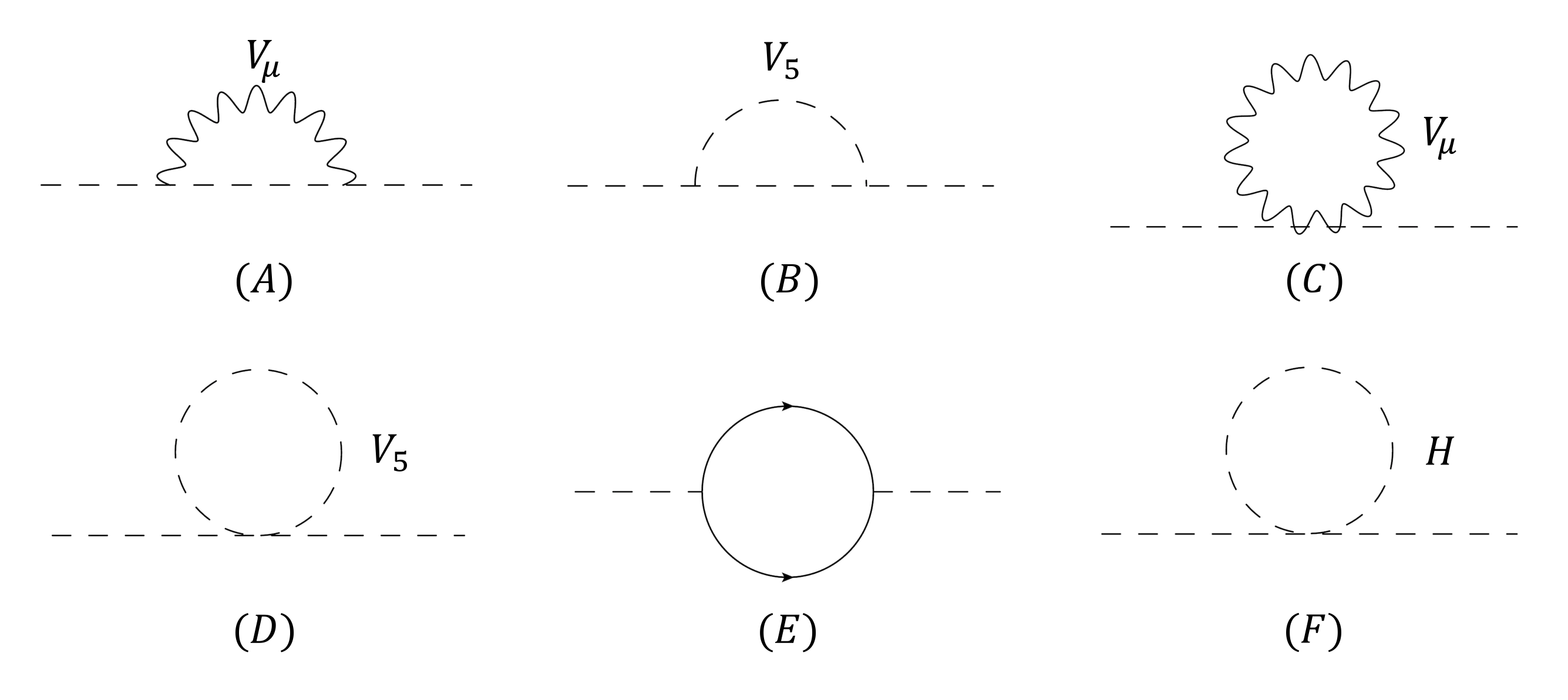}
\vspace{-1ex}
\mycaption{Generic Feynman diagrams for the one-loop contributions to the
self-energies of KK scalars. See caption of Fig.~\ref{fig:vv} for further 
explanations.
\label{fig:ss}}
\end{figure}

\begin{figure}[t]
\centering
\includegraphics[width=5.5in]{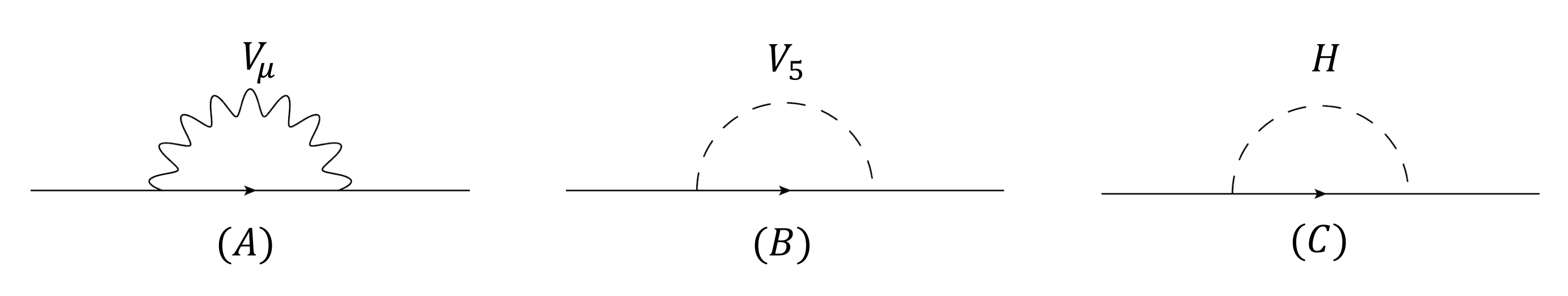}
\vspace{-2ex}
\mycaption{Generic Feynman diagrams for the one-loop contributions to the
self-energy of KK fermions. See caption of Fig.~\ref{fig:vv} for further 
explanations.
\label{fig:ff}}
\end{figure}

\noindent
In the following, results for the KK-mode mass corrections induced by boundary
terms are presented for a general theory with an unspecified non-Abelian gauge
interaction. The one-loop diagrams contributing to the masses of KK gauge
bosons, KK scalars and KK fermions are shown in Figs.~\ref{fig:vv}, \ref{fig:ss},
and \ref{fig:ff}, respectively.

The mass corrections obtained with the methods described in the previous section
read as follows. As before, we use the abbreviation $L_n \equiv \ln
(\Lambda^2/m_n^2)$.
\begin{align}
\bar{\delta} m^2_{V_n} &= m_n^2 \frac{g^2}{32\pi^2} \biggl [
 C(G)\biggl ( \frac{23}{3} L_n + \frac{154}{9}\biggr )
 - \sum_{i \in \rm scalars} (-1)^{P_i} T(r_i) \biggl ( \frac{1}{3} L_n - \frac{4}{9}\biggr )
\biggr ] \,, \label{eq:mv} \\
\bar{\delta} m^2_{S_{+n}} &= \overline{m}^2 + m_n^2 \frac{1}{32\pi^2} \biggl [
 C(r)g^2  ( 6 L_n + 16 )
 - \sum_{i \in \rm scalars} (-1)^{P_i} \lambda_i ( L_n + 1 ) \biggr ] \,, 
 \label{eq:ms} \\
\bar{\delta} m_{f_n} &= m_n \frac{1}{64\pi^2} \biggl [
 C(r) g^2 \biggl ( 9 L_n + 16 \biggr )
 - \sum_{i \in \rm scalars} (-1)^{P_i} h_i^2 ( L_n + 2 )
 \biggr ] \,. \label{eq:mf}
\end{align}
Here $m_{V_n}$ denotes the mass of the $n$-th KK excitations of a generic gauge
boson, where $C(G)$ is the quadratic Casimir invariants of the adjoint
representation.
Similarly, $m_{f_n}$ and $m_{S_{+n}}$ are the masses of a KK-fermion and $\mathbb{Z}_2$-even
KK-scalar, respectively, in the
representation $r$ with quadratic Casimir $C(r)$.
In a SU($N$) theory one has $C(G) = N$ and $C(r) = (N^2-1)/(2N)$ for the
fundamental representation. The sums run over the different scalar fields in the
theory, with $\mathbb{Z}_2$-parity $P_i$, Dynkin index $T(r_i)$, Yukawa coupling
$h_i$, and scalar 4-point coupling $\lambda_i$\footnote{The convention for the normalization of these couplings is the same as in appendix~\ref{sc:feynr}.}. Note that the two components of a complex
scalar field need to be counted separately in the sum.
As already pointed out in Ref.~\cite{cms},
fermion loops do not contribute to the self-energy boundary terms of gauge
bosons and scalars, due to a cancellation between $\mathbb{Z}_2$-even and -odd
fermion components.

A $\mathbb{Z}_2$-even scalar can also receive power-divergent contributions,
which can be written as a boundary mass term \cite{cms}
\begin{align}
{\cal L} \supset -\frac{\pi R}{2} \bigl [ \delta(x^5) + \delta(x^5+\pi R) \bigr ]
 \overline{m}^2 \Phi^\dagger \Phi\,.
\end{align}
This term produces a mass correction of $\overline{m}^2$ for the zero mode,
while the higher KK masses are shifted by $2\overline{m}^2$. Thus, relative to
the zero mode, the masses of the KK excitations receive a correction of
$\overline{m}^2$, see eq.~\eqref{eq:ms}. While naive dimensional analysis would
suggest that $\overline{m}^2 \sim {\cal O}(\Lambda^2)$, this is not consistent
with the presence of a light scalar in the spectrum, as is the case in the SM.
Instead, to generate the SM as a low-energy effective theory, one has to assume
that $\overline{m}^2$ is tuned to $\overline{m}^2 \ll R^{-2}$.

The logarithmic parts $\propto L_n$ in eqs.~\eqref{eq:mv}--\eqref{eq:mf}
can be compared to Ref.~\cite{cms}, but we find some discrepancies:  The
one-loop scalar mass corrections should be proportional to $C(r)$, instead of
$T(r)$ in Ref.~\cite{cms}, and the fermion mass contribution from Yukawa
couplings is a factor 3 smaller than reported there\footnote{Specifically, 
the $b_1$ terms in line (b) of Tab.~III in Ref.~\cite{cms} should have opposite
signs.}.
\label{trtocr}
As evident from the equations above, the non-logarithmic terms are smaller than
the terms proportional to $L_n \sim 4 \cdots 8$ (for $n \sim {\cal O}(1)$) by at
most a factor of a few. Thus their contribution is phenomenologically important.

\medskip
The KK mass corrections in MUED can be determined by simply substituting the
appropriate SM coupling constants and group theory factors in the formulae
above. For the gauge bosons this leads to
\begin{align}
\bar{\delta} m^2_{G_n} &= m_n^2 \frac{g_3^2}{32\pi^2} 
 \biggl ( 23 L_n + \frac{154}{3} \biggr )\,, \\
\bar{\delta} m^2_{W_n} &= m_n^2 \frac{g_2^2}{32\pi^2} 
 \biggl ( 15 L_n + \frac{104}{3} \biggr )\,, \\
\bar{\delta} m^2_{B_n} &= m_n^2 \frac{g_1^2}{16\pi^2} 
 \biggl ( -\frac{1}{6} L_n + \frac{2}{9} \biggr )\,,
\end{align}
while for the fermions one obtains
\begin{align}
\bar{\delta} m_{Q_n} &= m_n \frac{1}{16\pi^2} \biggl [
 g_3^2 \biggl ( 3 L_n+\frac{16}{3} \biggr ) 
 + g_2^2 \biggl ( \frac{27}{16} L_n+3 \biggr )  
 + g_1^2 \biggl ( \frac{1}{16}L_n + \frac{1}{9} \biggr ) \biggr ]\,, \\
\bar{\delta} m_{Q_{3n}} &= m_n \frac{1}{16\pi^2} \biggl [
 g_3^2 \biggl ( 3 L_n+\frac{16}{3} \biggr ) 
 + g_2^2 \biggl ( \frac{27}{16} L_n+3 \biggr )  
 + g_1^2 \biggl ( \frac{1}{16}L_n + \frac{1}{9} \biggr ) 
 - h_t^2 \biggl ( \frac{1}{4}L_n + \frac{1}{2} \biggr ) \biggr ]\,, \displaybreak[0] \\
\bar{\delta} m_{u_n} &= m_n \frac{1}{16\pi^2} \biggl [
 g_3^2 \biggl ( 3 L_n+\frac{16}{3} \biggr ) 
 + g_1^2 \biggl ( L_n + \frac{16}{9} \biggr ) \biggr ]\,, \\
\bar{\delta} m_{t_n} &= m_n \frac{1}{16\pi^2} \biggl [
 g_3^2 \biggl ( 3 L_n+\frac{16}{3} \biggr ) 
 + g_1^2 \biggl ( L_n + \frac{16}{9} \biggr )  
 - h_t^2 \biggl ( \frac{1}{2}L_n + {1} \biggr ) \biggr ]\,, \\
\bar{\delta} m_{d_n} &= m_n \frac{1}{16\pi^2} \biggl [
 g_3^2 \biggl ( 3 L_n+\frac{16}{3} \biggr ) 
 + g_1^2 \biggl ( \frac{1}{4}L_n + \frac{4}{9} \biggr ) \biggr ]\,, \displaybreak[0] \\
\bar{\delta} m_{L_n} &= m_n \frac{1}{16\pi^2} \biggl [
 g_2^2 \biggl ( \frac{27}{16} L_n+3 \biggr )  
 + g_1^2 \biggl ( \frac{9}{16}L_n + 1 \biggr ) \biggr ]\,, \\
\bar{\delta} m_{e_n} &= m_n \frac{1}{16\pi^2} 
 g_1^2 \biggl ( \frac{9}{4}L_n + 4 \biggr )\,,
\end{align}
where $Q_{3n}$ and $t_n$ denote the third generations of the KK excitations of
the left-handed and right-handed up-type quark fields, respectively. Finally,
the mass correction to the KK Higgs boson reads
\begin{align}
\bar{\delta} m^2_{H_{n}} &= \overline{m}_H^2 + m_n^2  \frac{1}{16\pi^2}  \biggl [
 g_2^2 \biggl ( \frac{9}{4}L_n + 6 \biggr ) +
 g_1^2 \biggl ( \frac{3}{4}L_n + 2 \biggr ) -
 \lambda_H (L_n+1) \biggr ]\,. \label{eq:mh}
\end{align}
In the above expressions, $g_{1,2,3}$ are the couplings of the SM U(1)$_{\rm Y}$, SU(2)$_{\rm
L}$ and SU(3)$_{\rm C}$ gauge groups, respectively, while $h_t$ is the top
Yukawa coupling and $\lambda_H$ the Higgs self-coupling.
The numerical impact of these corrections will be discussed in
section~\ref{sc:pheno}.


\section{KK-number violating interactions}
\label{sc:int}

\noindent
As is well-known, the Lorentz symmetry breaking from orbifolding 
leads to loop-induced boundary-localized interactions which can break KK number \cite{cms}.
From a phenomenological point of view, 2--0--0 interactions between a level-2 KK
mode and two zero modes are particularly interesting, since they can mediate 
single production and decay of a level-2 KK particle at colliders.

The logarithmically enhanced contributions, $\propto \ln(\Lambda R)$, to these
vertices have been considered in Refs.~\cite{cms,lev2dm}. Here, the
non-logarithmic contributions are also computed, which are important for two
reasons. On one hand, they can be numerically of similar order as the
$\ln(\Lambda R)$ term and thus lead to sizable corrections of the known
KK-number violating couplings. On the other hand, there are additional vertices
that are UV-finite but non-zero. Since these do not contain any $\ln(\Lambda R)$
terms, they have not been considered before, but they can lead to
phenomenologically relevant new production and decay channels.

\subsection{Approach}

\noindent
The calculation of the KK-number violating couplings can be relatively easily
performed by directly computing the $X_2$--$Y_0$--$Z_0$ vertices in the 4D
compactified theory. Here $X$, $Y$ and $Z$ stand for any MUED fields. Since all
leading-order vertices in MUED do conserve KK-number, one needs to consider only
level-1 KK modes inside the loop. 

As before, the authors have used {\sc FeynArts 3} \cite{feynarts} for the
amplitude generation, and {\sc FeynCalc} \cite{feyncalc} was employed for part of the Dirac and Lorentz algebra manipulations. Similar to the mass corrections
discussed above, UV divergences have been renormalized in the \msbar scheme
with the scale choice $\mu = \Lambda$.

Throughout this chapter, unless mentioned otherwise, contributions from
electroweak symmetry breaking (EWSB) have been neglected, since these are suppressed
by powers of $vR$. In particular, mixing between the KK-$Z$ boson and KK photon
or between the KK-top doublet and singlet has not been included for the
particles running inside the loops.

\subsection{Results}

\noindent
Let us begin by writing the results for a generic theory with arbitrary
non-Abelian gauge group and an arbitrary number of fermionic and scalar matter
fields. Detailed expressions for the specific field content and interactions of
MUED are listed in Appendix~\ref{app:int}.

\begin{figure}[t]
\centering
\includegraphics[width=5.5in]{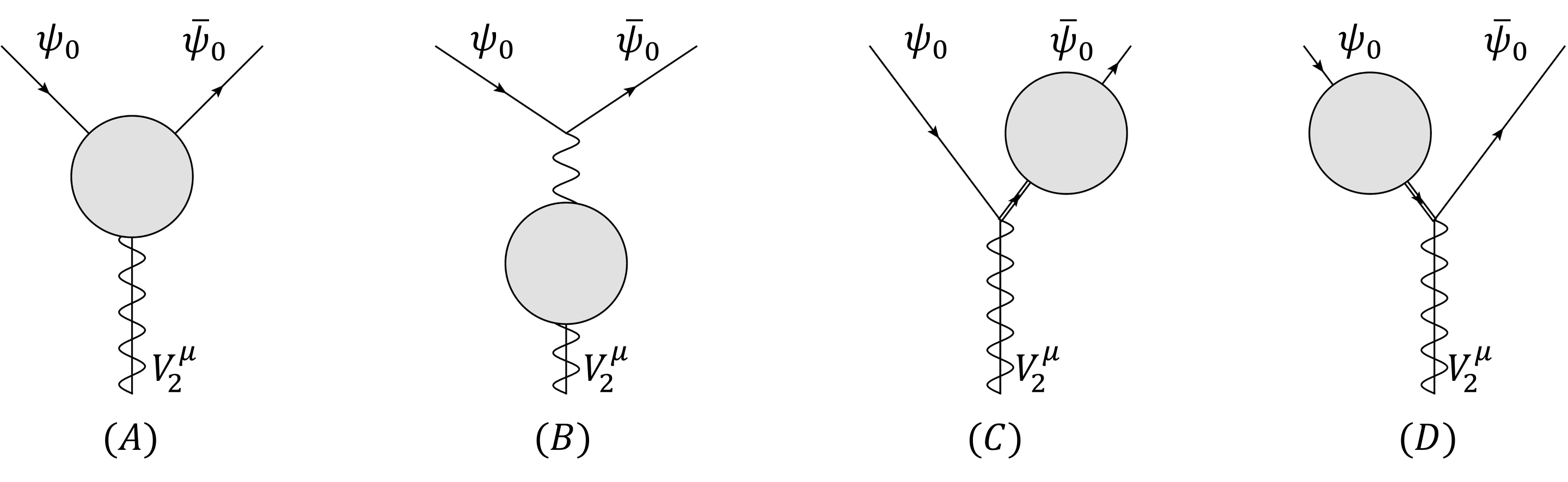}
\vspace{-2ex}
\mycaption{Contributions to the $\bar{\psi}_0$--$\psi_0$--$V_2$ vertex, where the 
blobs indicate one-loop corrections. Zero-mode vector and fermion propagators are depicted through
normal wavy and solid lines, respectively, whereas level-2 vector and fermion
propagators are shown as wavy-solid and double-solid lines, respectively.
\label{fig:f0f0v2}}
\end{figure}

\paragraph{\boldmath $\bar{\psi}_0$--$\psi_0$--$V_2^\mu$ coupling:} This vertex
can be written in the form
\begin{align}
-iC_{\psi_0\psi_0V_2} \gamma^\mu T^a P_\pm\,.
\end{align}
Here $P_\pm$ are right-/left-handed projectors and $T^a$ are the generators of the gauge group. For a U(1) group (such as
U(1)$_{\rm Y}$), $T^a$ is simply replaced by the charge (hypercharge $Y$). The
coefficient $C_{\psi_0\psi_0V_2}$ receives contributions from the vertex and
self-energy corrections shown in Fig.~\ref{fig:f0f0v2} and reads
\begin{align}
C_{\psi_0\psi_0V_2} &= \frac{\sqrt{2} g}{64\pi^2} \biggl [
 g^2\, C(G)_V \biggl ( \frac{23}{3} L_1 + \frac{157}{9} - 2 \pi^2 \biggr )
 + \sum_k g_k^2\, C(r_\psi)_k \biggl ( -9L_1-13 + \frac{7\pi^2}{4} \biggr ) \notag \\
 &\qquad\qquad + g^2\sum_{i\in\rm scalars} (-1)^{P_i} T(r_i)_V \biggl ( -\frac{1}{3} L_1 - \frac{2}{9}
\biggr ) \notag \\
 &\qquad\qquad + \sum_{i\in\rm scalars} (-1)^{P_i} h_i^2 \biggl (
 L_1-1+\frac{\pi^2}{4} + 2\frac{C_{\phi_{i0}\phi_{i0}V_0}}{C_{\psi_0\psi_0V_0}} 
 \biggr ) \biggr ]\,,
 \label{eq:vff}
\end{align}
where $C(G)_V$ is the adjoint Casimir of the gauge group of $V$, which has the
gauge coupling $g$, and $T(r_i)_V$ is the Dynkin index for the representation of
the scalar $i$ under the same group. The sum $\sum_k$ runs over all gauge groups
under which $\psi$ is charged, with gauge couplings $g_k$, and $C(r_\psi)_k$
being the Casimir of the representation of $\psi$ with respect to the gauge
group $k$. For U(1) groups, $T(r_i)$ and $C(r_\psi)_k$ get replaced by the
corresponding charges. $h_i$ is the Yukawa coupling between scalar $i$ and
$\psi$.

\begin{figure}[t]
\hfill \includegraphics[width=1.3in]{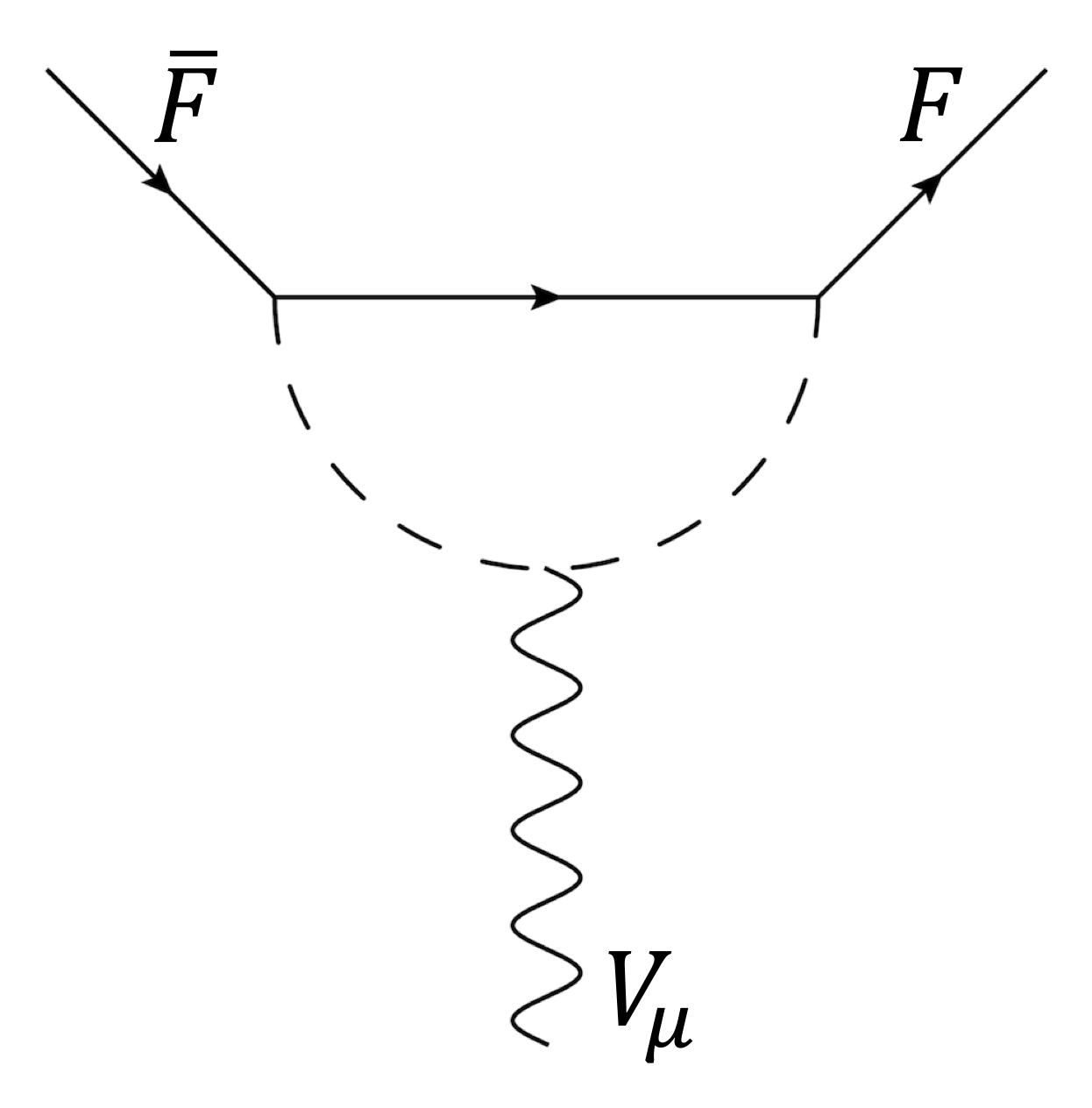} \hfill
\parbox[b]{10cm}{\mycaption{Vertex diagram contributing to KK-number violating
vector-boson--fermion couplings involving a KK-Higgs in the loop.
\label{fig:vertexyuk}}}
\vspace{2ex}
\end{figure}

Finally, eq.~\eqref{eq:vff} depends also on
$C_{\phi_{i0}\phi_{i0}V_0}/C_{\psi_0\psi_0V_0}$, the ratio of the
couplings of the scalar $i$ and the fermion $\psi$ to the gauge field $V$. Some
care must be taken when defining the signs of these couplings. 
The signs of the ratio should be $+1$ ($-1$) if $\psi_0$ and $\phi_{i0}$ have
the same representation or the same charge sign for the gauge group of $V$, and
they run in opposite directions (the same direction) in
Fig.~\ref{fig:vertexyuk}.

The logarithmic part of the first two lines in eq.~\eqref{eq:vff} agrees with
Ref.~\cite{cms}. The logarithmic part of the last line in eq.~\eqref{eq:vff} has
been computed in Ref.~\cite{lev2dm} for a U(1) group, but we obtain a different
result.

\paragraph{\boldmath $\bar{\psi}_2$--$\psi_0$--$V_0^\mu$ coupling:} There are two
form factors that can facilitate the single production and decay of a level-2
KK-quark. The first is a Dirac-type chiral interaction,
\begin{align}
-iC_{\psi_2\psi_0V_0} \gamma^\mu T^a P_\pm\,,
\label{eq:f2v}
\end{align}
whereas the second is a dipole-like interaction,
\begin{align}
-\tilde{D}_{\psi_2\psi_0V_0}\frac{\sigma^{\mu\nu}q_\nu}{2 m_{KK}} T^a
P_\pm\,,
\label{eq:dip}
\end{align}
where $q$ is the momentum $V_0^\mu$. Note that by only considering these two
expressions, we restrict ourselves
to \emph{transverse} polarization modes of the $V_0^\mu$ boson. If $V_0^\mu$ was
a massive $W$ or $Z$ boson, their longitudinal modes must be excluded when using
eq.~\eqref{eq:f2v}, since
they would receive contributions from an additional form factor proportional to
$(k_2-k_0)^\mu$, where $k_2$ and $k_0$ are the (incident) momenta of the
$\psi_2$ and $\psi_0$ fermion, respectively. The restriction to transverse gauge
boson polarizations is justified since the contribution of the longitudinal mode
of $W$ or $Z$ bosons is suppressed by $vR$.

\begin{figure}[t]
\centering
\includegraphics[width=5.5in]{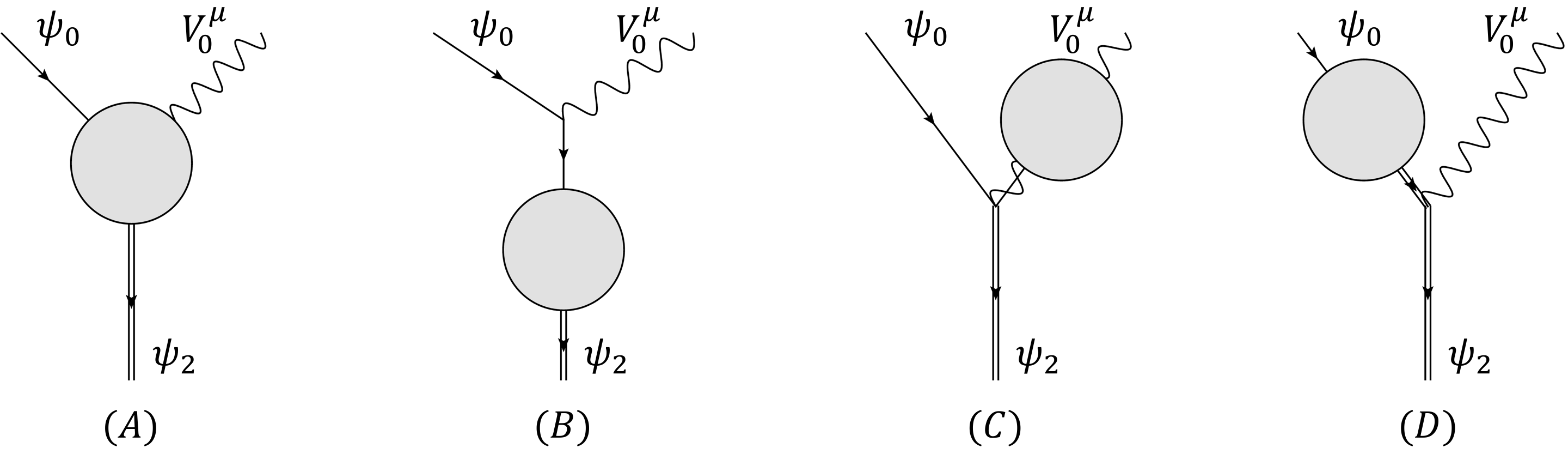}
\vspace{-2ex}
\mycaption{Contributions to the $\bar{\psi}_2$--$\psi_0$--$V_0$ vertex.
See caption of Fig.~\ref{fig:f0f0v2} for more explanations.
\label{fig:f2f0v0}}
\end{figure}

For the computation of the coefficient $C_{\psi_2\psi_0V_0}$ one needs to
consider the diagrams in Fig.~\ref{fig:f2f0v0}, which yield
\begin{align}
C_{\psi_2\psi_0V_0} &= \frac{\sqrt{2} g}{64\pi^2} \biggl [
 g^2\, C(G)_V \biggl ( \frac{\pi^2}{4} L_1 - 2 \biggr )
 + 4\sum_k g_k^2\, C(r_\psi)_k \biggr ]\,.
\end{align}
The dipole-like interaction in eq.~\eqref{eq:dip} is generated only by the
vertex diagrams in Fig.~\ref{fig:f2f0v0}~(A). The result reads
\begin{align}
D_{\psi_2\psi_0V_0} &= \frac{\sqrt{2} g}{64\pi^2} \biggl [
 g^2\, C(G)_V \bigl ( \pi^2-7 \bigr )
 + \sum_k g_k^2\, C(r_\psi)_k \biggl ( 3 - \frac{3\pi^2}{4} \biggr ) \notag \\
 &\qquad\qquad + \sum_{i\in\rm scalars} (-1)^{P_i} h_i^2 \biggl (
 \frac{\pi^2}{4}-1 + \frac{\pi^2-8}{2} \; \frac{C_{\phi_{i0}\phi_{i0}V_0}}{C_{\psi_0\psi_0V_0}} 
 \biggr ) \biggr ]\,.
\end{align}
As evident from these expressions, both $C_{\psi_2\psi_0V_0}$ and
$\tilde{D}_{\psi_2\psi_0V_0}$ are independent of $\ln(\Lambda R)$, and
have not been previously reported in the literature. 

\paragraph{\boldmath $V_2^{\mu,a}(p)$--$V_0^{\nu,b}(k_1)$--$V_0^{\rho,c}(k_2)$ 
coupling:} Restricting ourselves, as before, to transverse polarizations for
$V_0^{\nu,b}(k_1)$ and $V_0^{\rho,c}(k_2)$, this coupling can be written in the
form
\begin{align}
f_{abc} \Bigl\{
 &\bigl [ g_{\mu\nu}(p-k_1)_\rho +
          g_{\nu\rho}(k_1-k_2)_\mu +
	  g_{\rho\mu}(k_2-p)_\nu \bigr ]
  C_{V_2V_0V_0} \notag \\
 &+\bigl [ -g_{\mu\nu}k_{1,\rho} + g_{\rho\mu}k_{2,\nu} \bigr ] 
  D_{V_2V_0V_0} 
 + g_{\nu\rho}(k_1-k_2)_\mu \, E_{V_2V_0V_0} \Bigr\}\,.
\end{align}
Here $f_{abc}$ are the structure constants of the gauge group. Furthermore, $p$, $k_1$
and $k_2$ are the vector boson momenta, which are all taken to be incoming.

\begin{figure}[t]
\centering
\includegraphics[width=5.5in]{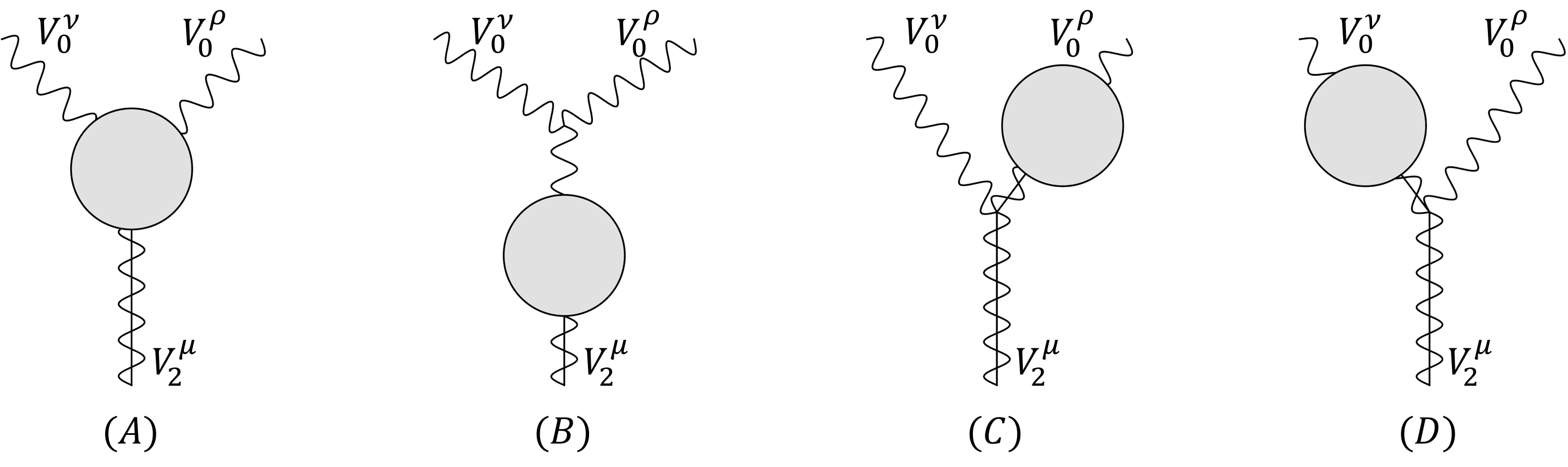}
\vspace{-2ex}
\mycaption{Contributions to the $V_2$--$V_0$--$V_0$ vertex.
See caption of Fig.~\ref{fig:f0f0v2} for more explanations.
\label{fig:v2v0v0}}
\end{figure}

The coefficient $C_{V_2V_0V_0}$ receives contributions from all diagrams in
Fig.~\ref{fig:v2v0v0}, whereas $D_{V_2V_0V_0}$ and $E_{V_2V_0V_0}$ are only
generated by the first diagram in the figure. At one-loop level, they are given
by
\begin{align}
C_{V_2V_0V_0} &= \frac{\sqrt{2} g^3}{64\pi^2} \biggl [
 C(G)_V \biggl ( -\frac{157}{9} + \frac{7\pi^2}{6} \biggr )
 - \sum_{i\in\rm scalars} (-1)^{P_i} T(r_i)_V 
 \biggl ( \frac{4}{9} - \frac{\pi^2}{18} \biggr ) \biggr ] \,,\\
D_{V_2V_0V_0} &= \frac{\sqrt{2} g^3}{64\pi^2} \biggl [
 C(G)_V \biggl ( \frac{91}{6}-\pi^2 \biggr )
 + \sum_{i\in\rm scalars} (-1)^{P_i} T(r_i)_V \, \frac{8-\pi^2}{12} \biggr ] 
 \,, \\
E_{V_2V_0V_0} &= \frac{\sqrt{2} g^3}{64\pi^2} \,C(G)_V 
 \biggl ( \frac{38}{3} - \frac{3\pi^2}{4} \biggr ) \,.
\end{align}
These results, which are also independent of $\ln(\Lambda R)$, are new.

\paragraph{\boldmath $\bar{T}_2$--$t_0$--$\Phi_0$ / $\bar{t}_2$--$t_0/b_0$--$\Phi_0$ coupling:} Level-2 KK top quarks can have loop-induced decays into zero-mode top or bottom and Higgs states, which are proportional to the top Yukawa coupling. Here any component of the Higgs doublet can appear in the final state, including the Higgs bosons as well as longitudinal $W$ and $Z$ bosons. The result, including strong and electroweak contributions, reads
\begin{align}
&\begin{aligned}
&\bar{t}_2{-}t_0{-}h_0: && +i\frac{m_t}{v}P_-\,C_{t_2t_0\Phi_0}\,, \\
&\bar{t}_2{-}t_0{-}Z_L: && -\frac{m_t}{v}P_-\,C_{t_2t_0\Phi_0}\,, \\
&\bar{t}_2{-}b_0{-}W_L^+: && -i\sqrt{2}\frac{m_t}{v}P_-\,C_{t_2t_0\Phi_0}\,, \\
\end{aligned} \\
&C_{t_2t_0\Phi_0} = \frac{\sqrt{2}}{64\pi^2} \biggl [ \begin{aligned}[t]
 & g_3^2 \biggl (
 \frac{8}{3}L_1+\frac{40}{3}-\pi^2 \biggr ) 
 + g_2^2 \biggl ( -3L - \frac{3}{2} + \frac{3\pi^2}{8} \biggr ) \\
 &+ g_1^2 \biggl ( \frac{23}{9}L_1 + \frac{83}{18} - \frac{5\pi^2}{24} \biggr ) 
 - 2h_t^2 L_1 + \lambda\frac{3}{2}(L_1+1) \biggr ], \end{aligned} \\[1ex]
&\begin{aligned}
&\bar{T}_2{-}t_0{-}h_0: && -i\frac{m_t}{v}P_+\,C_{T_2t_0\Phi_0}\,, \\
&\bar{T}_2{-}t_0{-}Z_L: && -\frac{m_t}{v}P_+\,C_{T_2t_0\Phi_0}\,, \\
\end{aligned} \\
&C_{T_2t_0\Phi_0} = \frac{\sqrt{2}}{64\pi^2} \biggl [ \begin{aligned}[t]
 & g_3^2 \biggl (
 \frac{8}{3}L_1+\frac{40}{3}-\pi^2 \biggr ) 
 + g_2^2 \biggl ( \frac{15}{4}L_1 + 9 - \frac{3\pi^2}{8} \biggr ) \\
 &+ g_1^2 \biggl ( -\frac{43}{36}L_1 + \frac{1}{9} + \frac{\pi^2}{24} \biggr ) 
 - 2h_t^2 L_1 + \lambda\frac{3}{2}(L_1+1) \biggr ]. \end{aligned}
\end{align}%

\paragraph{\boldmath $\Phi_2$ decay couplings:} The level-2 KK excitation of the SM Higgs
doublet can be decomposed into a neutral CP-even component $h_2$, a neutral
CP-odd component $\chi_2$, and a charged pair $\phi_2^\pm$,
\begin{align}
\Phi_2 = \begin{pmatrix} \phi^+_2 \\ \frac{1}{\sqrt{2}}(h_2+i\chi_2) \end{pmatrix}.
\end{align}
They have a rich
variety of loop-induced couplings to pairs of zero-mode particles. In this work
we do not attempt a comprehensive discussion of these channels, but only
present a few interesting aspects.

The leading decay channel of $h_2$ and $\chi_2$ is into $t\bar{t}$ pairs, which
is dominantly induced through QCD loops. The result is given by
\begin{align}
&\bar{t}_0t_0h_2: &&-i\frac{m_t}{v}C_{t_0t_0h_2}\,, \\
&\bar{t}_0t_0\chi_2: &&i\gamma_5\frac{m_t}{v}C_{t_0t_0h_2}\,, \\
&&&C_{t_0t_0h_2} = \frac{\sqrt{2} g_3^2}{64\pi^2} \, C_{\rm F} \biggl [
 -4L_1 -4 + \frac{\pi^2}{2} \biggr ].
\end{align}
The logarithmic part of this expression agrees with Ref.~\cite{lev2dm}.

$h_2$ does not have any decays into gluon pairs since there is a cancellation
between the $\mathbb{Z}_2$-even and $\mathbb{Z}_2$-odd KK-tops inside the vertex
loop. However, it can couple to electroweak gauge boson pairs via loops involving
level-1 KK-gauge and KK-Higgs bosons, although this effective
interaction is suppressed by $vR$. Nevertheless, this subdominant decay
channel is still interesting since it can lead to di-photon resonance signals.
We find that it can be written as
\begin{equation}
\mathcal{L}_{h_2V_0V_0}\supset 
 \sum_{j,k=0}^3
 \frac{i C_{jk}vR^2}{64\sqrt{2}\pi^2} h_2 F^j_{0,\mu\nu}F^{k,\mu\nu}_0\,,
\end{equation}
where the $j,k=0$ refers to the U(1) field $B_0^{\mu}$ and 
$j,k=1,2,3$ to the SU(2) gauge boson $W_0^{a,\mu}$. The coefficient $C_{jk}$ are
given by
\begin{align}
C_{00} &= \frac{g_1^2}{128} \bigl[
 g_1^2(8\pi^2-58)+24g_2^2(\pi^2-10)+48\lambda(2\pi^2-27) \notag \\
 &\qquad\quad +3(41g_1^2+93g_2^2-120\lambda)L_1\bigr], \\
C_{jj} &= -\frac{g_2^2}{128}\bigl[
 g_1^2(122-8\pi^2)+24\lambda(54-4\pi^2)+24g_2^2(10+3\pi^2) \notag \\
 &\qquad\qquad -15\left(5g_1^2+41g_2^2-24\lambda\right)L_1\bigr], 
 \qquad\qquad\qquad [j=1,2,3] \\
C_{03} &= -\frac{g_1g_2}{128}\bigl[
 2g_1^2(2\pi^2-65)-8\lambda(17-2\pi^2) +8g_2^2(35+3\pi^2) \notag \\ 
 &\qquad\qquad +3\left(5g_1^2+81g_2^2\right)L_1\bigr], \\
C_{01}&=C_{02} =C_{12}=C_{13}=C_{23}=0\,. 
\end{align}
In contrast to the CP-even component, the CP-odd $\chi_2$ can have a
loop-induced coupling to gluon pairs. The
$g^a_{0,\mu}(k_1)$-$g^a_{0,\nu}(k_2)$-$\chi_2$ has the form
\begin{align}
&\frac{-i}{v}C_{g_0g_0\chi_2}\epsilon^{\mu\nu k_1 k_2}, &
&C_{g_0g_0\chi_2} = \frac{\sqrt{2} g_3^2}{64}\, m_t R^2.
\end{align}
As for the $h_2$ decay into vector bosons, it is suppressed by $vR$, but may
still be relevant for the production of $\chi_2$ at hadron colliders.


\section{Phenomenological implications}
\label{sc:pheno}

\noindent
In this section, we study the mass spectrum of KK particles with improved one-loop
corrections, including finite (non-logarithmic) terms, 
and their decays and collider implications of KK-number violating
interactions.

\subsection{Mass hierarchy}
\label{sc:spectrum}

\begin{figure}[t!]
\centering
\includegraphics[width=0.47\columnwidth]{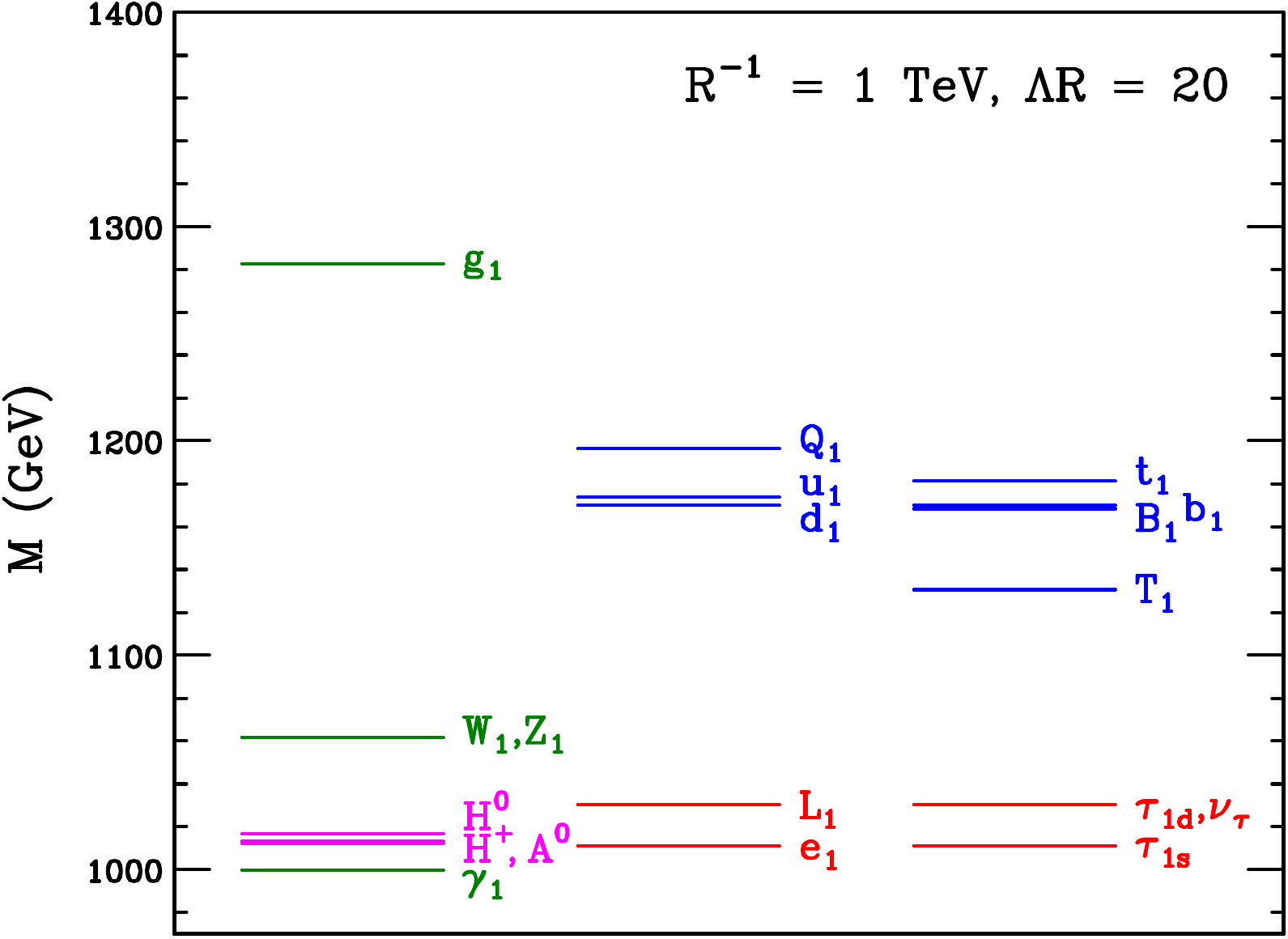} \hspace*{0.3cm}
\includegraphics[width=0.47\columnwidth]{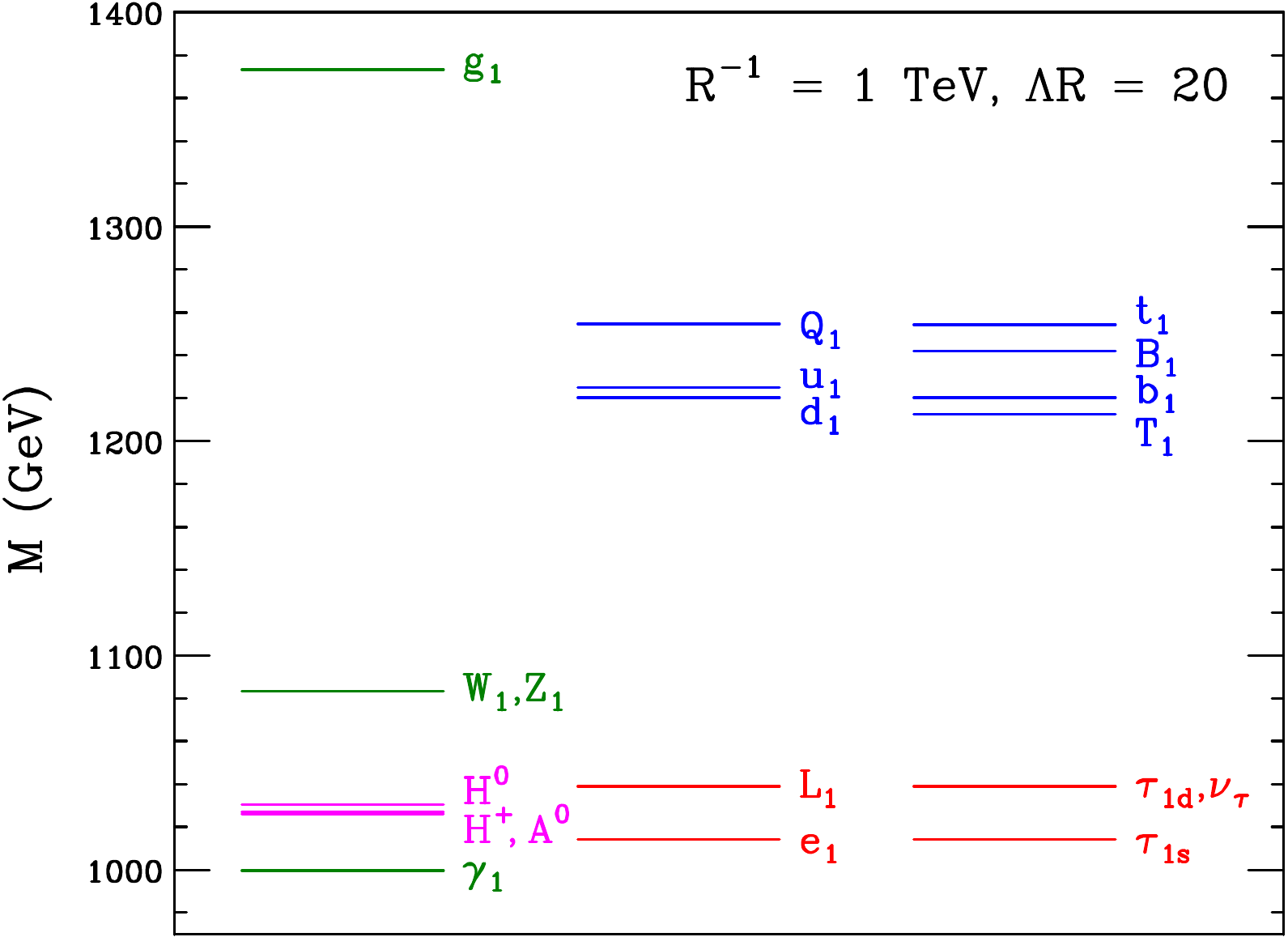}
\mycaption{Mass spectrum of KK particles at level-1 for $R^{-1}=$ 1 TeV and
 $\Lambda R=20$ without (left) / with (right) finite contributions. 
 \label{fig:mass}}
\end{figure}

\noindent
We begin our discussion with the mass spectrum of KK particles at level-1, which
is shown in Fig.~\ref{fig:mass} for $R^{-1}=$ 1 TeV and $\Lambda R=20$ without
(left) and with (right) finite contributions, respectively. KK bosons (either
spin-0 or spin-1) are shown in the left column, while KK fermions are in the
middle (for first two generations) and right column (for third generation).  In
general is the mass spectrum slightly broadened by the finite corrections. For example, the
mass splitting $\delta = \frac{m_{Q_1} - m_{\gamma_1}}{m_{\gamma_1}}$ between KK
quark ($Q_1$) and KK photon ($\gamma_1$) increases from $\sim$20\% to $\sim$25\%
(for $\Lambda R=20$), making the decay products harder in the cascade decays. Since
they become slightly heavier for a given value of $R^{-1}$, their production
cross sections of KK quarks would decrease slightly. Therefore it is worth
investigating the implications of finite terms to see which effect between the
increased efficiency and the reduced production cross section would make a more
pronounced difference. 

\begin{figure}[t]
\centering
\includegraphics[width=0.47\columnwidth]{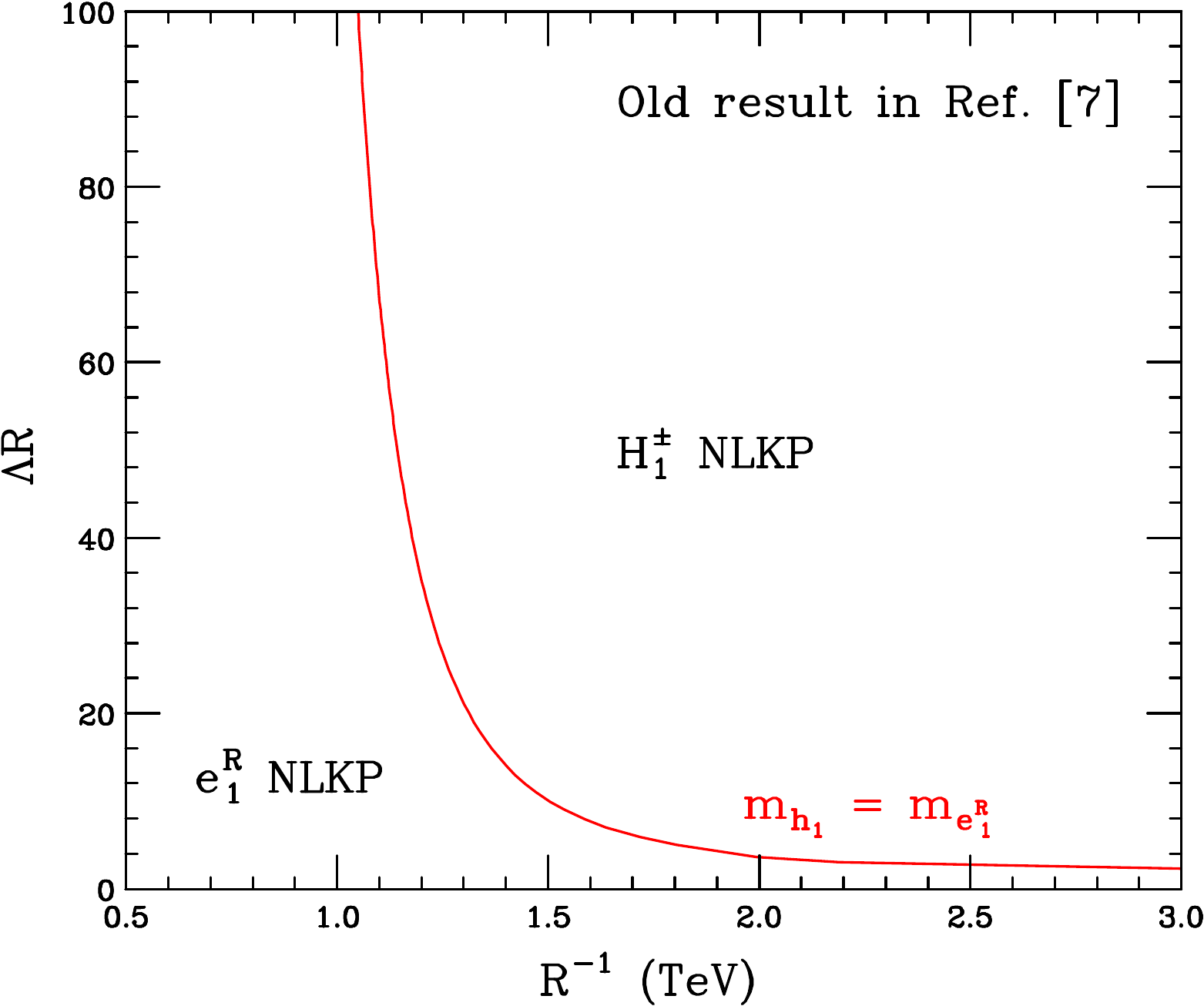}\hspace*{0.3cm}
\includegraphics[width=0.47\columnwidth]{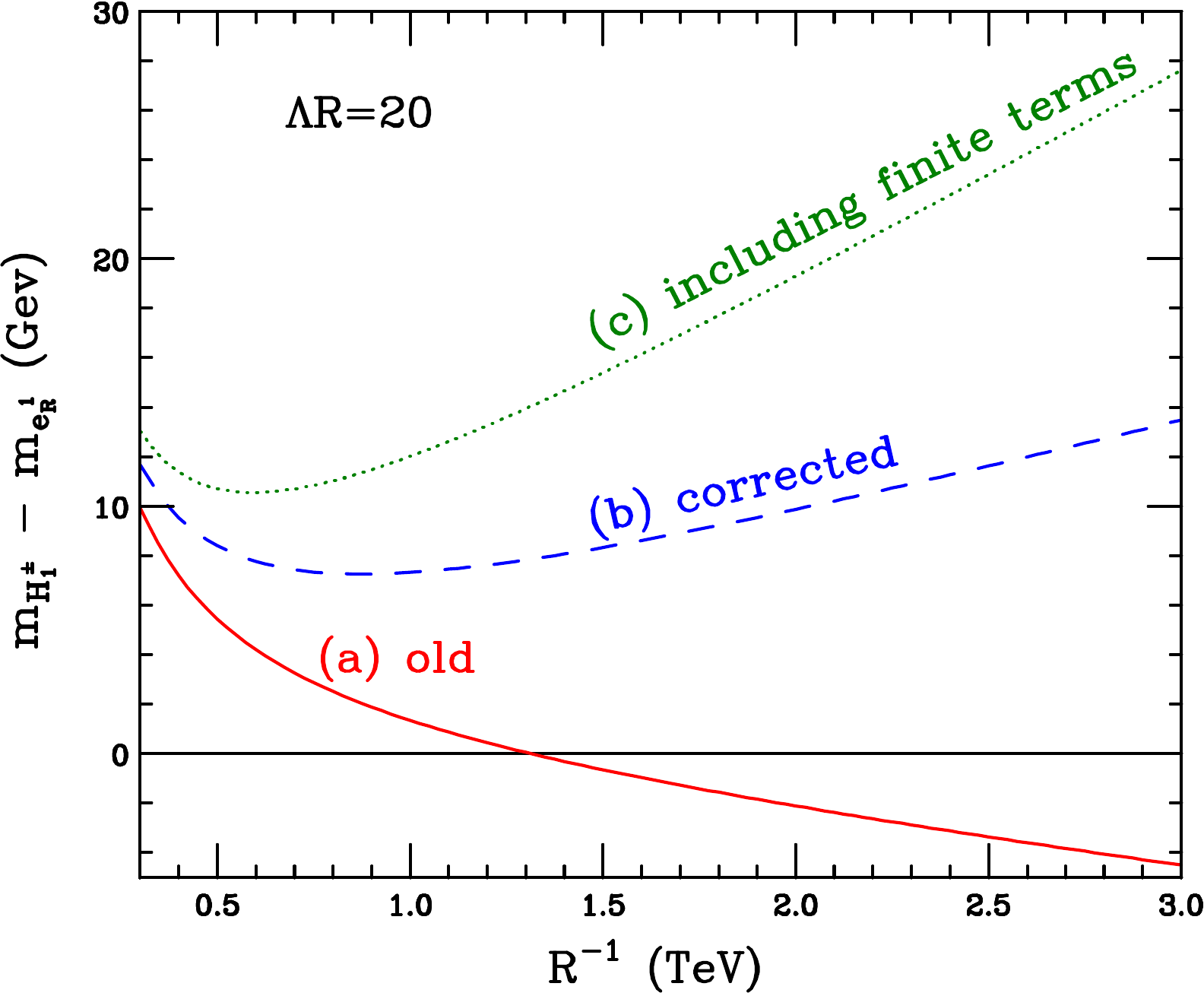}
\mycaption{\label{fig:nlkp}Left: The phase diagram in the ($R^{-1}$, $\Lambda R$) plane from Ref. \cite{Cembranos:2006gt} is reproduced using the incorrect numerical factor (see text), which shows that the KK Higgs could be the NLKP in MUED for a large value of $R^{-1}$. Right: Fixing $\Lambda R=20$, the old (incorrect) result is shown in (red, solid) as a function of $R^{-1}$. The correct result is shown in (blue, dashed), while the curve in (green, dotted) includes finite terms. We find that KK leptons are always the NLKP.}
\end{figure}

The dependence on $\Lambda R$ is logarithmic and we observe similar patterns in the mass hierarchy for a wide range in ($R^{-1}$, $\Lambda R$) space, with the
exception of the KK Higgs and KK leptons. The KK Higgs bosons masses (magenta in the left column) are highly degenerate
with the SU(2)$_{\rm L}$-singlet KK lepton masses ($e_1$, red in the middle column), as shown in the left panel of Fig.
\ref{fig:mass}, but finite corrections increase the mass difference between the two, as shown
in the right panel. In particular, this can affect the hierarchy of the lightest
and next-to-lightest level-1 KK particles, abbreviated as LKP and NLKP,
respectively. The (LKP, NLKP) structure has been studied in detail in
Ref.~\cite{Cembranos:2006gt}, and we reproduce some of their findings as shown
in the left panel of Fig.~\ref{fig:nlkp}. For a given value of $\Lambda R$, the
NLKP is the SU(2)$_{\rm L}$-singlet KK lepton if $R^{-1} < R^{-1}_\circ$, while
the NLKP is the charged KK Higgs for $R^{-1} < R^{-1}_\circ$, where
$R^{-1}_\circ$ is determined by $m_{H_1^\pm} \left ( R^{-1}_\circ, \Lambda R
\right ) = m_{e_1^R} \left ( R^{-1}_\circ, \Lambda R \right )$. The red curve in
the left panel of Fig.~\ref{fig:nlkp} is the solution of this equation. To study
this in detail, we plot the mass difference between them as a function of
$R^{-1}$ for $\Lambda R=20$. The corresponding result (red, solid) is labeled as
(a) in the right panel of Fig. \ref{fig:nlkp}.  Fixing a typo in the Higgs mass
correction of Ref. \cite{cms} ($\frac{3}{2}$ should be $\frac{9}{4}$, as already
mentioned in section~\ref{trtocr}), we obtain the (blue, dashed) curve, labeled
as (b).  With this correction, KK leptons are always the NLKP, in contrast to what is
shown in the left panel. Including the finite terms in eq.~\eqref{eq:mh},
we find an even larger mass splitting, shown by the (green, dotted) curve (labeled
as (c)).  This could have some impact on the computation of the KK-photon relic
abundance, since co-annihilation processes are important in this degenerate mass
spectrum \cite{lev2dm}.

\begin{figure}[t!]
\centering
\includegraphics[width=0.48\columnwidth]{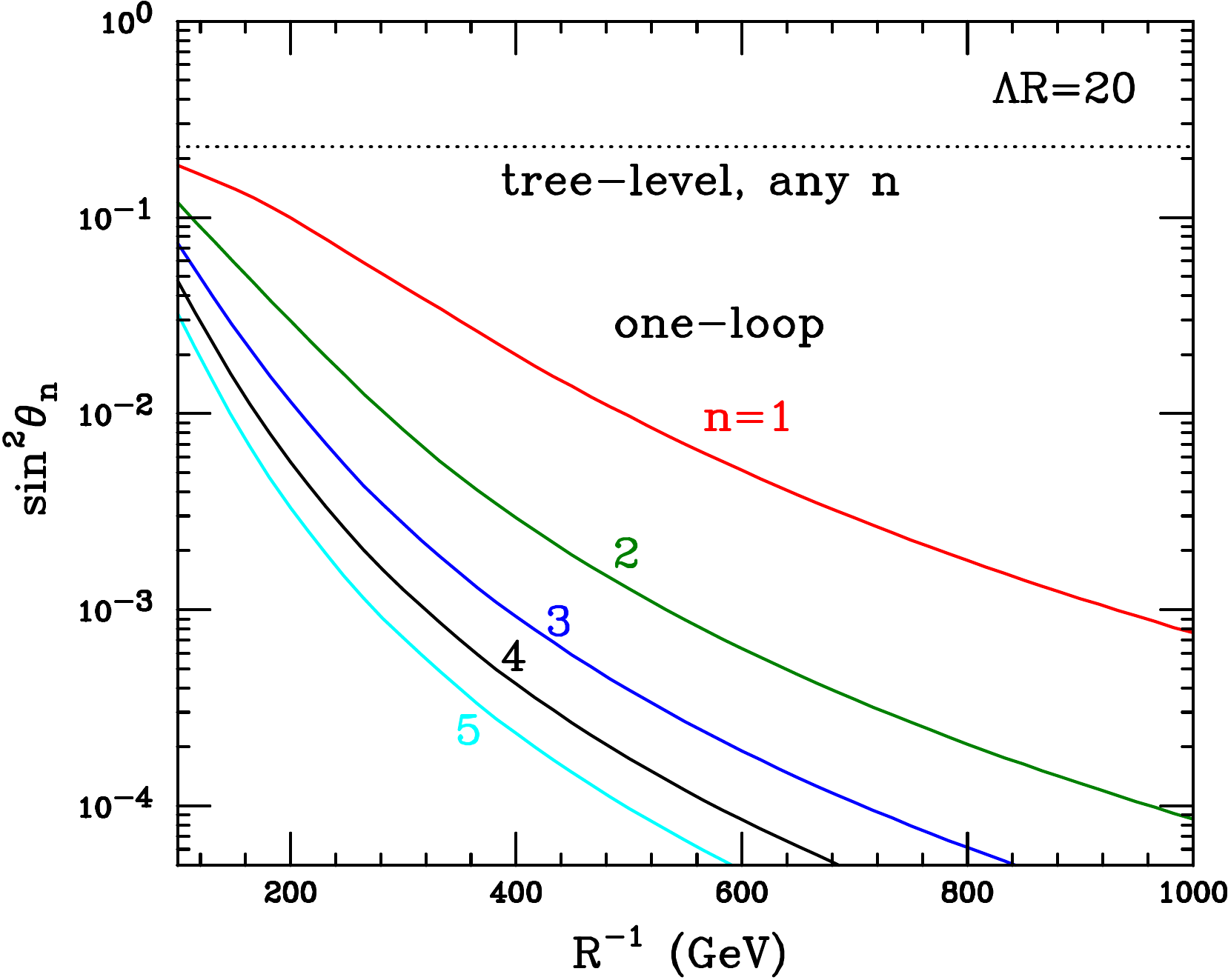}\hspace*{0.3cm}
\includegraphics[width=0.48\columnwidth]{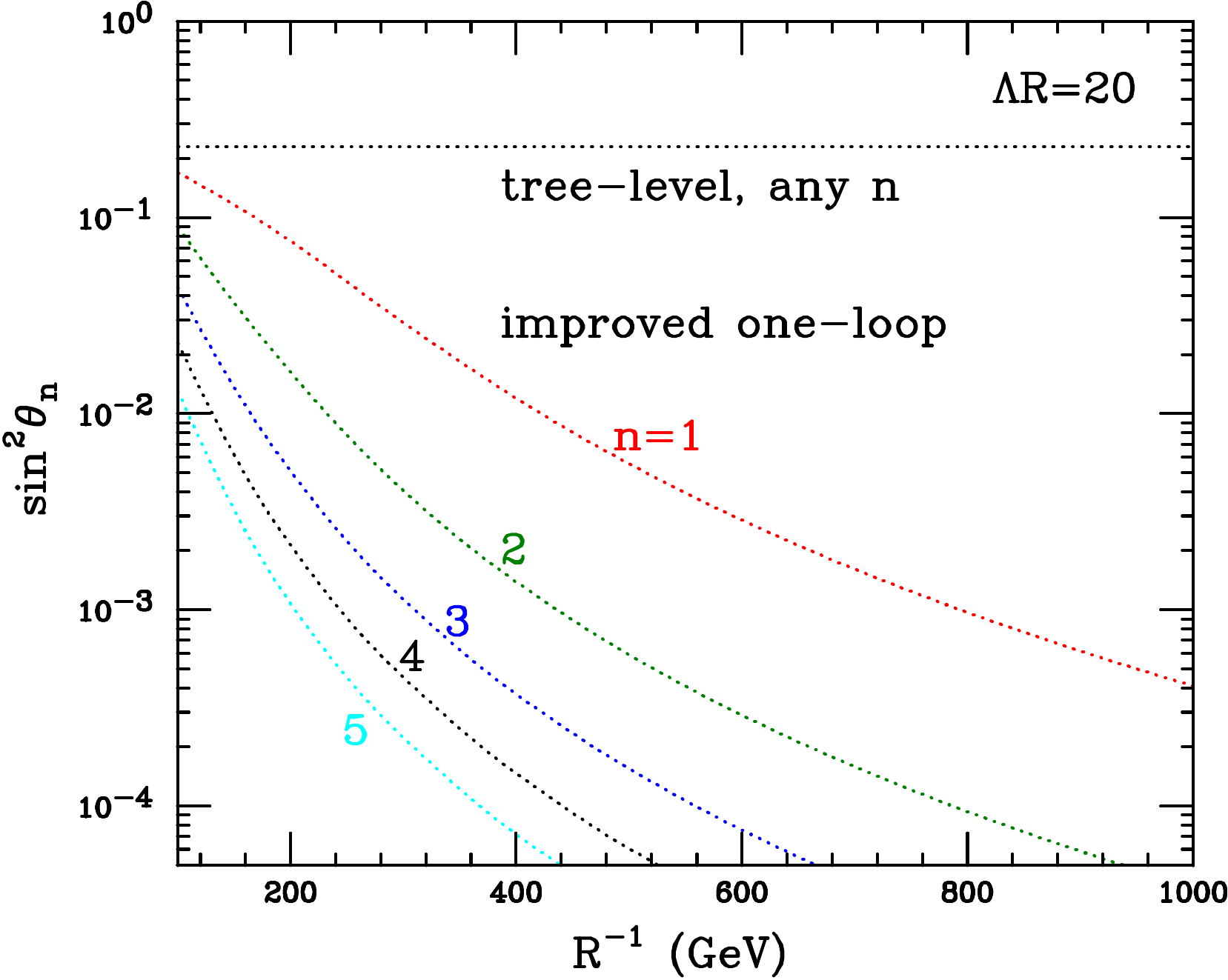} 
\mycaption{Dependence of the Weinberg angle $\theta_n$ for KK levels ($n=1, \cdots, 5$) on $R^{-1}$ for $\Lambda R=20$ with (right) / without (left) finite contributions. 
 \label{fig:weinberg}}
\end{figure}

Finally, we revisit the mass eigenstates of the KK photon and KK $Z$ boson.
In the weak eigenstate basis, the mass matrix is found to be
\begin{equation}
\left ( 
\begin{array}{cc}
\frac{n^2}{R^2} + \hat\delta m^2_{B_n} + \frac{1}{4} g_1^2 v^2& \frac{1}{4} g_1 g_2 v^2 \\
 \frac{1}{4} g_1 g_2 v^2 & \frac{n^2}{R^2} + \hat\delta m^2_{W_n} + \frac{1}{4} g_2^2 v^2 
\end{array} \label{eq:massv2}
\right )\, ,
\end{equation}
where $\hat\delta$ is the total one-loop correction, including both bulk and
boundary contributions. 
In Fig.~\ref{fig:weinberg}, we show the dependence of the Weinberg mixing angle
$\theta_n$ on $R^{-1}$ for the first five KK levels ($n=1, \dots, 5$) for $\Lambda
R=20$, without (left) and with (right) finite contributions respectively. As shown in the plots,
the Weinberg angles are further suppressed by the finite terms,
$\frac{\sin^2\theta^{\rm new}_n}{\sin^2\theta^{\rm old}_n} \lesssim 0.55$ for large
$R^{-1}$. Their dependence on $\Lambda R$ is weak and similar to that in
Ref.~\cite{cms}.

\subsection{Branching ratios of level-2 KK excitations}
\label{sc:br}

\noindent
The decay channels of level-1 KK particles are the same as before.
Although there are minor numerical changes due to the change in mass spectrum,
including finite terms, the main branching fractions remain the same as those in
Ref.~\cite{cms2}. In this section, we focus on the branching fractions of
level-2 KK particles.

Unlike the decay of $n=1$ KK particles, which always give rise to an invisible
stable KK particle in the decay, $n=2$ decays do not necessarily produce such
missing-particle signatures.
In fact, there are three decay channels of level-2 KK particles: 
(i) decay to two $n=0$ modes (denoted as $200$), 
(ii) decay to two $n=1$ modes ($211$), and 
(iii) decay to one $n=2$ and one $n=0$ modes ($220$).
Both the $220$ and the $211$ channels are phase space suppressed, since KK particles are
more or less degenerate around $m_n \sim n/R$, while $200$ decays are
suppressed by one loop. Therefore branching fractions of $n=2$ KK particles are
sensitive to details of the coupling structure and mass spectrum, which
illustrates the 
importance of computing the finite corrections. In the case of the $211$ decay
channel, each level-1 KK particle would then proceed through its own cascade decay and give
one missing particle at the end. Therefore single production of a level-2 KK
particle followed by a $211$ decay gives two missing particles, while pair production
of level-2 KK particles plus their subsequent $211$ decay gives four missing
particles at the end of their cascade decays. On the other hand, a KK particle that
decays via a $200$ channel will appear as a resonance, if both SM particles can be
reconstructed. 

\subsubsection{$\psi_2$ decays\label{sec:f2decay}}

\begin{figure}[t]
\centering
\includegraphics[width=0.47\columnwidth]{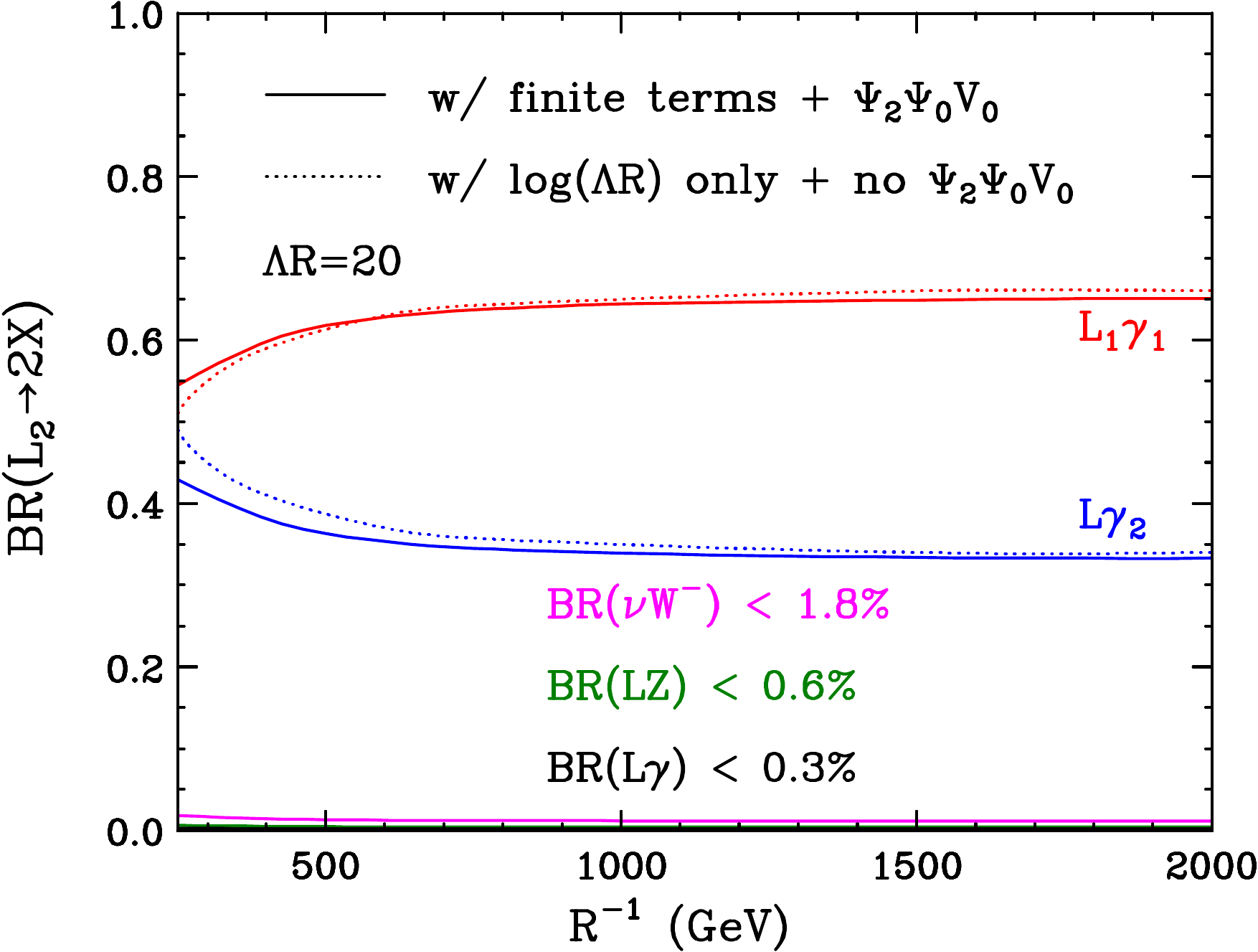}\hspace*{0.3cm}
\includegraphics[width=0.47\columnwidth]{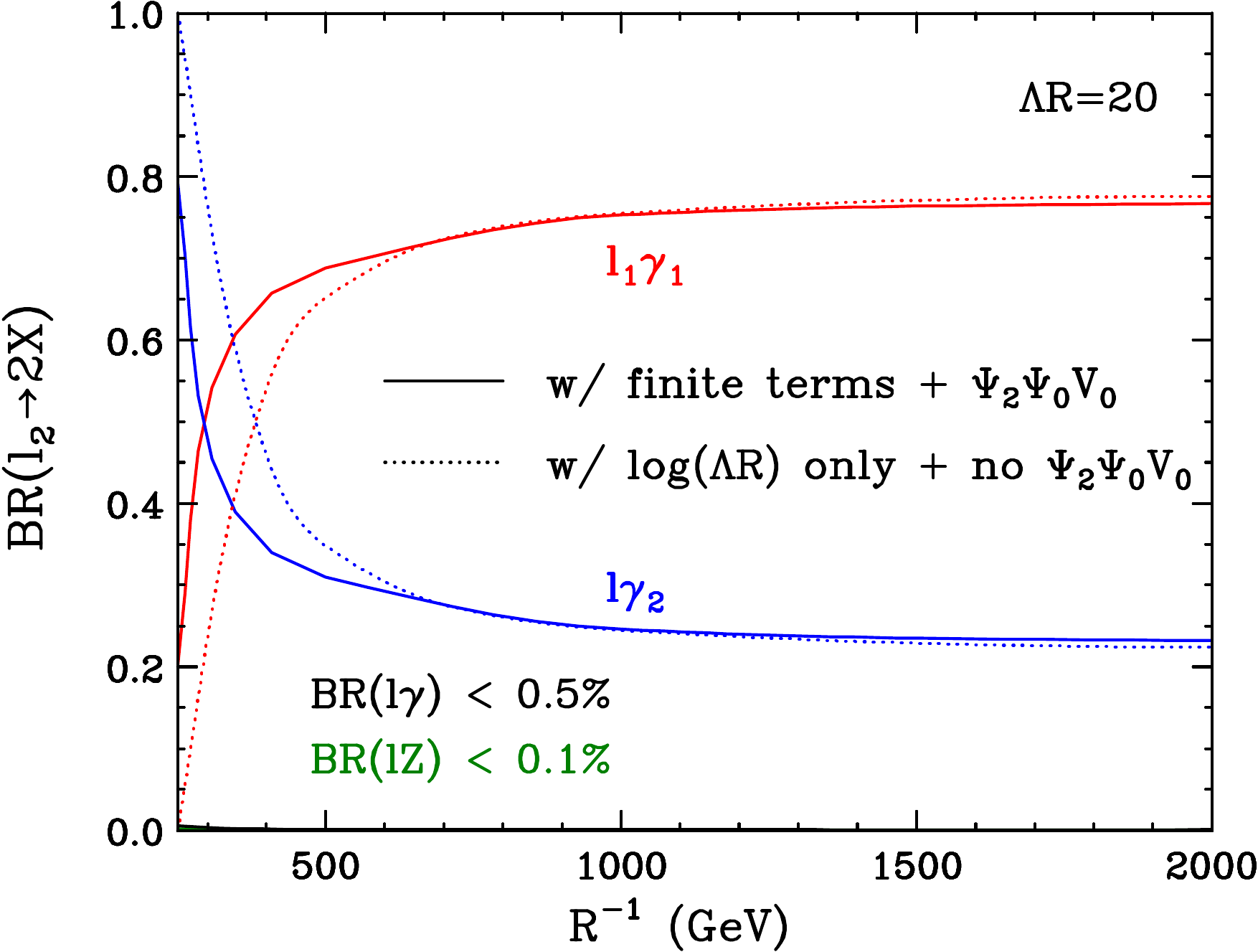} \\
\mycaption{Branching fractions of SU(2)$_{\rm L}$-doublet level-2 KK lepton (left) and charged SU(2)$_{\rm L}$-singlet level-2 KK lepton (right).
Solid curves include finite corrections and new decay channels, while dotted curves are old results. 
 \label{fig:eL2eR2}}
\end{figure}

\begin{figure}[th!]
\centering
\includegraphics[width=0.47\columnwidth]{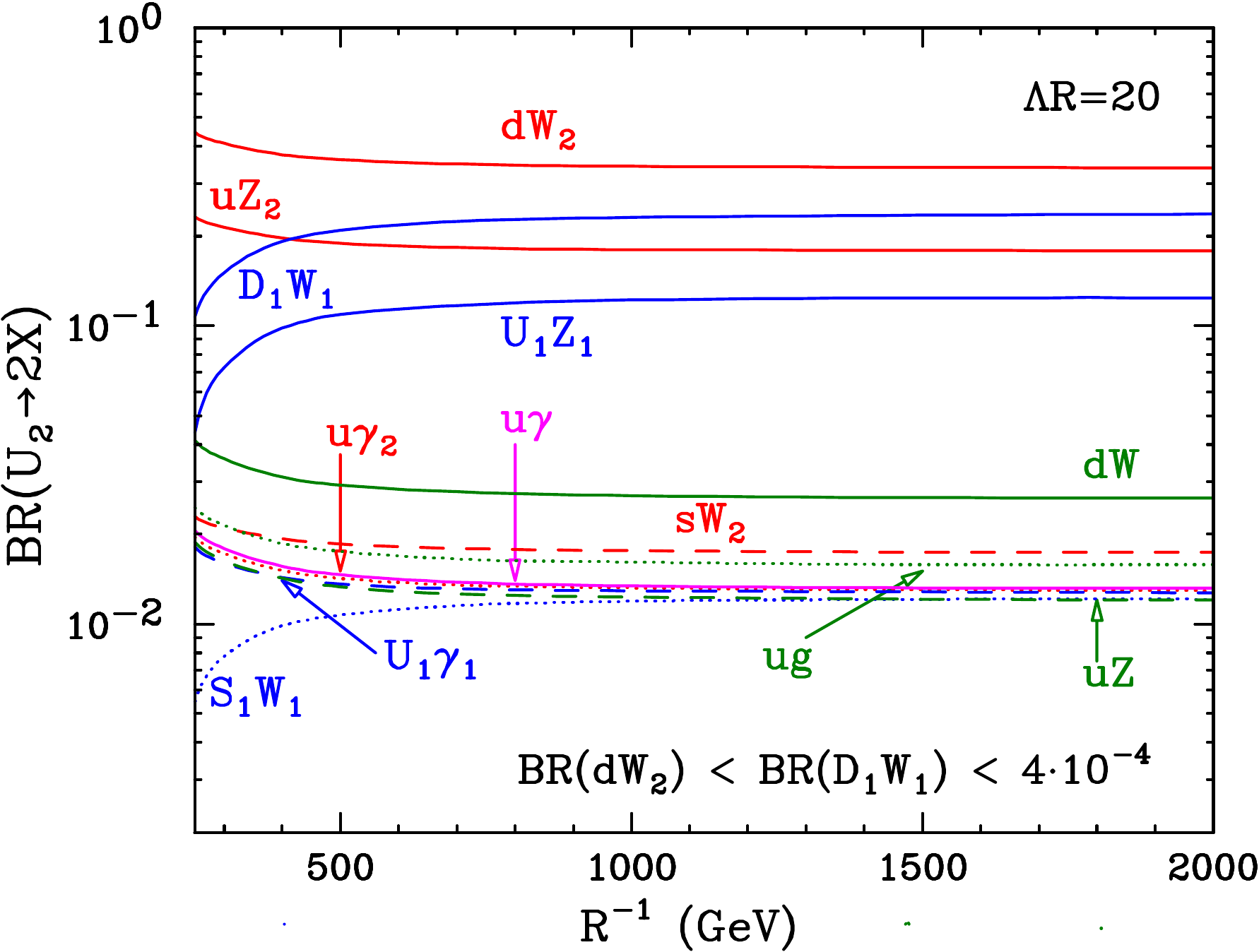} \hspace*{0.3cm}
\includegraphics[width=0.47\columnwidth]{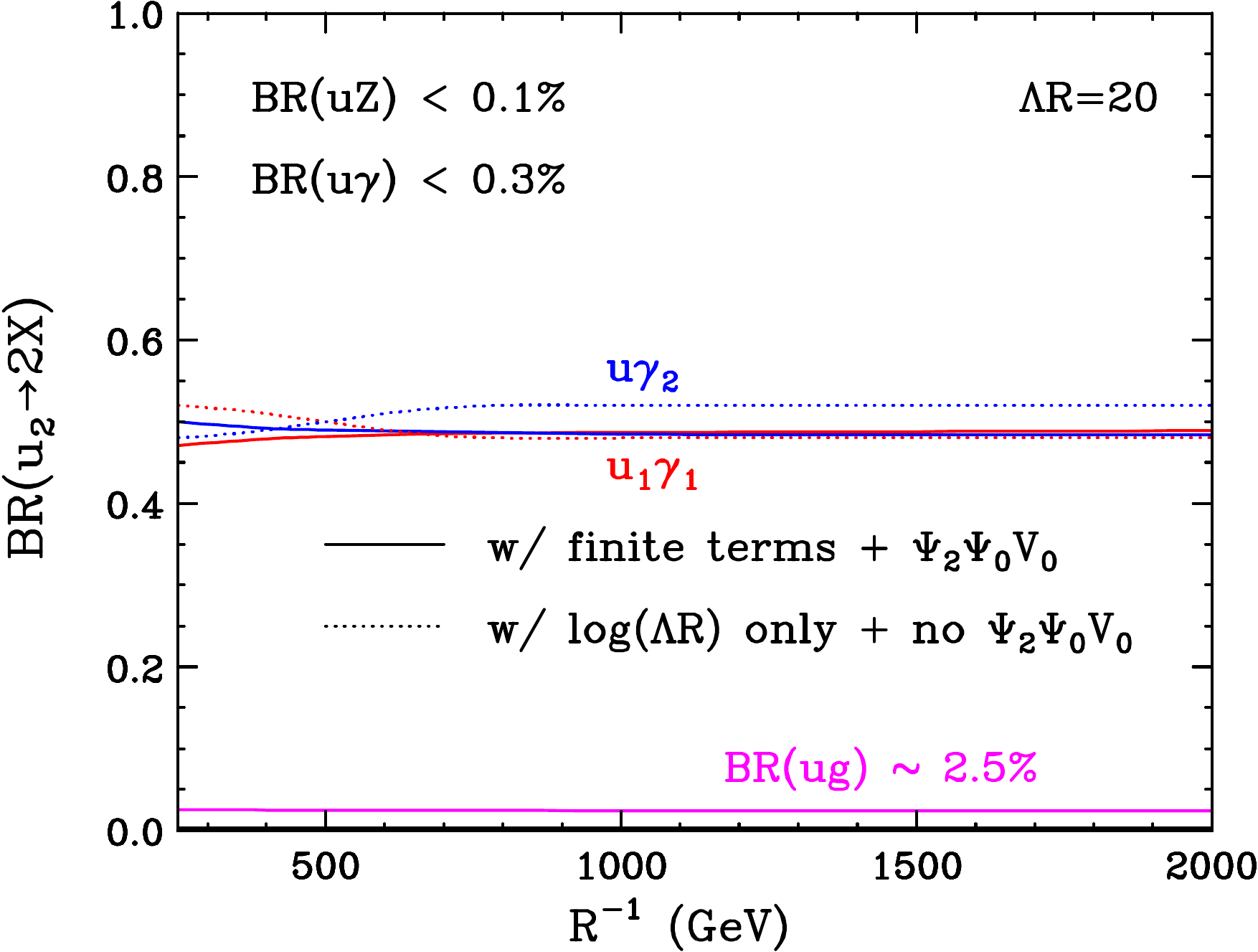}
\mycaption{Branching fraction of SU(2)$_{\rm L}$-doublet level-2 KK quark (left) and SU(2)$_{\rm L}$-singlet level-2 KK quark (right) for the up-type.
 \label{fig:uL2uR2}}
\end{figure}

\noindent
We first consider the branching fractions of level-2 KK fermions, which are shown in Figs.~\ref{fig:eL2eR2} and \ref{fig:uL2uR2} for KK quarks and KK leptons, respectively. 
The branching fractions for SU(2)$_{\rm L}$-doublet KK fermions are shown in the left panel, while those for SU(2)$_{\rm L}$-singlet KK fermions are on the right. 
In Fig.~\ref{fig:eL2eR2} and the right panel of Fig.~\ref{fig:uL2uR2}, results with finite corrections are shown in solid curves, whereas previous results from Ref.~\cite{uedlevel2} are shown in dotted curves. 
While one observes no significant changes in existing decay channels, there are new ones based on our findings as explained in the previous section.
SU(2)$_{\rm L}$ KK leptons have the new $\nu W^-$, $\ell Z$, and $\ell\gamma$ channels, which contribute with 0.1\% to 2\%, while the branching fraction of $u_2$ to $u g$ is as big as 2.5\%. 
In the case of KK lepton decays, EWSB effects are important, {\it i.e.} a sizable mixing between KK photon and KK $Z$ is expected for low values of $R^{-1}$ (see eq. \ref{eq:massv2}).
It turns out that $m_{\ell_2} - m_{\gamma_1} - m_{\ell_1}$ approaches $0$  as $R^{-1} \to v$, which is why the $\ell\gamma_2$ branching fraction becomes larger. This effect is more pronounced for the SU(2)$_{\rm L}$ singlet lepton. This pattern does not appear for KK quarks, since mass corrections to KK quarks are larger than those to KK gauge bosons (see Fig. \ref{fig:uL2uR2}). 

Branching fractions of SU(2)$_{\rm L}$ doublet quarks are rather complicated as shown in the left panel of Fig. \ref{fig:uL2uR2}. New decay channels $dW$, $ug$ and $uZ$, show branching fractions of $2.5$--$4.5\%$, $1.5$--$2.5\%$ and $1$--$2\%$, respectively.

Branching fractions for neutral KK leptons and the down-type KK quarks are similar.
For example, looking at the $d_2$ decay we find the $d \gamma_2$ and $d_1 \gamma_1$ channels to be dominant with BR $\sim 45\%$, but the branching fraction into $d g$ is slightly higher at about 8\%. This is due to the different hypercharge couplings between the up-type and the down-type quarks. Branching fractions into $d Z$ and $d \gamma$ are negligible as before.

\begin{figure}[t!]
\centering
\includegraphics[width=0.47\columnwidth]{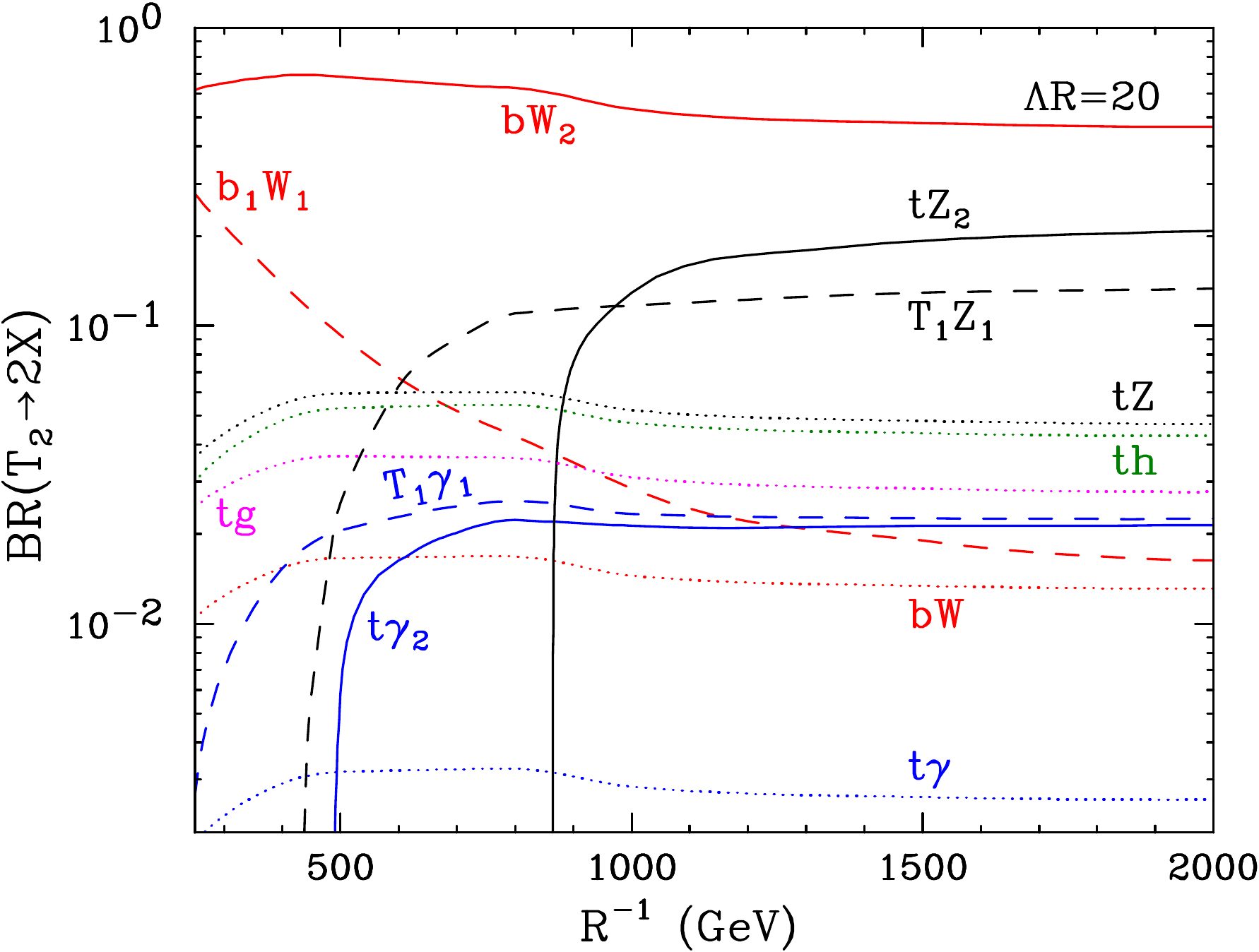}\hspace*{0.3cm}
\includegraphics[width=0.47\columnwidth]{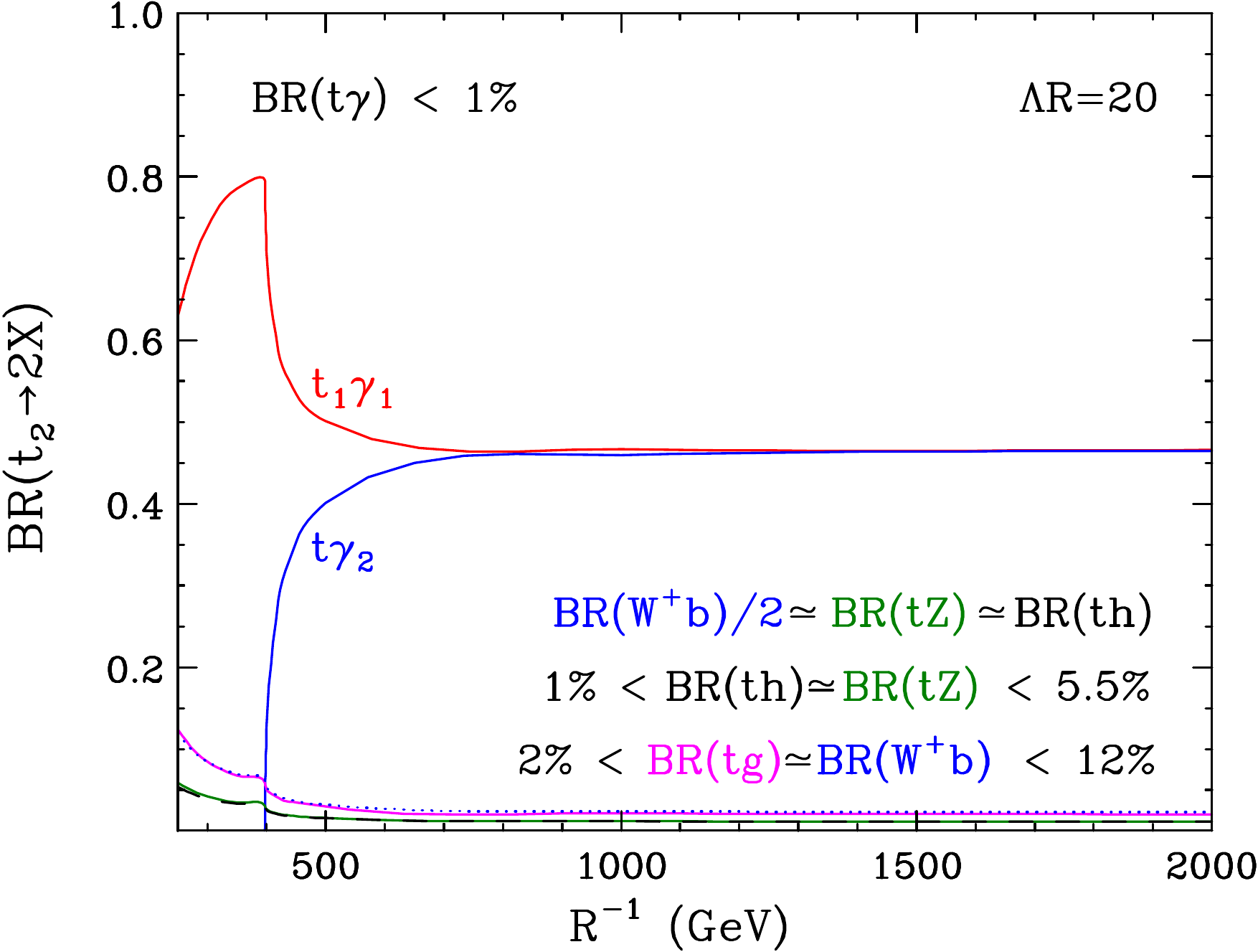} \\
\mycaption{Branching fraction of SU(2)$_{\rm L}$-doublet KK top quark (left) and SU(2)$_{\rm L}$-singlet KK top quark (right).
 \label{fig:ut2}}
\end{figure}

Finally we show the branching fractions of level-2 KK top quarks in Fig.~\ref{fig:ut2}. In this case we find that the branching fractions of the SU(2)$_{\rm L}$ doublet KK top into $t h$ or $t Z$ are 3--6\% each.
Other 200 decay modes into $t g$ and $bW^+$ show branching fractions of about 2--4\% and 1--2\%, respectively. 
The $t\gamma$ branching fraction is below one percent, which implies that the KK top decays directly to two SM particles $\sim$ 8--15\% of the time. 
The SU(2)$_{\rm L}$ singlet KK top does not have decays to SU(2)$_{\rm L}$ gauge bosons, and branching fractions for $th$ and $t g$ are of order 1--5.5\% and 2--12\%, respectively, for $ 250 \gev < R^{-1} < 2 \tev$.
Due to the Yukawa correction to the KK top mass, two-body decays of level-2 KK top into $W^+_1 B_1$, $t_1 Z_1$ and $t \gamma_2$ are suppressed for low values of $R^{-1}$.

\subsubsection{$V_2$ decays\label{sec:v2decay}}

\begin{figure}[t!]
\centering
\includegraphics[width=0.47\columnwidth]{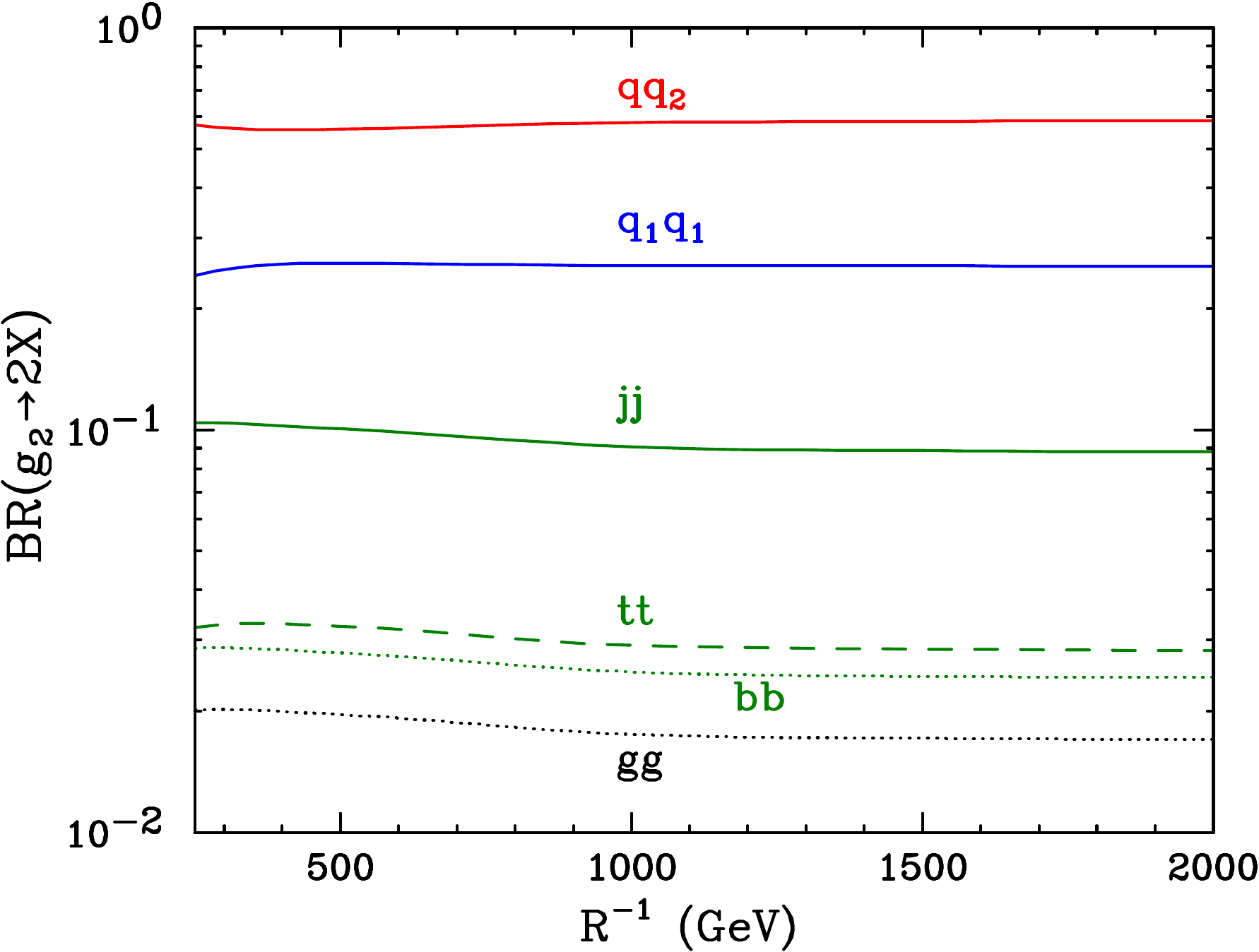}\hspace*{0.1cm}
\includegraphics[width=0.47\columnwidth]{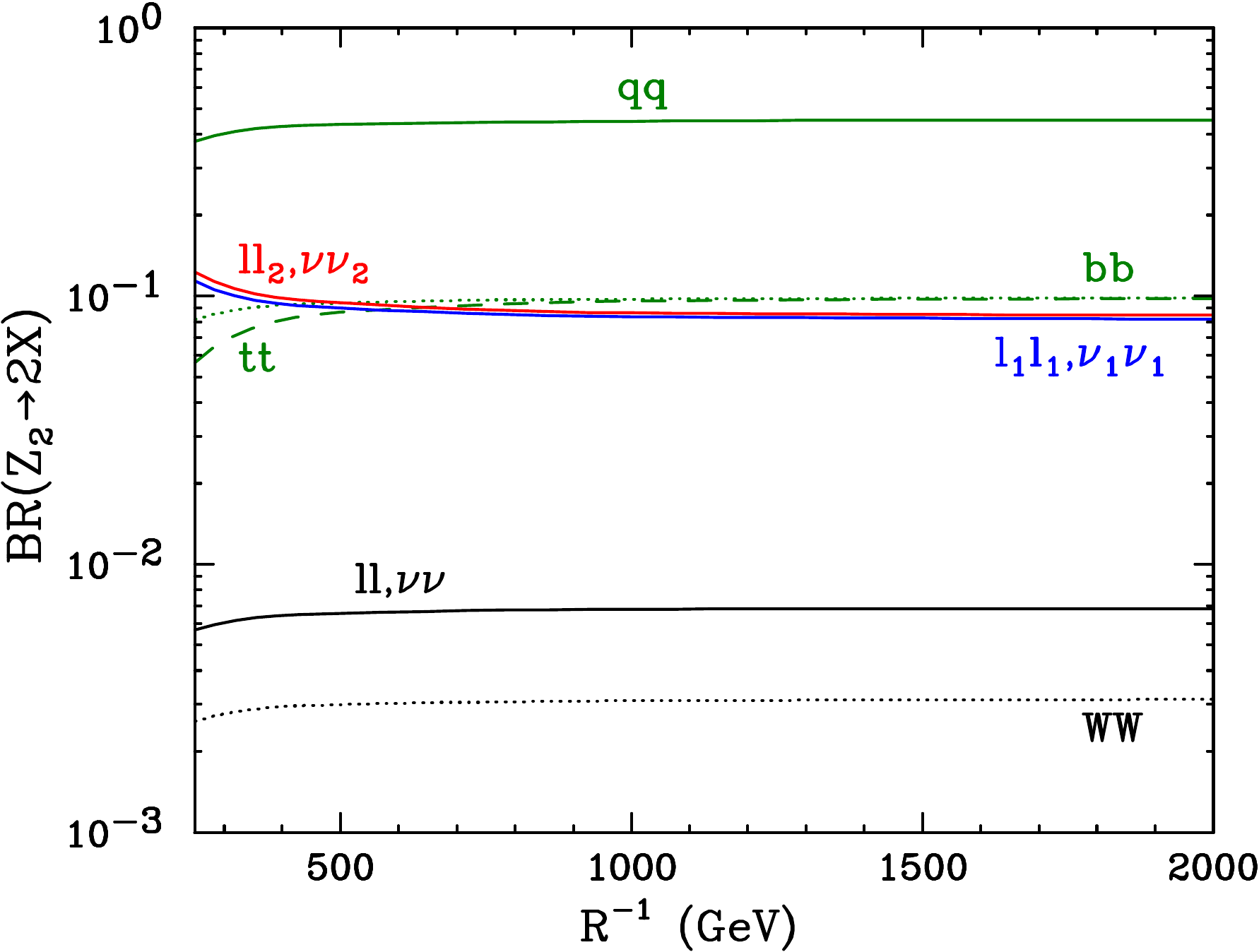} \\ \vspace{0.2cm}
\includegraphics[width=0.47\columnwidth]{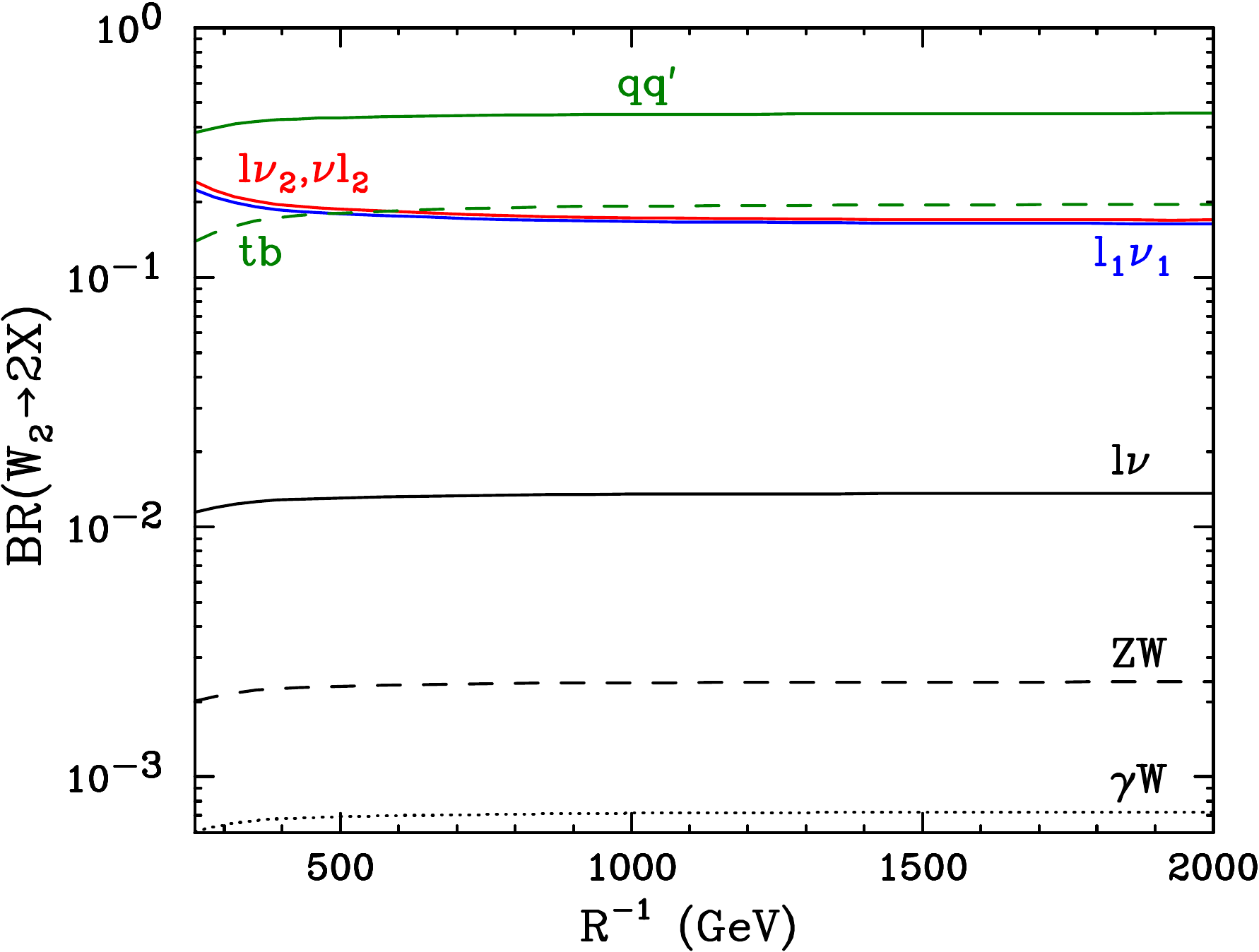}\hspace*{0.1cm}
\includegraphics[width=0.47\columnwidth]{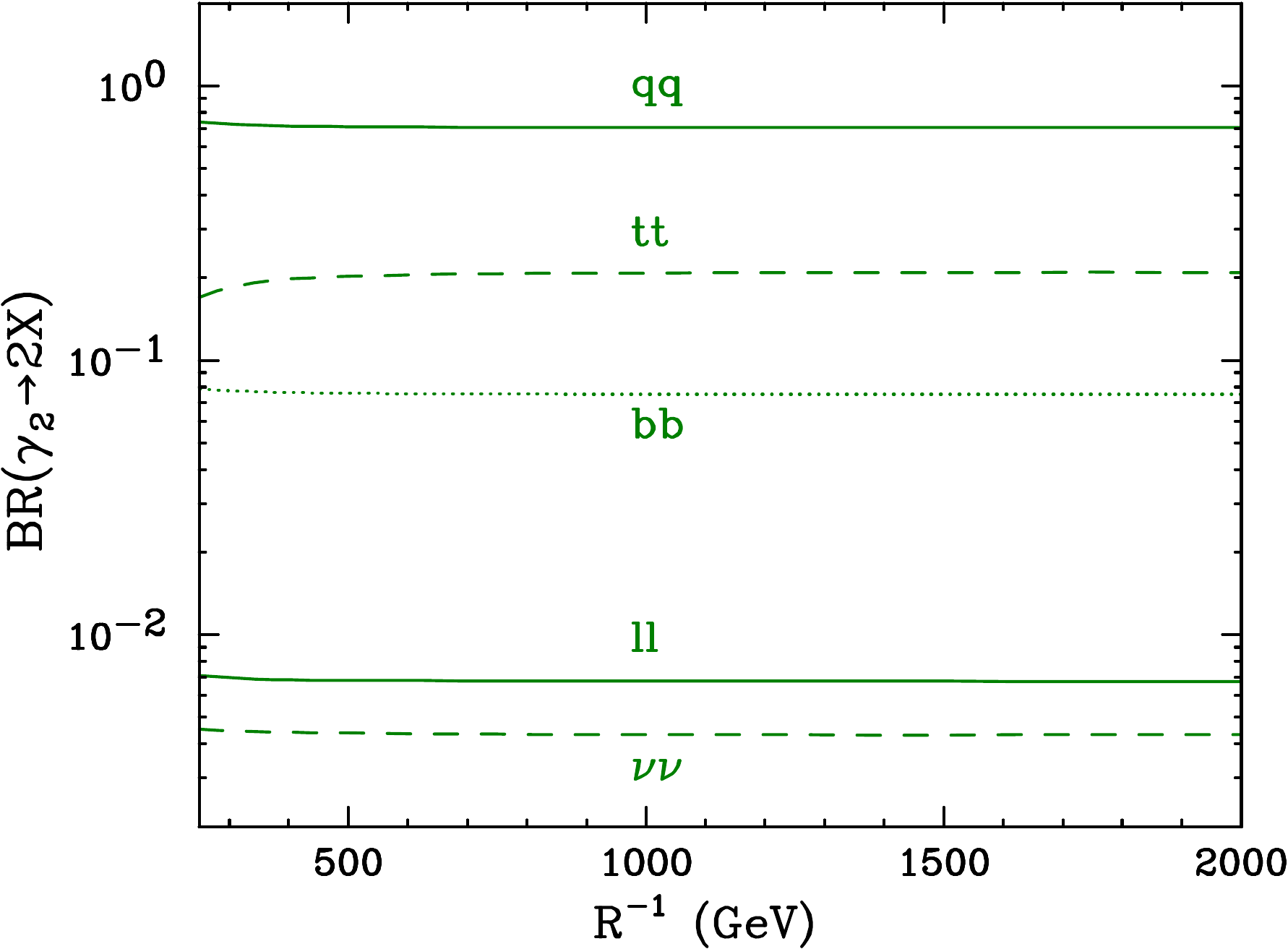} 
\mycaption{Branching fractions of $\gamma_2$, $Z_2$, $W_2^\pm$ and $g_2$ for $\Lambda R=20$.
 \label{fig:v2}}
\end{figure}

\noindent
Fig.~\ref{fig:v2} shows the branching fractions of $n=2$ KK gauge bosons as a function of $R^{-1}$. Overall, we find our results are similar to those in Ref. \cite{uedlevel2} with a few notable changes.

\medskip
Firstly we considered the new decay channels $g_2 \to gg$, $Z_2 \to W^+ W^-$,  $W^\pm_2 \to Z W^\pm$ and $W^\pm_2 \to \gamma W^\pm$. Their rather moderate branching fractions contribute with 1--2\%,  0.3\%, 2--3\%, and 0.7\%, respectively. 
The leptonic branching fractions of $Z_2$ and $\gamma_2$ become smaller with finite corrections and are now about 0.7\%. 

\begin{figure}[t!]
\centering
\includegraphics[width=0.47\columnwidth]{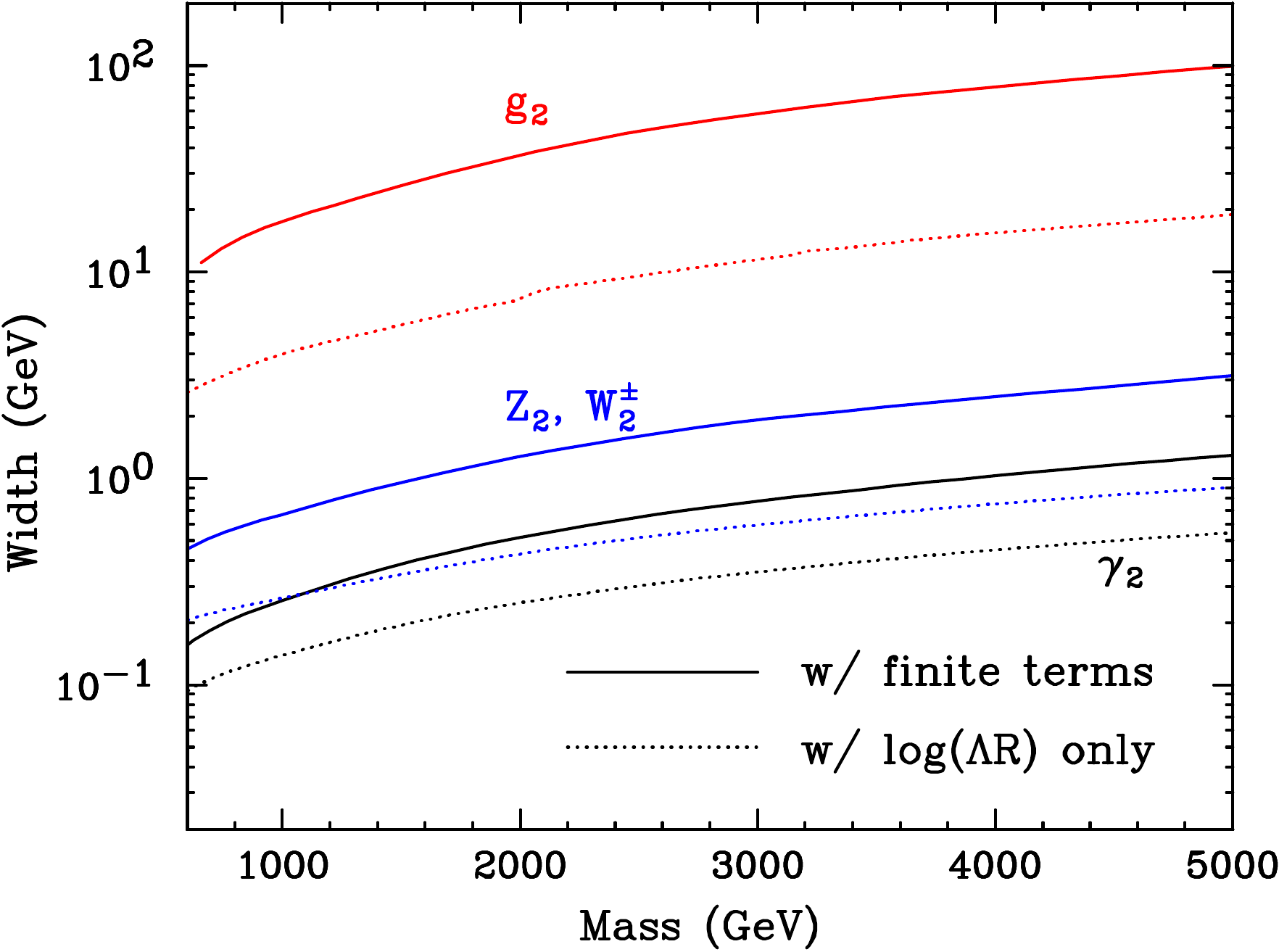} 
\mycaption{The decay width of level-2 gauge bosons as a function of $R^{-1}$ for $\Lambda R=20$. Solid curves include finite corrections, while dotted curves are old results.
 \label{fig:v2width}}
\end{figure}

Due to larger mass corrections for strongly interacting particles (KK-gluon and KK-quarks), only the $n=2$ KK gluon can decay to KK-quarks ($q q_2$ or $q_1 q_1$), while two-body decays of KK $Z$ and $W$ gauge bosons into KK quarks are kinematically closed. With finite corrections and additional decay channels, the total decay widths of level-2 bosons increase by a factor of $\sim 2$ for electroweak gauge bosons and $\sim 5$ for KK gluon as shown in Fig.~\ref{fig:v2width}. However, their decay widths are still very small due to the phase space suppression of 220 and 211 decays and loop-suppression of 200 decays, as mentioned at the beginning. For electroweak gauge bosons, $\Gamma_{V_2}/m_{V_2} \lesssim 10^{-3}$, and $\Gamma_{g_2}/ m_{g_2}  \sim 0.02$ for KK gluons.  

\subsection{Cross-sections and signatures}
\label{sc:xsection}

\noindent
Single production of level-1 KK particles is forbidden due to KK-parity and therefore 
they must be produced in pairs from collisions of two SM particles or from the  decay of level-2 KK particles. However, both single and pair productions are possible for level-2 KK particles.
Single production cross sections are suppressed by a loop factor, while pair production cross sections are suppressed by phase space.

All cross sections are calculated at tree level considering five partonic quark flavors in the proton along with the gluon at the 14 TeV LHC. We sum over the final state quark flavors and include charge-conjugated contributions. We used CTEQ6L parton distributions \cite{Pumplin:2002vw} and chose the scale of the strong coupling constant to be equal to the parton-level center-of-mass energy. All results are obtained using {\sc CalcHEP} \cite{Belyaev:2012qa} based on the  implementation of the MUED model from Ref.~\cite{uedcomphep}.
Since the particle content and KK number conserving interactions remain the same, we only modified the KK mass spectrum and KK number violating interactions in the existing implementation which is based on Ref.~\cite{cms}. We also implemented the new interactions which are described throughout our paper.

\begin{figure}[t!]
\centering
\includegraphics[width=0.47\columnwidth]{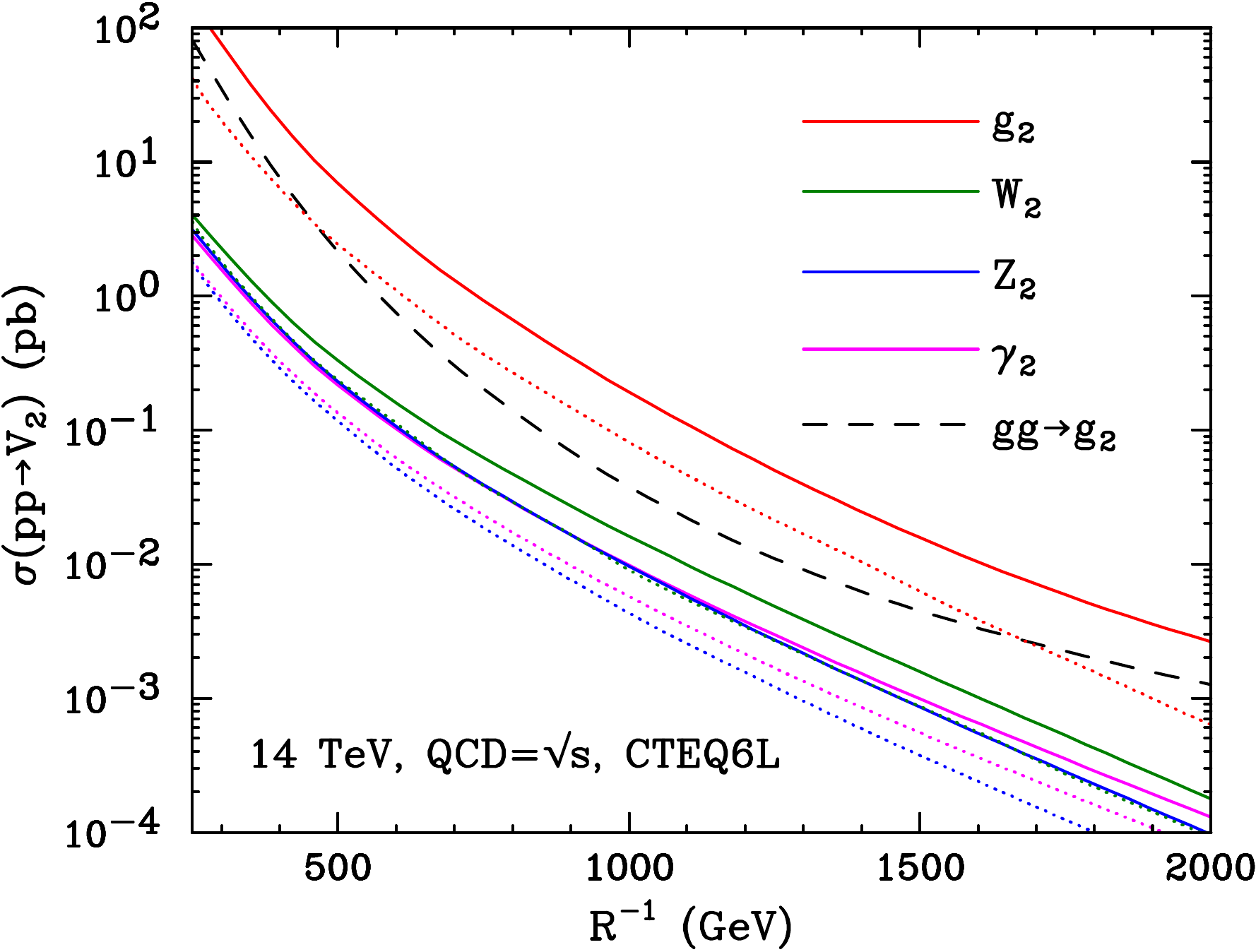}\hspace*{0.3cm}
\includegraphics[width=0.47\columnwidth]{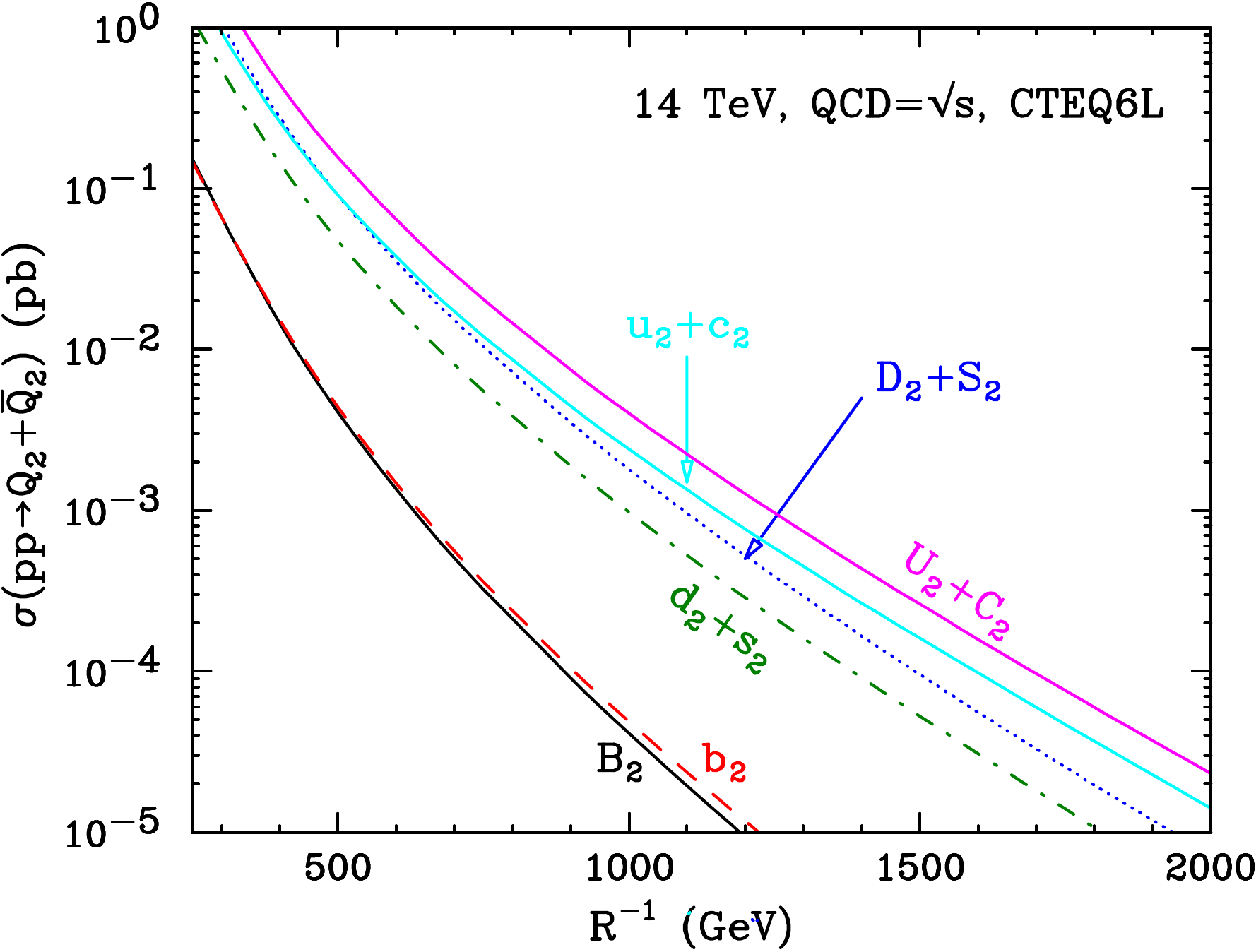} \\  
\mycaption{Single production cross section of level-2 KK gauge bosons (left) and level-2 KK fermions (right) as a function of $R^{-1}$. Dotted curves (left) are results from Ref. \cite{uedlevel2} and solid curves are new results including finite terms. Level-2 fermion cross sections and $\sigma( gg\to g_2 ) $ have been computed first time. The cut-off scale has been set to $\Lambda R=20$.
 \label{fig:single}}
\end{figure}

\begin{figure}[t!]
\centering
\includegraphics[width=0.47\columnwidth]{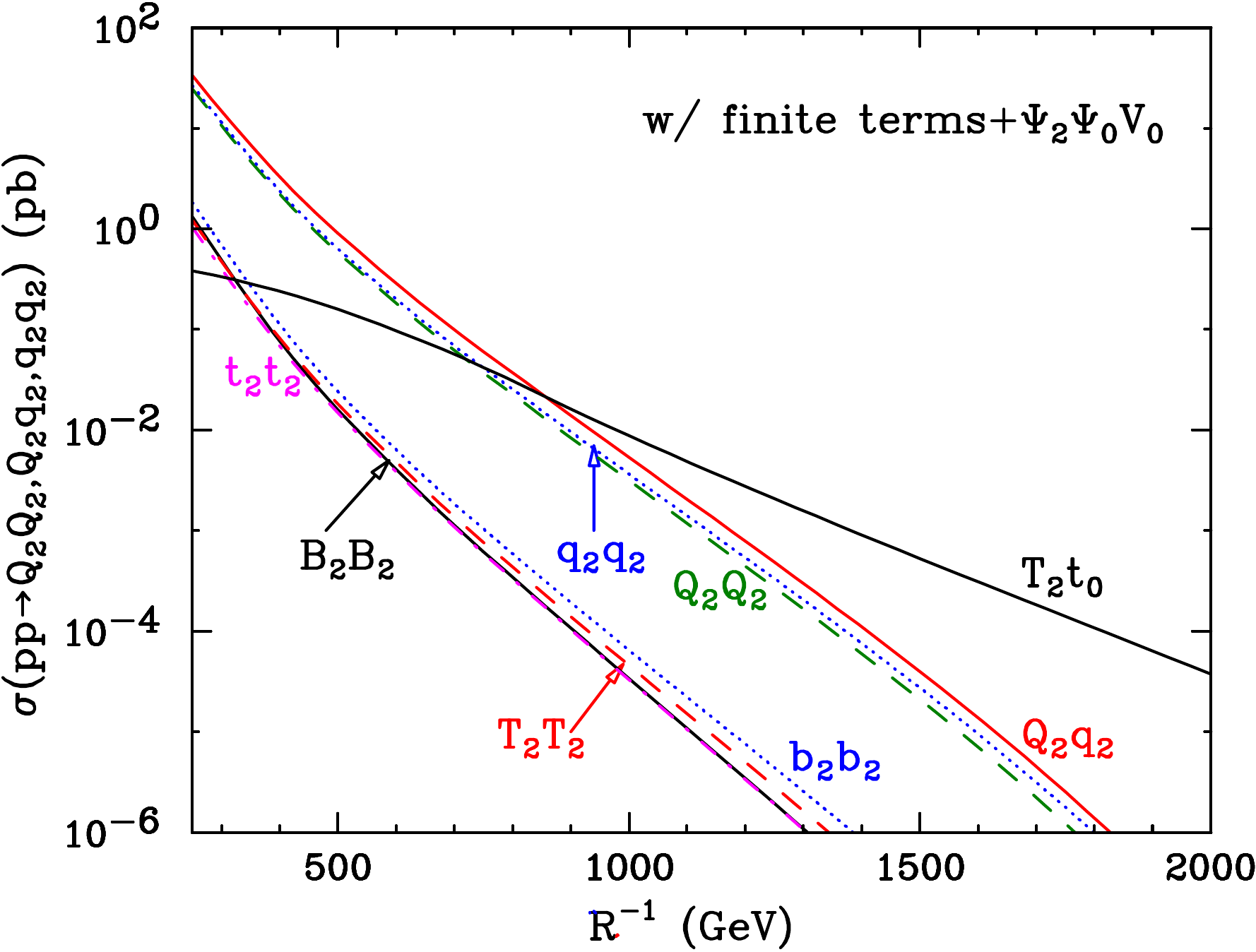}\hspace*{0.1cm}
\includegraphics[width=0.47\columnwidth]{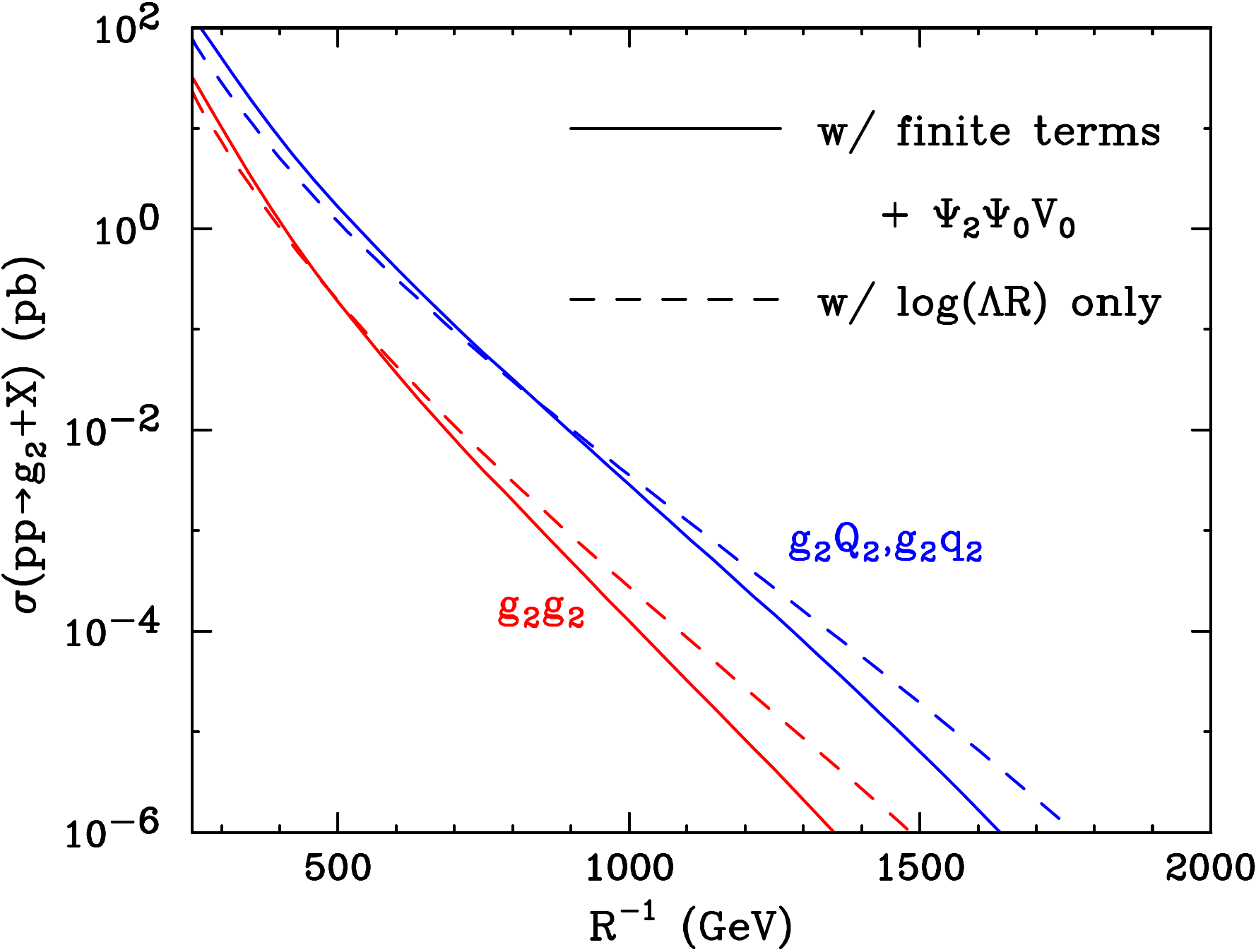} \\  
\mycaption{
Strong production of $n=2$ KK particles at the 14 TeV LHC. The left panel shows KK-quark pair production, while the right panel shows KK-quark/KK-gluon associated production and KK gluon pair production.
Updated results (solid curves) are similar to old results (in dashed curves from Ref. \cite{uedlevel2}). The cut-off scale has been set to $\Lambda R=20$.
 \label{fig:pair}}
\end{figure}

We summarize single production cross sections of $n=2$ KK gauge bosons (left) and $n=2$ KK quarks (right) in Fig.~\ref{fig:single}. 
While overall one observes a slight increase in production cross sections for the KK-gauge bosons, the $gg \to g_2$ production channel has been computed for the first time here and contributes at a sizable level. 
All KK-fermion single production cross sections presented here had also not been considered previously. 
Fig.~\ref{fig:pair} shows the pair production of KK quarks (left) and associated production of KK quark and KK gluon (right), respectively. 

Another interesting channel is associated production of KK top with SM top quark.
$pp \to T_2 \bar{t} + t \bar{T}_2$ is shown as a (black, solid) curve, labeled as `$T_2 t_0$' in the left panel of Fig. \ref{fig:pair}. Since $T_2$ has a large branching fraction into $th$ and a sizable branching fraction into $t \gamma$, $bW$ or $tZ$, this production could be constrained by cross section measurements of SM processes such as $t\bar t \gamma$, $t \bar t$, $t \bar t h$ and $t \bar t$.

\begin{figure}[t!]
\centering
\includegraphics[width=0.48\columnwidth]{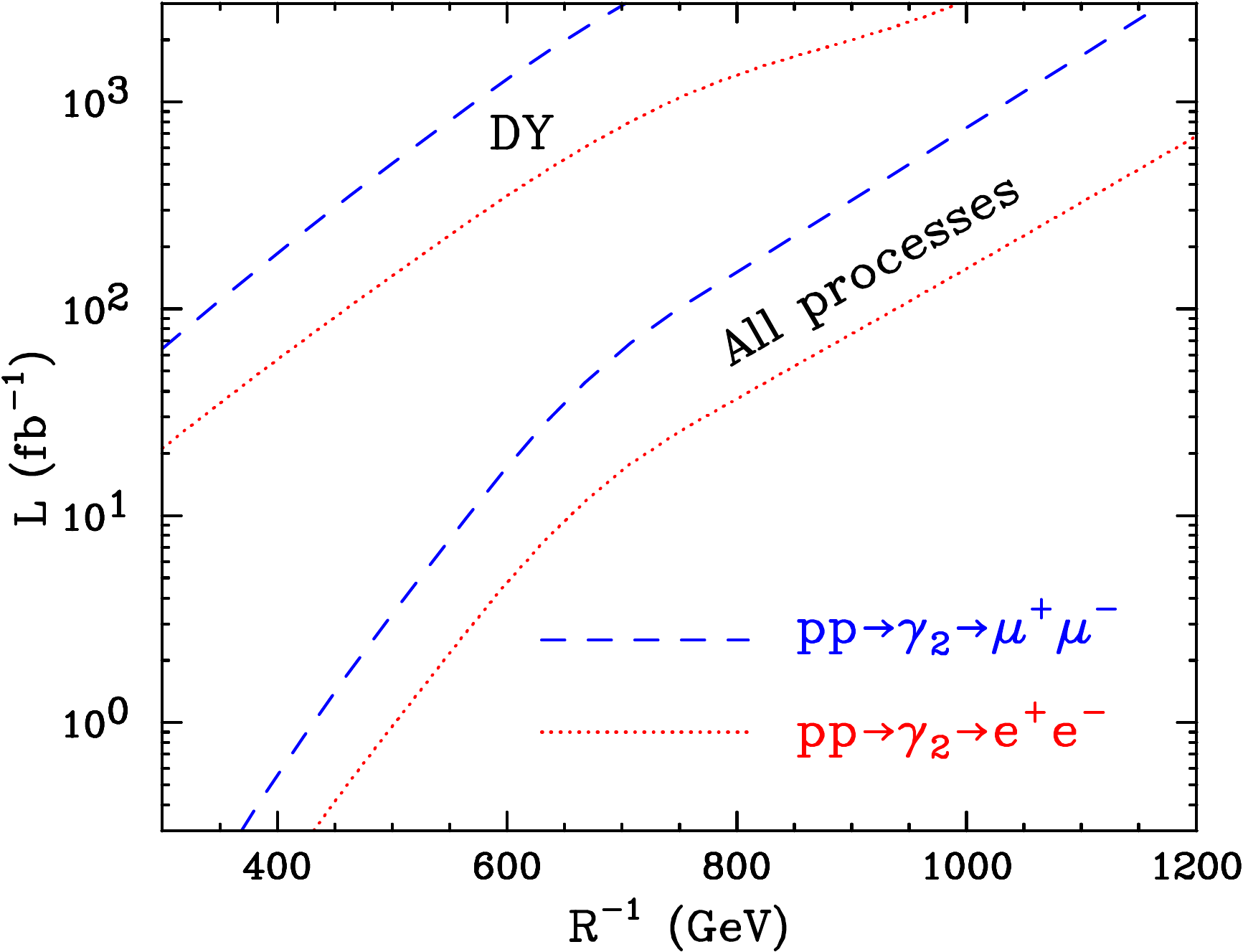}\hspace*{0.3cm}
\includegraphics[width=0.48\columnwidth]{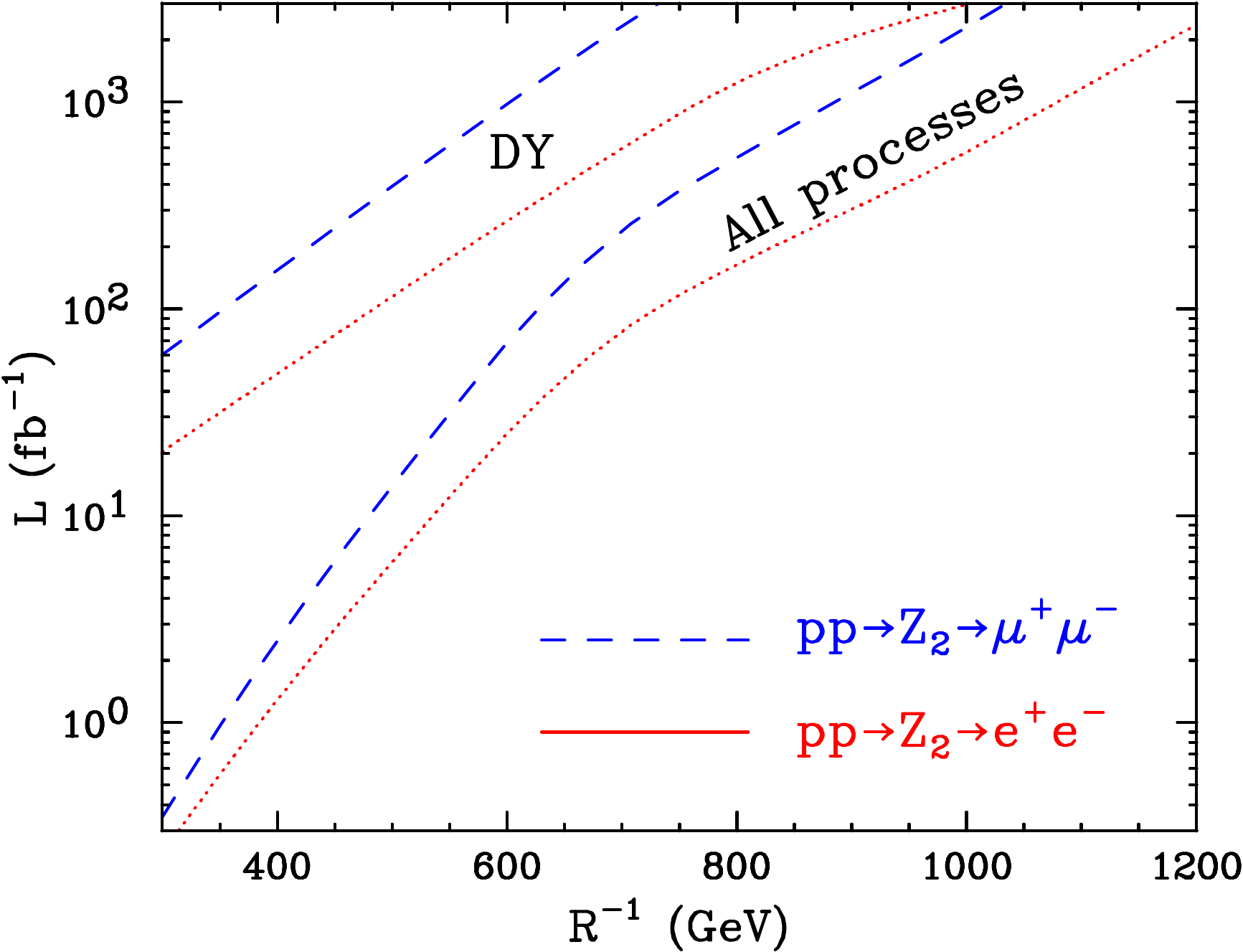} \\  
\mycaption{5$\sigma$ discovery reach for $\gamma_2$ (left) and $Z_2$ (right). We show the total integrated luminosity ${\cal L}$ (in fb$^{-1}$) required for a 5$\sigma$ excess of signal over SM backgrounds in the di-electron (red, dotted) and di-muon (blue, dashed) channels. In each plot, the upper set of curves labeled as `DY' make use of the single production of $\gamma_2$ or $Z_2$ (from Fig.~\ref{fig:single}), while the lower set of curves labeled as `All processes' includes indirect $\gamma_2$ and $Z_2$ productions from $n=2$ KK quarks (see Fig.~\ref{fig:pair}). We assumed the same signal and background efficiencies used in Ref.~\cite{uedlevel2} and combined with our updated cross sections and branching fractions.
 \label{fig:reach}}
\end{figure}

Finally, we plot the total integrated luminosity ${\cal L}$ (in fb$^{-1}$) required for a 5$\sigma$ excess of signal over background in the di-electron (red, dotted) or di-muon (blue, dashed) channel, as a function of $R^{-1}$ (in GeV). We have used the same backgrounds, basic cuts and detector resolutions as described in Ref.~\cite{uedlevel2}. In each panel of Fig.~\ref{fig:reach}, the upper set of lines labeled `DY' only utilizes the single $V_2$ productions from Fig.~\ref{fig:single}. The lower set of lines (labeled `All processes') includes in addition indirect $\gamma_2$ and $Z_2$ production from the cascade decays of  level-2 KK quarks to level-2 KK gauge bosons from Fig.~\ref{fig:pair}. We do not include contributions from single production of level-2 KK quarks, so as to compare more directly against results in Ref.~\cite{uedlevel2}. They would make a small contribution to the total luminosity as shown in the right panel of Fig.~\ref{fig:single}.

For both di-electron and di-muon channel, we observe no significant change in the required luminosity compared to results from Ref.~\cite{uedlevel2}, although
we notice a slight reduction or increase in the luminosity, depending on the value of $R^{-1}$.
This is due to the interplay between improved results on cross sections and branching fractions. 
Overall, production cross sections are increased as shown in Figs.~\ref{fig:single} and \ref{fig:pair}, while branching fractions decrease as shown Fig.~\ref{fig:v2}. 
The high-luminosity LHC with 3 ab$^{-1}$ would be able to probe the level-2 KK photon up to $R^{-1}\sim1.2$ TeV in the $\mu^+\mu^-$ and $R^{-1} \sim 1.5$ TeV in the $e^+e^-$ channel. The corresponding reach for the level-2 KK $Z$ boson is lower due to the relevant branching fractions.


\section{Conclusions}
\label{sc:concl}

\noindent
In this article we presented the one-loop corrected mass spectrum and KK-number violating decays of level-2 KK states into SM particles in models with universal extra dimensions. 
As a concrete framework we chose to add one additional universal extra dimension to the SM, which is compactified on a circle with a $\mathbb{Z}_2$ orbifold. Due to its non-renormalizability the model is regarded as an effective low-energy theory with a hard cutoff scale  $\Lambda$ at which an unspecified UV-completion is expected to describe the physics. This enables us to write down sensible \msbar counterterms with a logarithmic sensitivity to the cutoff. All calculations were performed in the 4D effective theory using publicly available software supplemented by in-house routines.

The self energy diagrams giving rise to the mass corrections contain an infinite tower of states in the loop, whose summation requires additional regularization. To this end we employed the Poisson summation identity to identify the divergent pieces in winding number space and remove them. The results can be divided into logarithmically divergent boundary terms and finite bulk contributions.

The low cutoff scale ($\Lambda R \lesssim 50$, considering perturbativity and unitarity \cite{ued,uedcutoff}) implies that the leading (logarithmic) terms in the one-loop corrections to KK masses are not as large as those in supersymmetry, and finite contributions could play an important role phenomenologically. 

When including the new finite (non-logarithmic) corrections, the mass spectrum broadens and each KK particle becomes heavier, which implies that the pair production cross sections of level-1 KK particles at colliders would be reduced but their acceptance rate would increase. We also examined the nature of the NLKP, and confirmed that it is always the right-handed KK lepton, which is different from what has been stated in the literature, where the NLKP was thought to be the KK-Higgs for a large KK scale. The KK Weinberg angles are further reduced such that weak eigenstates are basically mass eigenstates.

Using the same methodology, we have calculated finite corrections to the decays of level-2 KK states into SM particles, including previously unknown couplings. Since the interactions violate KK number, only a finite number of diagrams contribute to the vertices and no additional regularization is necessary.
We then revisited the computation of branching fractions for level-2 KK particles.
For KK fermions the basic features remain the same as before with the addition of new decay modes opening up at the few-percent level. The largest effects appear in the decay of level-2 KK top quarks, {\it i.e.,} the branching fraction of the left-handed KK top quark into $t h$ is about 20--30\%. Branching fractions of level-2 gauge bosons are also updated. Overall, the decay widths of level-2 particles are observed to increase when these effects are included, but they are still narrower than the detector resolution.

Finally, we would like to make a few comments about other potentially interesting collider and dark matter phenomenology. In this paper, we showed results for the production of level-2 KK gauge bosons at the LHC. It is desirable to study these with a more detailed simulation, including single and pair production of level-2 KK fermions, and set bounds on ($R^{-1}$, $\Lambda R$) from various resonance searches, such as $V_2$ decays to $\ell \bar \ell$, $jj$, $W^+ W^-$, $W^\pm Z$, $Z Z$, $t \bar t$.
Here one can make use of boosted $W$, $Z$ and $t$ event topologies.

The collider phenomenology of singly produced level-2 KK fermions provides interesting signatures. For instance, searches for excited quarks in various final states would constrain level-2 decays such as $p p \to Q_2 / q_2 \to q^\prime V$, where $V$ = $\gamma$, $Z$, $W$ or $g$. $Q_2$ or $q_2$ could appear as a single three-jet resonance via $q^\prime V_2$ with $V_2 \to f_0 f_0^\prime$. 
Other interesting topologies involve the top quark and the Higgs. They may not provide the best sensitivity in a search for this particular model since certain signal-to-background ratios may be small. However they could serve as a benchmark model for various searches and provide useful search grounds. We list a few examples below. 
\begin{itemize}
\item $p p \to t_2 \bar{t}_2 $ with $t_2 \to t h$, $t_2 \to t g $ or $t_2 \to t \gamma$
\item $p p \to T_2 \bar{T}_2 $ with $T_2 \to t h$, $T_2 \to Z t$, $T \to t g $, $T \to t \gamma$ or $T_2 \to t \gamma_2$ $ (\gamma_2 \to t\bar t)$

\item $p p \to B_2 \bar{B}_2$ with $B_2 \to Z b$ or $B_2 \to W t$ (and small branching fractions to $B_2 \to g b$, $B_2 \to b \gamma$)
\item $p p \to b_2 \bar{b}_2$ with $b_2 \to g b$ (and small branching fractions to $b_2 \to \gamma b$, and $b_2 \to Z b$)

\item $p p \to Q_2,\,q_2,\,Q_2 \bar{Q}_2,\, q_2 \bar{q}_2$ (both single and pair production) with $Q_2 \to q g$ or $Q_2 \to q \gamma$

\item$T_2 \bar{T}_2 \to t \bar{t} h + X$ (inclusive $t\bar t h$ production)
\end{itemize}

As discussed earlier, level-1 KK particles are always produced in pairs due to KK parity and lead to signals with missing transverse momentum. Final states with jets + leptons + missing transverse momentum are known to provide stringent bounds on $R^{-1}$ (see Ref. \cite{uedlhc}). It is worth revisiting these analyses with our improved mass spectrum, since the broader mass pattern will lead to signal efficiency gains while at the same time the increased masses will reduce the production cross sections.

The computation of the relic abundance of KK dark matter has a rather long history \cite{ueddm,lev2dma,lev2dm}. Ref. \cite{lev2dm} includes both coannihilation and resonance effects, which play a crucial role in increasing the preferred mass scale of the KK photon. Our results imply that a slightly broader mass spectrum would reduce effective cross sections in the coannihilation processes 
(which are suppressed by $\frac{ e^{- x_f^\prime (m^\prime_1 - m^\prime_{\gamma_1})} } {e^{- x_f (m_1 - m_{\gamma_1})}}   \approx   e^{ -x_f  (m^\prime_1 -m_1 ) }$ where $x_f \approx x_f^\prime$ is the freeze-out temperature and $m_1$ is the mass of the coannihilating particle, $m_1^\prime$ is the improved mass, and $m_{\gamma_1}\approx m_{\gamma_1}^\prime$ is the mass of KK photon), pushing $m_{\gamma_1}$ to a lower value. However, $11 \to 2 0$ processes with the level-2 particle decaying to two zero modes would increase the effective annihilation cross section efficiently, increasing the preferred value for $m_{\gamma_1}$. This is a highly non-trivial and complicated exercise and we postpone it to a follow-up study.

We hope that our results will be useful for investigations of the phenomenology of universal extra dimensions and also provide interesting event topologies for various collider searches \cite{Craig:2016rqv}.

\noindent


\section*{Acknowledgments}

\noindent
The authors thank H.~C.~Cheng and M. Schmaltz for useful correspondence and
A. Pukhov for advice regarding questions about {\sc CalcHEP}.
This work has been supported in part by the National Science Foundation under
grant no.\ PHY-1519175 and in part by US Department of Energy under grant no.\ DE-SC0017965.


\appendix

\section{Feynman rules of MUED}
\label{sc:feynr}

\noindent
This appendix lists a complete set of Standard Model Lagrangian in a
universal extra dimensions model in 5 dimensions with a $S^1/\mathbb{Z}_2$
orbifold compactification. The conventions are chosen such that Greek indices
take values $0,1,2,3$, assigned to the uncompactified dimensions, while capital
Latin indices describe the full 5D theory, where the extra spatial dimension is
denoted as $x^5$ where necessary. The 5D coupling constants are labeled with a
superscript $(5)$ and are related to the 4D effective couplings through
\begin{align}
g &= \frac{g^{(5)}}{\sqrt{\pi R}}\,, &
h_i &= \frac{h_i^{(5)}}{\sqrt{\pi R}}\,, & 
\lambda &= \frac{\lambda^{(5)}}{\pi R}\,,
\end{align}
for the gauge, Yukawa and Higgs self coupling respectively.

Furthermore we have to define the conventions for the extension of the Clifford
algebra,
\begin{align}
\Gamma^M &= \left(\gamma^\mu, i\gamma^5\right)\,, 
\quad \text{such that} \quad \left\{\Gamma^M,\Gamma^N\right\} = 2 g^{MN},
\end{align}
where $g^{MN}$ is the 5D metric tensor
\begin{align}
g_{MN} &= 
\begin{pmatrix}
g_{\mu\nu} & 0 \\
0          & -1
\end{pmatrix} ,
\end{align}
and $g^{\mu\nu} = \text{diag}\{+---\}$ the usual 4D metric.
It is also helpful to define an extended set of $\Delta$ symbols
\cite{uedcomphep}
\begin{align}
&\Delta_{mnl}^1  = \delta_{l,m+n} + \delta_{n,l+m} + \delta_{m,l+n} \nonumber \, ,\\
&\Delta_{mnlk}^2 = \delta_{k,l+m+n} + \delta_{l,m+n+k} + \delta_{m,n+k+l}+ \delta_{n,k+l+m} +\delta_{k+m,l+n} + \delta_{k+l,m+n} +\delta_{k+n,l+m} \nonumber\, ,\\
&\Delta_{mnlk}^3 = - \delta_{k,l+m+n} - \delta_{l,m+n+k} - \delta_{m,n+k+l} -\delta_{n,k+l+m} +\delta_{k+l,m+n} + \delta_{k+m,l+n} +\delta_{k+n,l+m}\nonumber  \, ,\\
&\Delta_{mnl}^4 = -\delta_{l,m+n} + \delta_{n,l+m} + \delta_{m,l+n} \nonumber\, ,\\
&\Delta_{mnlk}^5 = - \delta_{k,l+m+n} - \delta_{l,m+n+k} + \delta_{m,n+k+l}+ \delta_{n,k+l+m} - \delta_{k+l,m+n} + \delta_{k+m,l+n} +\delta_{k+n,l+m} \, .
\end{align}

\subsection{The Gauge Sector}

\noindent
As a generic example, we show the gauge sector Lagrangian for a single vector field in
the adjoint representation of SU($N$), which contains a four-component vector
$V_\mu\left(x,x^5\right)$ and a fifth component $V_5\left(x,x^5\right)$, which takes the role of a Goldstone
boson. Additionally we require the ghost field $c\left(x,x^5\right)$.
After compactification the Lagrangians read
\begin{align}
\mathcal{L}_{\rm Gauge} =& \inte \left\{-\frac{1}{4} F^a_{MN}F^{a,MN}\right\} = \inte \left\{-\frac{1}{4} F^a_{\mu\nu}F^{a,\mu\nu}-\frac{1}{2} F^a_{5\nu}F^{a,5\nu}\right\} \nonumber\, ,\\
\mathcal{L}_{\rm GF} =& \inte \left\{-\frac{1}{2\xi}\left(\partial^\mu V^a_\mu\left(x,x^5\right) - \xi\partial_5 V_5^a\left(x,x^5\right)\right)^2\right\} \nonumber\, ,\\
\mathcal{L}_{\rm Ghost} =& \inte \left\{\overline{c}^a\left(x,x^5\right) \left(-\partial^\mu \partial_\mu +\xi \partial_5^2\right)c^a\left(x,x^5\right) \nonumber \right. \\
&\left. + g^{(5)} f^{abc}\left(-\partial^\mu\overline{c}^a\left(x,x^5\right) V_\mu^c\left(x,x^5\right) c^b\left(x,x^5\right)+\xi\partial_5\overline{c}^a\left(x,x^5\right)V_5^c c^b\left(x,x^5\right)\right)\right\} \; ,
\end{align}
with $\xi$ being the gauge parameter in the generalized $R_\xi$ gauge.
After decomposing the 5D fields into Fourier modes, according to 
\begin{align}
V_\mu \left(x,x^5\right) &= \frac{1}{\sqrt{\pi R}} \biggl [ V_\mu^{0}(x) + \sqrt{2} \summ
V_\mu^{n} (x) \,\cos \frac{n x^5}{R} \biggr  ] \, , \nonumber\\
V_5 \left(x,x^5\right) &= \sqrt{\frac{2}{\pi R}} \summ V_5^n (x) \,\sin \frac{n x^5}{R} \nonumber\, ,\\
c^a\left(x,x^5\right) &= \frac{1}{\sqrt{\pi R}} \biggl [ c^{0,a}(x) + \sqrt{2} \summ  c^{n,a}
(x) \,\cos \frac{n x^5}{R}  \biggr  ] \, ,
\end{align}
and performing the integral over the fifth dimension one obtains the effectively
4D pure Yang-Mills pieces, given by
\begin{align}
\inte \left ( F^a_{5\nu} F^{a5\nu} \right ) &= -\summ \left ( \frac{n}{R}V^{n,a}_\mu + \partial_\mu V^n_5 - g C^{abc}V^{n,b}_5 V^{0,c}_\mu  \right )^2  \nonumber\\
	&+ \summt \sqrt{2} g C^{ade} \left (\frac{m}{R}V^{m,a}_\mu +\partial_\mu V^{m,a}_5 -g C^{abc} V^{m,b}_5 V^{0,c}_\mu \right ) V^{n,d}_5 V^{l,e\mu} \Delta^4_{mnl}  \nonumber \\
	&- \frac{g^2}{2} C^{abc}C^{ade} \summf V^{m,b}_5 V^{n,c}_\mu V^{l,d}_5 V^{k,e\mu} \Delta^5_{mlnk}  \, , \displaybreak[0] \\
\inte \left ( F^a_{\mu\nu} F^{a\mu\nu} \right ) &= F^{0,a}_{\mu\nu} F^{0,a\mu\nu} + 2g C^{abc} \summ F^{0,a}_{\mu\nu} V^{n,b\mu} V^{n,c\nu}
		+ \summ \left ( \partial_\mu V^{n,a}_\nu - \partial_\nu V^{n,a}_\mu \right )^2   \nonumber \\
	&+ 2g C^{abc} \summ \left ( \partial_\mu V^{n,a}_\nu - \partial_\nu V^{n,a}_\mu \right ) \left ( V^{0,b\mu} V^{n,c\nu} + V^{0,c\nu} V^{n,b\mu} \right ) \nonumber \\
	&+ \sqrt{2} g C^{abc} \summt \left ( \partial_\mu V^{m,a}_\nu - \partial_\nu V^{m,a}_\mu \right ) V^{n,b\mu} V^{l,c\nu} \Delta^1_{mnl}  \nonumber \\
	&+ \summ g^2 (C^{abc} ( V^{0,b}_\mu V^{n,c}_\nu + V^{0,c}_\nu V^{n,b}_\mu ))^2 \nonumber \\
	&+ \sqrt{2} g^2 \summt C^{abc}C^{ade} \left ( V^{0,b}_\mu V^{m,c}_\nu + V^{0,c}_\nu V^{m,b}_\mu \right ) V^{n,d\mu} V^{l,e\nu} \Delta^1_{mnl}  \nonumber\\
	&+ \summf \frac{g^2}{2} C^{abc}C^{ade} V^{m,b}_\mu V^{n,c}_\nu V^{l,d\mu} V^{k,e\nu} \Delta^2_{mnlk}  ,
\end{align}
a gauge fixing part given by
\begin{equation}
\mathcal{L}_{\rm GF} = -\frac{1}{2\xi}\biggl[\left(\partial^\mu V_\mu^0\right)^2
+ \summ\Bigl(\partial^\mu V_\mu^n - \frac{\xi n }{R}V_5^n\Bigr)^2\biggr],
\end{equation}
and finally the ghost piece 
\begin{align}
\mathcal{L}_{\rm Ghost} =& -\overline{c}^{0,a}\partial^\mu\partial_\mu c^{0,a} - g f^{abc} \partial^\mu\overline{c}^{0,a}V_\mu^{0,c}c^{0,b}- \summ\left[\overline{c}^{n,a}\partial^\mu\partial_\mu c^{n,a} +\xi \frac{n^2}{R^2}\overline{c}^{n,a}c^{n,a}\right]\nonumber\\
&-gf^{abc} \summ\left[\partial^\mu\overline{c}^{0,a} V^{n,c}_\mu c^{n,b} +\partial^\mu\overline{c}^{n,a} V^{0,c}_\mu c^{n,b} +\partial^\mu\overline{c}^{n,a} V_\mu^{n,c} c^{0,b} +\xi\frac{n}{R}\overline{c}^{n,a} V_5^{n,b} c^{0,c}\right]\nonumber\\
&-\frac{g}{\sqrt{2}}f^{abc}\summt\left[\partial^\mu\overline{c}^{l,a} V^{m,c}_\mu c^{n,b}\Delta^1_{lmn} +\xi\frac{l}{R}\overline{c}^{l,a} V^{m,c}_5 c^{n,b}\Delta^4_{lmn} \right] .
\end{align}

\subsection{The Fermion Sector}

\noindent
Analogously, we show the Lagrangian for a fermion in the fundamental
representation coupled to a generic  SU($N$) gauge field. The structure of the
SM makes it necessary to distinguish between Fermions $\Psi$ that transform as
doublets under SU(2) and those that are singlets $\psi$. Their respective
decomposition is
\begin{align}
\Psi\left(x,x^5\right) &= \frac{1}{\sqrt{\pi R}} \left \{ \Psi_L (x) + \sqrt{2} \summ 
\left [ P_- \Psi_L^n (x) \,\cos \frac{n x^5}{R} + P_+ \Psi_R^n (x) \,\sin
\frac{n x^5}{R} \right ] \right \} , \nonumber \\
\psi\left(x,x^5\right) &= \frac{1}{\sqrt{\pi R}} \left \{ \psi_R (x) + \sqrt{2} \summ 
\left [ P_+ \psi_R^n (x) \,\cos \frac{n x^5}{R} + P_- \psi_L^n (x) \,\sin
\frac{n x^5}{R} \right ] \right \} .
\end{align}
The generic Lagrangian for either of the Fermions coupling to $V_M\left(x,x^5\right)$ can be
written as 
\begin{align}
{\cal L}_{\Psi}&=\inte \left \{ i \bar{\Psi}\left(x,x^5\right) \Gamma^M \left [ \partial_M  + i{g^{(5)}} V_M\left(x,x^5\right) \right ] \Psi\left(x,x^5\right) \right \} \nonumber  \\
&= i \bar{\Psi}_L \g^\mu \left ( \partial_\mu + i{g} V^0_\mu \right ) \Psi_L - \summ\frac{n}{R} \left [ \bar{\Psi}^n_R \Psi^n_L + \bar{\Psi}^n_L \Psi^n_R \right ] \nonumber \\
&+ \summ \left [ i \bar{\Psi}^n_R \g^\mu \left ( \partial_\mu + i{g} V^0_\mu \right ) \Psi^n_R +  i \bar{\Psi}^n_L \g^\mu \left ( \partial_\mu +  i{g} V^0_\mu \right ) \Psi^n_L -g \bar{\Psi}_L \g^\mu V^n_\mu \Psi^n_L + g\bar{q}_L i\g^5 V^n_5 \Psi^n_R\right ]\nonumber \\
&- \frac{g}{\sqrt{2}} \summt \left [ \bar{\Psi}^m_L \g^\mu V^n_\mu \Psi^l_L \Delta^1_{mnl} +
	\bar{\Psi}^m_R \g^\mu V^n_\mu \Psi^l_R \Delta^4_{mln} + \bar{\Psi}^m_L i \g^5 V^n_5 \Psi^l_R \Delta^4_{lnm} \right ] .
\end{align}

\subsection{The Higgs Sector}

\noindent
Due to the somewhat complicated structure of the four-point interactions between
the Higgs and electroweak gauge bosons, we here show the Higgs Lagrangian not
just for a generic
gauge group, but write the explicit Lagrangian for a Higgs doublet coupling to
the U(1)$_{\rm Y}$ field $B_M\left(x,x^5\right)$ and the SU(2)$_{\rm L}$ field $W_M\left(x,x^5\right)$.

The Higgs doublet $\Phi\left(x,x^5\right)$ is decomposed as 
\begin{align}
\Phi\left(x,x^5\right) &= \frac{1}{\sqrt{\pi R}} \left \{ \Phi_0(x) + \sqrt{2} \summ  
 \Phi_n (x) \, \cos \frac{n x^5}{R}  \right  \} 
\end{align}
and inserted into the Lagrangian
\begin{align}
{\cal L}_{\rm Higgs} &= \inte \left [ \left ( D_M \Phi\left(x,x^5\right) \right )^\dagger \left ( D^M \Phi\left(x,x^5\right) \right ) + \mu^2 \Phi^\dagger \left(x,x^5\right)
	\Phi\left(x,x^5\right) - \lambda \left ( \Phi^\dagger \left(x,x^5\right) \Phi\left(x,x^5\right) \right )^2 \right ] 
	 \nonumber \displaybreak[0] \\
&= \left [ \left ( \partial_\mu + i {g_2} W^0_\mu + \frac{i {g_1}}{2} B^0_\mu \right ) \Phi_0 \right ]^\dagger
	\left [ \left ( \partial^\mu + i {g_2} W^{0\mu} + \frac{i {g_1}}{2} B^{0\mu} \right )\Phi_0 \right ]\nonumber  \\
&+ \summ \left [ \left ( \partial_\mu + i {g_2} W^0_\mu + \frac{i {g_1}}{2} B^0_\mu \right ) \Phi_n \right ]^\dagger
	\left [ \left ( \partial^\mu + i {g_2} W^{0\mu} + \frac{i {g_1}}{2} B^{0\mu} \right ) \Phi_n \right ] 
	 \nonumber \displaybreak[0] \\
&+ i {g_2} \summ \left [ ( \partial^\mu \Phi_0 )^\dagger W^n_\mu \Phi_n +
	( \partial^\mu \Phi_n )^\dagger W^n_\mu \Phi_0 - \Phi_n^\dagger {W^n_\mu}^\dagger (\p^\mu \Phi_0) -
	\Phi_0^\dagger {W^n_\mu}^\dagger (\p^\mu \Phi_n ) \right ] \nonumber \\
&+ i {g_2} \summt \left [ ( \partial^\mu \Phi_m )^\dagger W^n_\mu \Phi_l -
	\Phi_l^\dagger {W^n_\mu}^\dagger (\p^\mu \Phi_m) \right ] \Delta^1_{mnl} \nonumber \\
&+ \frac{i {g_1}}{2} \summ \left [ ( \p^\mu \Phi_0 )^\dagger B^n_\mu \Phi_n +
	( \p^\mu \Phi_n )^\dagger B^n_\mu \Phi_0 - \Phi_n^\dagger B^n_\mu (\p^\mu \Phi_0) -
	\Phi_0^\dagger B^n_\mu (\p^\mu \Phi_n ) \right ] \nonumber \\
&+ \frac{i {g_1}}{2\sqrt{2}} \summt \left [ ( \p^\mu \Phi_m )^\dagger B^n_\mu \Phi_l -
	\Phi_l^\dagger B^n_\mu (\p^\mu \Phi_m) \right ] \Delta^1_{mnl} 
\nonumber \displaybreak[0] \\
&+ g_2^2 \summ \left [ \Phi_0^\dagger {W^0_\mu}^\dagger W^{n\mu} \Phi_n +
	\Phi_n^\dagger {W^0_\mu}^\dagger W^{n\mu} \Phi_0 + \Phi_0^\dagger {W^{n\mu}}^\dagger W^0_\mu \Phi_n +
	\Phi_0^\dagger {W^{n\mu}}^\dagger W^n_\mu \Phi_0 + \Phi_n^\dagger {W^{n\mu}}^\dagger W^0_\mu \Phi_0 \right ] \nonumber \\
&+ \frac{g_2^2}{\sqrt{2}} \summt \left [ \Phi_m^\dagger {W^0_\mu}^\dagger W^{n\mu} \Phi_l +
	\Phi_0^\dagger {W^m_\mu}^\dagger W^{l\mu} \Phi_n + \Phi_m^\dagger {W^n_\mu}^\dagger W^{0\mu} \Phi_l +
	\Phi_m^\dagger {W^n_\mu}^\dagger W^{l\mu} \Phi_0 \right ] \Delta^1_{mnl} \nonumber \\
&+ \frac{g_2^2}{2} \summf \left [ \Phi_m^\dagger {W^n_\mu}^\dagger W^{l\mu} \Phi_k \right ] \Delta^2_{mnlk} \nonumber
 \displaybreak[0] \\
&+ \frac{{g_1} {g_2}}{2} \summ \left [ \Phi_0^\dagger {W^0_\mu}^\dagger B^{n\mu} \Phi_n + \Phi_n^\dagger {W^0_\mu}^\dagger B^{n\mu} \Phi_0
	+ \Phi_0^\dagger {W^n_\mu}^\dagger B^{0\mu} \Phi_n + \Phi_0^\dagger {W^n_\mu}^\dagger B^{n\mu} \Phi_0
	+ \Phi_n^\dagger {W^n_\mu}^\dagger B^{0\mu} \Phi_0 \right. \nonumber \\
	&\hspace{2.5 cm}+ \Phi_n^\dagger B^{n\mu} W^0_\mu \Phi_0 + \Phi_0^\dagger B^{n\mu} W^0_\mu \Phi_n
	+ \Phi_n^\dagger B^{0\mu} W^n_\mu \Phi_0 + \Phi_0^\dagger B^{n\mu} W^n_\mu \Phi_0 + \Phi_0^\dagger B^{0\mu} W^n_\mu \Phi_n \left. \right ] \nonumber \\
&+ \frac{{g_1} {g_2}}{2\sqrt{2} } \summt \left [ \Phi_m^\dagger {W^0_\mu}^\dagger B^{n\mu} \Phi_l
	+ \Phi_0^\dagger {W^m_\mu}^\dagger B^{l\mu} \Phi_n + \Phi_m^\dagger {W^n_\mu}^\dagger B^{0\mu} \Phi_l
	+ \Phi_m^\dagger {W^n_\mu}^\dagger B^{l\mu} \Phi_0 \right . \nonumber \\
	& \hspace{2.5 cm} + \Phi_l^\dagger B^{n\mu} W^0_\mu \Phi_m + \Phi_n^\dagger B^{l\mu} W^m_\mu \Phi_0
	+ \Phi_l^\dagger B^{0\mu} W^n_\mu \Phi_m + \Phi_0^\dagger B^{l\mu} W^n_\mu \Phi_m \left. \right ] \Delta^1_{mnl}\nonumber \\
&+ \frac{{g_1} {g_2}}{4 } \summf \left [ \Phi_m^\dagger {W^n_\mu}^\dagger B^{l\mu} \Phi_k
	+ \Phi_k^\dagger B^{l\mu} W^n_\mu \Phi_m \right ] \Delta^2_{mnlk} \notag 
 \displaybreak[0] \\
&+ \frac{g_1^2}{4} \summ \left [ 2 \Phi_0^\dagger B^0_\mu B^{n\mu} \Phi_n + 2 \Phi_n^\dagger B^0_\mu B^{n\mu} \Phi_0 +
	 \Phi_0^\dagger B^n_\mu B^{n\mu} \Phi_0 \right ] \nonumber \\
&+ \frac{g_1^2}{4\sqrt{2} } \summt \left [ 2 \Phi_m^\dagger B^0_\mu B^{n\mu} \Phi_l + \Phi_0^\dagger B^m_\mu B^{l\mu} \Phi_n +
	 \Phi_m^\dagger B^n_\mu B^{l\mu} \Phi_0 \right ] \Delta^1_{mnl} \nonumber \\
&+ \frac{g_1^2}{4} \summf \left [ \Phi_m^\dagger B^n_\mu B^{l\mu} \Phi_k \right ] \Delta^2_{mnlk} 
\nonumber \displaybreak[0] \\
&+ \mu^2 \left [ \Phi_0^\dagger \Phi_0 + \summ \Phi_n^\dagger \Phi_n \right ]
	- \summ \left ( \frac{n}{R}\right )^2 \Phi_n^\dagger \Phi_n 
\nonumber \displaybreak[0] \\
&+ \frac{1}{\sqrt{2}} \summt \frac{n}{R} \Phi_n^\dagger \left ( \frac{i{g_1}}{2} B^m_5 + i{g_2} {W^m_5} \right ) \Phi_l \Delta^4_{mnl} \nonumber \\
&- \frac{1}{\sqrt{2}} \summt \frac{m}{R} \Phi_l^\dagger \left ( \frac{i{g_1}}{2} B^n_5 + i{g_2} W^n_5 \right ) \Phi_m \Delta^4_{mnl}
\nonumber \displaybreak[0] \\
&- \frac{1}{2} \summf \Phi_l^\dagger \left ( \frac{i{g_1}}{2} B^m_5 + i{g_2} {W^m_5}^\dagger \right )
	\left ( \frac{i{g_1}}{2} B^n_5 + i{g_2} W^n_5 \right ) \Phi_k \Delta^5_{mnlk} \nonumber \\
& -\summ\left[\Phi_0^\dagger\left(g_2 W_5^n + \frac{g_1}{2}B_5^n\right)^2\Phi_0 + \frac{n}{R}\Phi_0^\dagger\left(g_2 W_5^n + \frac{g_1}{2}B_5^n\right)\Phi_n - \frac{n}{R}\Phi_n^\dagger\left(g_2 W_5^n + \frac{g_1}{2}B_5^n\right)\Phi_0 \right]\nonumber \displaybreak[0] \\
&-\frac{1}{\sqrt{2}}\summt\left[\Phi_0^\dagger\left(g_2 W_5^k +
\frac{g_1}{2}B_5^k\right)\left(g_2 W_5^l + \frac{g_1}{2}B_5^l\right)\Phi_m
 \right. \notag \\
 &\hspace{2.5 cm}\left. +\Phi_m^\dagger\left(g_2 W_5^k + \frac{g_1}{2}B_5^k\right)\left(g_2 W_5^l + \frac{g_1}{2}B_5^l\right)\Phi_0\right]\Delta^4_{klm}
 \nonumber\displaybreak[0] \\
&- \lambda\left ( \Phi_0^\dagger \Phi_0 \right )^2 - \lambda\summ \left [\left(\Phi_0^\dagger \Phi_n+ \Phi_n^\dagger \Phi_0\right)^2 +2 \Phi_0^\dagger \Phi_0\Phi_n^\dagger \Phi_n\right ]\nonumber \\
&- \sqrt{2}\lambda\summt \Phi_m^\dagger \Phi_n \left(\Phi_l^\dagger \Phi_0 + \Phi_0^\dagger \Phi_l\right)\Delta^1_{mnl}
	- \frac{\lambda}{2} \summf \Phi_m^\dagger \Phi_n \Phi_l^\dagger \Phi_k \Delta^2_{mnlk} \, .
\end{align}

\subsection{The Yukawa Sector}

\noindent
To ensure that the SM Fermions acquire a mass through EWSB one has
to consider the Yukawa couplings to the Higgs field. For a down-type fermion  
they are described by
\begin{align}
\mathcal{L}_{\rm Yukawa} &= \inte \left \{ h^{(5)}_i \bar{\Psi}\left(x,x^5\right) \psi\left(x,x^5\right)\Phi\left(x,x^5\right) \right \} 
 \notag \\
&= h_i \bar{\Psi}_L \psi_R \Phi_0
+ h_i\summ \left [ \bar{\Psi}^n_L \psi^n_R \Phi_0 + \bar{\Psi}^n_R \psi^n_L
\Phi_0 \right ] 
+ h_i \summ \left [ \bar{\Psi}_L \psi^n_R \Phi_n + \bar{\Psi}^n_L \psi_R  \Phi_n\right ] \nonumber\\
&+ \frac{h_i}{\sqrt{2}} \summt \left [ \bar{\Psi}^n_L \psi^m_R \Phi_l\Delta^1_{mnl} +
	\bar{\Psi}^n_R \psi^m_L \Phi_l \Delta^4_{mnl}  \right ] .
\end{align}
and for an up-type fermion they can be constructed in complete analogy.


\section{KK-number violating couplings in MUED}
\label{app:int}

\noindent
In this appendix, the KK-number violating couplings discussed in
section~\ref{sc:int} are shown for the MUED extension of the SM. Here 
$g_{1,2,3}$ are the couplings of the SM U(1)$_{\rm Y}$, SU(2)$_{\rm
L}$ and SU(3)$_{\rm C}$ gauge groups, respectively, while $h_t$ is the top
Yukawa coupling and $\lambda_H$ the Higgs self-coupling. The $L_n$ is defined as $L_n \equiv \ln (\Lambda^2/m_n^2)$.

\paragraph{\boldmath $\bar{\psi}_0$--$\psi_0$--$V_2^\mu$ coupling:}
$-iC_{\psi_0\psi_0V_2} \gamma^\mu T^a P_\pm\,$
\begin{align}
C_{Q_0Q_0G_2} &= \frac{\sqrt{2} g_3}{64\pi^2} \biggl [
\begin{aligned}[t]
 &g_3^2 \biggl ( 11L_1 + 35 - \frac{11\pi^2}{3} \biggr ) +
  g_2^2 \biggl ( -\frac{27}{4}L_1 - \frac{39}{4} + \frac{21\pi^2}{16} \biggr ) \\
 &+g_1^2 \biggl ( -\frac{1}{4}L_1 - \frac{13}{36} + \frac{7\pi^2}{144} \biggr )
  \biggr ]
\end{aligned} \\
C_{t_{L0}t_{L0}G_2} &= C_{b_{L0}b_{L0}G_2} \notag \\
&= \frac{\sqrt{2} g_3}{64\pi^2} \biggl [
\begin{aligned}[t]
 &g_3^2 \biggl ( 11L_1 + 35 - \frac{11\pi^2}{3} \biggr ) +
  g_2^2 \biggl ( -\frac{27}{4}L_1 - \frac{39}{4} + \frac{21\pi^2}{16} \biggr ) \\
 &+g_1^2 \biggl ( -\frac{1}{4}L_1 - \frac{13}{36} + \frac{7\pi^2}{144} \biggr ) +
  h_t^2 \biggl ( L_1 -1 + \frac{\pi^2}{4} \biggr )
  \biggr ]
\end{aligned} \displaybreak[0] \\
C_{u_0u_0G_2} &= \frac{\sqrt{2} g_3}{64\pi^2} \biggl [
 g_3^2 \biggl ( 11L_1 + 35 - \frac{11\pi^2}{3} \biggr ) +
 g_1^2 \biggl ( -4L_1 - \frac{52}{9} + \frac{7\pi^2}{9} \biggr ) \biggr ] \\
C_{t_{R0}t_{R0}G_2} &= \frac{\sqrt{2} g_3}{64\pi^2} \biggl [
\begin{aligned}[t]
 &g_3^2 \biggl ( 11L_1 + 35 - \frac{11\pi^2}{3} \biggr ) +
  g_1^2 \biggl ( -4L_1 - \frac{52}{9} + \frac{7\pi^2}{9} \biggr ) \\
 &+ h_t^2 \biggl ( 2L_1 - 2 + \frac{\pi^2}{2} \biggr )\biggr ] 
\end{aligned} \\
C_{d_0d_0G_2} &= \frac{\sqrt{2} g_3}{64\pi^2} \biggl [
 g_3^2 \biggl ( 11L_1 + 35 - \frac{11\pi^2}{3} \biggr ) +
 g_1^2 \biggl ( -L_1 - \frac{13}{9} + \frac{7\pi^2}{36} \biggr ) \biggr ] 
 \displaybreak[0] \\
C_{Q_0Q_0Z_2} &= \frac{\sqrt{2} g_2}{64\pi^2} \biggl [
\begin{aligned}[t]
 &g_3^2 \biggl ( -12L_1 - \frac{52}{3} + \frac{7\pi^2}{3} \biggr ) +
  g_2^2 \biggl ( \frac{33}{4}L_1 + \frac{299}{12} - \frac{43\pi^2}{16} \biggr ) \\
 &+g_1^2 \biggl ( -\frac{1}{4}L_1 - \frac{13}{36} + \frac{7\pi^2}{144} \biggr )
  \biggr ]
\end{aligned} \displaybreak[0] \\
C_{t_{L0}t_{L0}Z_2} &= C_{b_{L0}b_{L0}Z_2} \notag \\
&= \frac{\sqrt{2} g_2}{64\pi^2} \biggl [
\begin{aligned}[t]
 &g_3^2 \biggl ( -12L_1 - \frac{52}{3} + \frac{7\pi^2}{3} \biggr ) +
  g_2^2 \biggl ( \frac{33}{4}L_1 + \frac{299}{12} - \frac{43\pi^2}{16} \biggr ) \\
 &+g_1^2 \biggl ( -\frac{1}{4}L_1 - \frac{13}{36} + \frac{7\pi^2}{144} \biggr ) +
  h_t^2 \biggl ( L_1 - 3 + \frac{\pi^2}{4} \biggr )
  \biggr ]
\end{aligned} \displaybreak[0] \\
C_{L_0L_0Z_2} &= \frac{\sqrt{2} g_2}{64\pi^2} \biggl [
 g_2^2 \biggl ( \frac{33}{4}L_1 + \frac{299}{12} - \frac{43\pi^2}{16} \biggr ) +
 g_1^2 \biggl ( -\frac{9}{4}L_1 - \frac{13}{4} + \frac{7\pi^2}{16} \biggr ) 
 \biggr ] \displaybreak[0] \\
C_{Q_0Q_0B_2} &= \frac{\sqrt{2} g_1}{64\pi^2} \biggl [
\begin{aligned}[t]
 &g_3^2 \biggl ( -12L_1 - \frac{52}{3} + \frac{7\pi^2}{3} \biggr ) +
  g_2^2 \biggl ( -\frac{27}{4}L_1 - \frac{39}{4} + \frac{21\pi^2}{16} \biggr ) \\
 &+g_1^2 \biggl ( -\frac{7}{12}L_1 - \frac{7}{12} + \frac{7\pi^2}{144} \biggr )
  \biggr ]
\end{aligned} \\
C_{t_{L0}t_{L0}B_2} &= C_{b_{L0}b_{L0}B_2} \notag \\
&= \frac{\sqrt{2} g_1}{64\pi^2} \biggl [
\begin{aligned}[t]
 &g_3^2 \biggl ( -12L_1 - \frac{52}{3} + \frac{7\pi^2}{3} \biggr ) +
  g_2^2 \biggl ( -\frac{27}{4}L_1 - \frac{39}{4} + \frac{21\pi^2}{16} \biggr ) \\
 &+g_1^2 \biggl ( -\frac{7}{12}L_1 - \frac{7}{12} + \frac{7\pi^2}{144} \biggr ) +
  h_t^2 \biggl ( L_1 + 5 + \frac{\pi^2}{4} \biggr )
  \biggr ]
\end{aligned} \displaybreak[0] \\
C_{u_0u_0B_2} &= \frac{\sqrt{2} g_1}{64\pi^2} \biggl [
 g_3^2 \biggl ( -12L_1 - \frac{52}{3} + \frac{7\pi^2}{3} \biggr ) +
 g_1^2 \biggl ( -\frac{13}{3}L_1 - 6 + \frac{7\pi^2}{9} \biggr ) \biggr ] \\
C_{t_{R0}t_{R0}B_2} &= \frac{\sqrt{2} g_1}{64\pi^2} \biggl [
\begin{aligned}[t]
 &g_3^2 \biggl ( -12L_1 - \frac{52}{3} + \frac{7\pi^2}{3} \biggr ) +
  g_1^2 \biggl ( -\frac{13}{3}L_1 - 6 + \frac{7\pi^2}{9} \biggr ) \\
 &+ h_t^2 \biggl ( 2L_1 - 5 + \frac{\pi^2}{2} \biggr ) \biggr ] 
\end{aligned} \displaybreak[0] \\
C_{d_0d_0B_2} &= \frac{\sqrt{2} g_1}{64\pi^2} \biggl [
 g_3^2 \biggl ( -12L_1 - \frac{52}{3} + \frac{7\pi^2}{3} \biggr ) +
 g_1^2 \biggl ( -\frac{4}{3}L_1 - \frac{5}{3} + \frac{7\pi^2}{36} \biggr ) \biggr ] 
\displaybreak[0] \\
C_{L_0L_0B_2} &= \frac{\sqrt{2} g_1}{64\pi^2} \biggl [
 g_2^2 \biggl ( -\frac{27}{4}L_1 - \frac{39}{4} + \frac{21\pi^2}{16} \biggr ) 
 +g_1^2 \biggl ( -\frac{31}{12}L_1 - \frac{125}{36} + \frac{7\pi^2}{16} \biggr )
  \biggr ] \\
C_{e_0e_0B_2} &= \frac{\sqrt{2} g_1^3}{64\pi^2} 
 \biggl ( -\frac{28}{3}L_1 - \frac{119}{9} + \frac{7\pi^2}{4} \biggr ) \biggr ]
\end{align}

\paragraph{\boldmath $\bar{\psi}_2$--$\psi_0$--$V_0^\mu$ coupling:}
 $-i\tilde{C}_{\psi_2\psi_0V_0}\gamma^\mu P_\pm\quad$ [$V_0$ transverse]

\vspace{2ex}\noindent
Note that $\tilde{C}$ is defined without $T^a$, in contrast to
eq.~\eqref{eq:f2v}. In the expressions below, $A$ is an SU(3) color index.
\begin{align}
\tilde{C}_{Q_2Q_0G_0} &= \frac{\sqrt{2} g_3}{64\pi^2} \,T^A \biggl [
 g_3^2 \biggl ( -\frac{2}{3} + \frac{3\pi^2}{4} \biggr ) + 
 3g_2^2 + \frac{g_1^2}{9} \biggr ] \quad \text{[including $Q=T,B$]} \\
\tilde{C}_{u_2u_0G_0} &= \frac{\sqrt{2} g_3}{64\pi^2} \,T^A \biggl [
 g_3^2 \biggl ( -\frac{2}{3} + \frac{3\pi^2}{4} \biggr ) + 
 \frac{16}{9}g_1^2 \biggr ] \qquad\;\; \text{[including $u=t$]} \\
\tilde{C}_{d_2d_0G_0} &= \frac{\sqrt{2} g_3}{64\pi^2} \,T^A \biggl [
 g_3^2 \biggl ( -\frac{2}{3} + \frac{3\pi^2}{4} \biggr ) + 
 \frac{4}{9}g_1^2 \biggr ] \displaybreak[0] \\
\tilde{C}_{Q_2Q_0Z_0} &= \frac{\sqrt{2} (\pm\frac{1}{2}g_2 {c}_{\rm W}-\frac{1}{6}g_1 s_{\rm W})}{64\pi^2} \biggl [
 \frac{16}{3}g_3^2 + 
 3g_2^2 +
 \frac{g_1^2}{9} \biggr ] \pm 
 \frac{\sqrt{2} g_2^3 c_{\rm W}}{128\pi^2} \biggl (\frac{\pi^2}{2}-4 \biggr ) 
\label{eq:ffz} \\
&\hspace{20em} \text{[including $Q=T,B$]} \notag \displaybreak[0] \\[1ex]
\tilde{C}_{Q_2Q_0\gamma_0} &= \frac{\sqrt{2} (\pm\frac{1}{2}g_2 s_{\rm W}+
 \frac{1}{6}g_1 c_{\rm W})}{64\pi^2} \biggl [
 \frac{16}{3}g_3^2 + 
 3g_2^2 +
 \frac{g_1^2}{9} \biggr ] \pm 
 \frac{\sqrt{2} g_2^3 s_{\rm W}}{128\pi^2} \biggl (\frac{\pi^2}{2}-4 \biggr ) 
 \\
&\hspace{20em} \text{[including $Q=T,B$]} \notag \displaybreak[0] \\[1ex]
-\frac{1}{s_{\rm W}}\tilde{C}_{u_2u_0Z_0} &= \frac{1}{c_{\rm W}}\tilde{C}_{u_2u_0\gamma_0}
 = \frac{\sqrt{2} g_1}{96\pi^2} \biggl [
 \frac{16}{3}g_3^2 + 
 \frac{16}{9}g_1^2 \biggr ]  \qquad\;\; \text{[including $u=t$]} \\
-\frac{1}{s_{\rm W}}\tilde{C}_{d_2d_0Z_0} &= \frac{1}{c_{\rm W}}\tilde{C}_{d_2d_0\gamma_0}
 = -\frac{\sqrt{2} g_1}{192\pi^2} \biggl [
 \frac{16}{3}g_3^2 + 
 \frac{4}{9}g_1^2 \biggr ] \displaybreak[0] \\
\tilde{C}_{L_2L_0Z_0} &= \frac{\sqrt{2} (\pm\frac{1}{2}g_2 c_{\rm W}+\frac{1}{2}g_1 s_{\rm W})}{64\pi^2} \bigl [
 3g_2^2 +
 g_1^2 \bigr ] \pm 
 \frac{\sqrt{2} g_2^3 c_{\rm W}}{128\pi^2} \biggl (\frac{\pi^2}{2}-4 \biggr ) \\
\tilde{C}_{L_2L_0\gamma_0} &= \frac{\sqrt{2} (\pm\frac{1}{2}g_2 s_{\rm W}-\frac{1}{2}g_1 c_{\rm W})}{64\pi^2} \bigl [
 3g_2^2 +
 g_1^2 \bigr ] \pm 
 \frac{\sqrt{2} g_2^3 s_{\rm W}}{128\pi^2} \biggl (\frac{\pi^2}{2}-4 \biggr ) \\
-\frac{1}{s_{\rm W}}\tilde{C}_{e_2e_0Z_0} &= \frac{1}{c_{\rm W}}\tilde{C}_{e_2e_0\gamma_0}
 = -\frac{\sqrt{2} g_1^3}{16\pi^2} 
\end{align}
In eqs.~\eqref{eq:ffz}~ff., the $\pm$ signs indicate the upper/lower entry of a fermion
doublet.

\paragraph{\boldmath $\bar{\psi}_2$--$\psi_0$--$V_0^\mu$ coupling:}
 $-\tilde{D}_{\psi_2\psi_0V_0}\frac{\sigma^{\mu\nu}q_\nu}{2 m_{KK}}
P_\pm$

\vspace{2ex}\noindent
Note that $\tilde{D}$ is defined without $T^a$, in contrast to
eq.~\eqref{eq:dip}. In the expressions below, $A$ is an SU(3) color index.
\begin{align}
\tilde{D}_{Q_2Q_0G_0} &= \frac{\sqrt{2} g_3}{64\pi^2} \,T^A \biggl [
 g_3^2 ( -17 + 2\pi^2 ) + 
 g_2^2 \biggl ( \frac{9}{4} - \frac{9\pi^2}{16} \biggr ) +
 g_1^2 \biggl ( \frac{1}{12} - \frac{\pi^2}{48} \biggr ) \biggr ] \\
\tilde{D}_{T_2t_{L0}G_0} &= \tilde{D}_{B_2b_{L0}G_0} = 
 \frac{\sqrt{2} g_3}{64\pi^2}\,T^A \biggl [
 g_3^2 ( -17 + 2\pi^2 ) + 
 g_2^2 \biggl ( \frac{9}{4} - \frac{9\pi^2}{16} \biggr ) \notag \\
 &\hspace{11em} +
 g_1^2 \biggl ( \frac{1}{12} - \frac{\pi^2}{48} \biggr ) +
 h_t^2 \biggl ( \frac{\pi^2}{4}-1 \biggr ) \biggr ] \\[1ex]
\tilde{D}_{u_2u_0G_0} &= \frac{\sqrt{2} g_3}{64\pi^2}\,T^A \biggl [
 g_3^2 ( -17 + 2\pi^2 ) + 
 g_1^2 \biggl ( \frac{4}{3} - \frac{\pi^2}{3} \biggr ) \biggr ] \\
\tilde{D}_{t_2t_{R0}G_0} &= \frac{\sqrt{2} g_3}{64\pi^2}\,T^A \biggl [
 g_3^2 ( -17 + 2\pi^2 ) + 
 g_1^2 \biggl ( \frac{4}{3} - \frac{\pi^2}{3} \biggr ) +
 h_t^2 \biggl ( \frac{\pi^2}{2}-2 \biggr ) \biggr ] \\
\tilde{D}_{d_2d_0G_0} &= \frac{\sqrt{2} g_3}{64\pi^2}\,T^A \biggl [
 g_3^2 ( -17 + 2\pi^2 ) + 
 g_1^2 \biggl ( \frac{1}{3} - \frac{\pi^2}{12} \biggr ) \biggr ]
 \displaybreak[0] \\
\label{eq:ffz2}
\tilde{D}_{Q_2Q_0Z_0} &= \frac{\sqrt{2} (\pm\frac{1}{2}g_2 c_{\rm W}-\frac{1}{6}g_1 s_{\rm W})}{64\pi^2} \biggl [
 g_3^2 ( 4-\pi^2 ) + 
 g_2^2 \biggl ( \frac{9}{4} - \frac{9\pi^2}{16} \biggr ) +
 g_1^2 \biggl ( \frac{1}{12} - \frac{\pi^2}{48} \biggr ) \biggr ] \notag \\
&\quad \pm \frac{\sqrt{2} g_2^3 c_{\rm W}}{128\pi^2} (-14+2\pi^2)  \\[1ex]
\tilde{D}_{T_2t_{L0}Z_0}/\tilde{D}_{B_2b_{L0}Z_0} &= \frac{\sqrt{2} (\pm\frac{1}{2}g_2 c_{\rm W}-\frac{1}{6}g_1 s_{\rm W})}{64\pi^2} \biggl [
 g_3^2 ( 4-\pi^2 ) + 
 g_2^2 \biggl ( \frac{9}{4} - \frac{9\pi^2}{16} \biggr ) +
 g_1^2 \biggl ( \frac{1}{12} - \frac{\pi^2}{48} \biggr ) \biggr ] \notag \\
&\quad \pm \frac{\sqrt{2} g_2^3 c_{\rm W}}{128\pi^2} (-14+2\pi^2) \notag \\[1ex]
&\quad + \frac{\sqrt{2} h_t^2}{64\pi^2} \biggl [ 
 \pm \frac{g_2c_{\rm W}}{2} \biggl ( 3 - \frac{\pi^2}{4} \biggr ) -
 \frac{g_1s_{\rm W}}{6} \biggl ( -13 + \frac{7\pi^2}{4} \biggr )
 \biggr ] \displaybreak[0] \\[1ex]
\tilde{D}_{Q_2Q_0\gamma_0} &= \frac{\sqrt{2} (\pm \frac{1}{2}g_2 s_{\rm W}+\frac{1}{6}g_1 c_{\rm W})}{64\pi^2} \biggl [
 g_3^2 ( 4-\pi^2 ) + 
 g_2^2 \biggl ( \frac{9}{4} - \frac{9\pi^2}{16} \biggr ) +
 g_1^2 \biggl ( \frac{1}{12} - \frac{\pi^2}{48} \biggr ) \biggr ] \notag \\
&\quad \pm \frac{\sqrt{2} g_2^3 s_{\rm W}}{128\pi^2} (-14+2\pi^2)\\[1ex]
\tilde{D}_{T_2t_{L0}\gamma_0}/\tilde{D}_{B_2b_{L0}\gamma_0} &= \frac{\sqrt{2} (\pm \frac{1}{2}g_2 s_{\rm W}+\frac{1}{6}g_1 c_{\rm W})}{64\pi^2} \biggl [
 g_3^2 ( 4-\pi^2 ) + 
 g_2^2 \biggl ( \frac{9}{4} - \frac{9\pi^2}{16} \biggr ) +
 g_1^2 \biggl ( \frac{1}{12} - \frac{\pi^2}{48} \biggr ) \biggr ] \notag \\
&\quad \pm \frac{\sqrt{2} g_2^3 s_{\rm W}}{128\pi^2} (-14+2\pi^2) \notag \\
&\quad + \frac{\sqrt{2} h_t^2}{64\pi^2} \biggl [ 
 \pm \frac{g_2s_{\rm W}}{2} \biggl ( 3 - \frac{\pi^2}{4} \biggr ) +
 \frac{g_1c_{\rm W}}{6} \biggl ( -13 + \frac{7\pi^2}{4} \biggr )
 \biggr ] \displaybreak[0] \\[1ex]
-\frac{1}{s_{\rm W}}\tilde{D}_{u_2u_0Z_0} &= \frac{1}{c_{\rm W}}\tilde{D}_{u_2u_0\gamma_0}
 = \frac{\sqrt{2} g_1}{96\pi^2} \biggl [
 g_3^2 ( 4-\pi^2 ) + 
 g_1^2 \biggl ( \frac{4}{3} - \frac{\pi^2}{3} \biggr ) \biggr ] \\
-\frac{1}{s_{\rm W}}\tilde{D}_{t_2t_{R0}Z_0} &= 
 \frac{1}{c_{\rm W}}\tilde{D}_{t_2t_{R0}\gamma_0}
 = \frac{\sqrt{2} g_1}{96\pi^2} \biggl [
 g_3^2 ( 4-\pi^2 ) + 
 g_1^2 \biggl ( \frac{4}{3} - \frac{\pi^2}{3} \biggr ) +
 h_t^2 \biggl ( - \frac{\pi^2}{4} \biggr ) \biggr ] \\
-\frac{1}{s_{\rm W}}\tilde{D}_{d_2d_0Z_0} &= \frac{1}{c_{\rm W}}\tilde{D}_{d_2d_0\gamma_0}
 = -\frac{\sqrt{2} g_1}{192\pi^2} \biggl [
 g_3^2 ( 4-\pi^2 ) + 
 g_1^2 \biggl ( \frac{1}{3} - \frac{\pi^2}{12} \biggr ) \biggr ] 
 \displaybreak[0] \\
\tilde{D}_{L_2L_0Z_0} &= \frac{\sqrt{2} (\pm\frac{1}{2}g_2 c_{\rm W}+\frac{1}{2}g_1 s_{\rm W})}{64\pi^2} \biggl [
 g_2^2 \biggl ( \frac{9}{4} - \frac{9\pi^2}{16} \biggr ) +
 g_1^2 \biggl ( \frac{3}{4} - \frac{3\pi^2}{16} \biggr ) \biggr ] \notag \\
&\quad \pm \frac{\sqrt{2} g_2^3 c_{\rm W}}{128\pi^2} (-14+2\pi^2) \\
\tilde{D}_{L_2L_0\gamma_0} &= \frac{\sqrt{2} (\pm\frac{1}{2}g_2 s_{\rm
W}-\frac{1}{2}g_1 c_{\rm W})}{64\pi^2} \biggl [
 g_2^2 \biggl ( \frac{9}{4} - \frac{9\pi^2}{16} \biggr ) +
 g_1^2 \biggl ( \frac{3}{4} - \frac{3\pi^2}{16} \biggr ) \biggr ] \notag \\
&\quad \pm \frac{\sqrt{2} g_2^3 s_{\rm W}}{128\pi^2} (-14+2\pi^2)
 \displaybreak[0] \\
-\frac{1}{s_{\rm W}}\tilde{D}_{e_2e_0Z_0} &= \frac{1}{c_{\rm W}}\tilde{D}_{e_2e_0\gamma_0}
 = -\frac{\sqrt{2} g_1^3}{64\pi^2} \biggl ( 3 - \frac{3\pi^2}{4} \biggr ) 
\end{align}
In eqs.~\eqref{eq:ffz2}~ff., the $\pm$ signs indicate the upper/lower entry of a fermion
doublet.

\paragraph{\boldmath $V_2^{\mu,a}(p)$--$V_0^{\nu,b}(k_1)$--$V_0^{\rho,c}(k_2)$ 
coupling:}
\begin{align}
f_{abc} \Bigl\{
 &\bigl [ g_{\mu\nu}(p-k_1)_\rho +
          g_{\nu\rho}(k_1-k_2)_\mu +
	  g_{\rho\mu}(k_2-p)_\nu \bigr ]
  C_{V_2V_0V_0} \notag \\
 &+\bigl [ -g_{\mu\nu}k_{1,\rho} + g_{\rho\mu}k_{2,\nu} \bigr ] 
  D_{V_2V_0V_0} 
 + g_{\nu\rho}(k_1-k_2)_\mu \, E_{V_2V_0V_0} \Bigr\} \qquad
 [\text{all momenta incoming}] \notag
\end{align}
\begin{align}
C_{G_2G_0G_0} &= \frac{3\sqrt{2} g_3^3}{64\pi^2}
 \biggl ( -\frac{157}{9} + \frac{7\pi^2}{6}  \biggr ) \\
D_{G_2G_0G_0} &= \frac{3\sqrt{2} g_3^3}{64\pi^2}
 \biggl ( \frac{91}{6}-\pi^2 \biggr ) \\
E_{G_2G_0G_0} &= \frac{3\sqrt{2} g_3^3}{64\pi^2}
 \biggl ( \frac{38}{3} - \frac{3\pi^2}{4} \biggr ) \displaybreak[0] \\
C_{Z_2W_0^+W_0^-} &= \frac{1}{c_{\rm W}}C_{W_2^-W_0^+Z_0}
 = -\frac{1}{s_{\rm W}}C_{W_2^-W_0^+\gamma_0} = \frac{i\sqrt{2} g_2^3}{64\pi^2}
 \biggl ( - \frac{316}{9} + \frac{85\pi^2}{36} \biggr ) \\
D_{Z_2W_0^+W_0^-} &= \frac{1}{c_{\rm W}}D_{W_2^-W_0^+Z_0}
 = -\frac{1}{s_{\rm W}}D_{W_2^-W_0^+\gamma_0} = \frac{i\sqrt{2} g_2^3}{64\pi^2}
 \biggl ( \frac{92}{3} - \frac{49\pi^2}{24} \biggr ) \\
E_{Z_2W_0^+W_0^-} &= \frac{1}{c_{\rm W}}E_{W_2^-W_0^+Z_0}
 = -\frac{1}{s_{\rm W}}E_{W_2^-W_0^+\gamma_0} = \frac{i\sqrt{2} g_2^3}{64\pi^2}
 \biggl ( \frac{38}{3} - \frac{3\pi^2}{4} \biggr )
\end{align}



\begin{thebibliography}{99}
\frenchspacing

\bibitem{ued}
  T.~Appelquist, H.~C.~Cheng and B.~A.~Dobrescu,
  Phys.\ Rev.\ D {\bf 64}, 035002 (2001)
  [hep-ph/0012100].


\bibitem{ueddm}
  H.~C.~Cheng, J.~L.~Feng and K.~T.~Matchev,
  Phys.\ Rev.\ Lett.\  {\bf 89}, 211301 (2002)
  [hep-ph/0207125];
  G.~Servant and T.~M.~P.~Tait,
  Nucl.\ Phys.\ B {\bf 650}, 391 (2003)
  [hep-ph/0206071];
  F.~Burnell and G.~D.~Kribs,
  Phys.\ Rev.\ D {\bf 73}, 015001 (2006)
  [hep-ph/0509118];
  K.~Kong and K.~T.~Matchev,
  JHEP {\bf 0601}, 038 (2006)
  [hep-ph/0509119].
  S.~Arrenberg, L.~Baudis, K.~Kong, K.~T.~Matchev and J.~Yoo,
  Phys.\ Rev.\ D {\bf 78}, 056002 (2008)
  [arXiv:0805.4210 [hep-ph]].

\bibitem{lev2dma}
  M.~Kakizaki, S.~Matsumoto, Y.~Sato and M.~Senami,
  Phys.\ Rev.\ D {\bf 71}, 123522 (2005)
  [hep-ph/0502059];
  M.~Kakizaki, S.~Matsumoto, Y.~Sato and M.~Senami,
  Nucl.\ Phys.\ B {\bf 735}, 84 (2006)
  [hep-ph/0508283].

\bibitem{lev2dm}
  M.~Kakizaki, S.~Matsumoto and M.~Senami,
  Phys.\ Rev.\ D {\bf 74}, 023504 (2006)
  [hep-ph/0605280];
  G.~Belanger, M.~Kakizaki and A.~Pukhov,
  JCAP {\bf 1102}, 009 (2011)
  [arXiv:1012.2577 [hep-ph]].



\bibitem{cms2}
  H.~C.~Cheng, K.~T.~Matchev and M.~Schmaltz,
  Phys.\ Rev.\ D {\bf 66}, 056006 (2002)
  [hep-ph/0205314].

\bibitem{uedpheno}
  T.~G.~Rizzo,
  Phys.\ Rev.\ D {\bf 64}, 095010 (2001)
  [hep-ph/0106336];
  C.~Macesanu, C.~D.~McMullen and S.~Nandi,
  Phys.\ Rev.\ D {\bf 66}, 015009 (2002)
  [hep-ph/0201300];
  J.~M.~Smillie and B.~R.~Webber,
  JHEP {\bf 0510}, 069 (2005)
  [hep-ph/0507170].

\bibitem{Cembranos:2006gt}
  J.~A.~R.~Cembranos, J.~L.~Feng and L.~E.~Strigari,
  Phys.\ Rev.\ D {\bf 75}, 036004 (2007)
  [hep-ph/0612157].

\bibitem{kkgluon}
  A.~Freitas and D.~Wiegand,
  JHEP {\bf 1709}, 058 (2017)
  [arXiv:1706.09442 [hep-ph]].


\bibitem{uedlhc}
  G.~Aad {\it et al.} [ATLAS Collaboration],
  JHEP {\bf 1504}, 116 (2015)
  [arXiv:1501.03555 [hep-ex]];
  N.~Deutschmann, T.~Flacke and J.~S.~Kim,
  Phys.\ Lett.\ B {\bf 771}, 515 (2017)
  [arXiv:1702.00410 [hep-ph]];
  J.~Beuria, A.~Datta, D.~Debnath and K.~T.~Matchev,
  arXiv:1702.00413 [hep-ph].


\bibitem{cms}
  H.~C.~Cheng, K.~T.~Matchev and M.~Schmaltz,
  Phys.\ Rev.\ D {\bf 66}, 036005 (2002)
  [hep-ph/0204342].

\bibitem{uedlevel2}
  A.~Datta, K.~Kong and K.~T.~Matchev,
  Phys.\ Rev.\ D {\bf 72}, 096006 (2005)
  [Erratum: Phys.\ Rev.\ D {\bf 72}, 119901 (2005)]
  [hep-ph/0509246].


\bibitem{uedmass}
  H.~Georgi, A.~K.~Grant and G.~Hailu,
  Phys.\ Lett.\ B {\bf 506}, 207 (2001)
  [hep-ph/0012379].


\bibitem{uedcutoff} 
  Z.~Chacko, M.~A.~Luty and E.~Ponton,
  JHEP {\bf 0007}, 036 (2000)
  [hep-ph/9909248].
  G.~Bhattacharyya, A.~Datta, S.~K.~Majee and A.~Raychaudhuri,
  Nucl.\ Phys.\ B {\bf 760} (2007) 117
  [hep-ph/0608208].
  R.~S.~Chivukula, D.~A.~Dicus and H.~J.~He,
  Phys.\ Lett.\ B {\bf 525} (2002) 175
  [hep-ph/0111016].

  
\bibitem{uedreview} 
  D.~Hooper and S.~Profumo,
  Phys.\ Rept.\  {\bf 453}, 29 (2007)
  [hep-ph/0701197].


\bibitem{nmued}
  T.~Flacke, A.~Menon and D.~J.~Phalen,
  Phys.\ Rev.\ D {\bf 79}, 056009 (2009)
  [arXiv:0811.1598 [hep-ph]];
  T.~Flacke, K.~Kong and S.~C.~Park,
  Mod.\ Phys.\ Lett.\ A {\bf 30}, 1530003 (2015)
  [arXiv:1408.4024 [hep-ph]].
  T.~Flacke, K.~Kong and S.~C.~Park,
  JHEP {\bf 1305}, 111 (2013)
  [arXiv:1303.0872 [hep-ph]].

\bibitem{uedcomphep}
  A.~Datta, K.~Kong and K.~T.~Matchev,
  New J.\ Phys.\  {\bf 12}, 075017 (2010)
  [arXiv:1002.4624 [hep-ph]].

\bibitem{Belyaev:2012ai} 
  A.~Belyaev, M.~Brown, J.~Moreno and C.~Papineau,
  JHEP {\bf 1306}, 080 (2013)
  [arXiv:1212.4858 [hep-ph]].

\bibitem{feynarts}
  T.~Hahn,
  Comput.\ Phys.\ Commun.\  {\bf 140}, 418 (2001)
  [hep-ph/0012260].

\bibitem{feyncalc}
  V.~Shtabovenko, R.~Mertig and F.~Orellana,
  Comput.\ Phys.\ Commun.\  {\bf 207}, 432 (2016)
  [arXiv:1601.01167 [hep-ph]].

\bibitem{Pumplin:2002vw} 
  J.~Pumplin, D.~R.~Stump, J.~Huston, H.~L.~Lai, P.~M.~Nadolsky and W.~K.~Tung,
  JHEP {\bf 0207}, 012 (2002)
  [hep-ph/0201195].

\bibitem{Belyaev:2012qa} 
  A.~Belyaev, N.~D.~Christensen and A.~Pukhov,
  Comput.\ Phys.\ Commun.\  {\bf 184}, 1729 (2013)
  [arXiv:1207.6082 [hep-ph]].
  
    
 
\bibitem{Craig:2016rqv} 
  N.~Craig, P.~Draper, K.~Kong, Y.~Ng and D.~Whiteson,
  arXiv:1610.09392 [hep-ph].
  
\end{thebibliography}
\end{document}